\documentclass[12pt,onepage,oneside]{article}
\usepackage{fullpage}
\usepackage{amsmath,amsthm}
\usepackage{color,soul}
\usepackage{natbib}
\usepackage{hyperref}
\usepackage{booktabs}
\usepackage{amsfonts}
\usepackage{graphicx}
\usepackage[export]{adjustbox}
\graphicspath{ {figures/} }
\usepackage{cleveref}
\usepackage{pdflscape}
\usepackage{multirow}
\usepackage[small,bf]{caption}
\usepackage{subcaption}
\usepackage[stable]{footmisc}
\usepackage{bm}
\usepackage{booktabs}
\usepackage{lineno}
\usepackage{makecell}
\usepackage{subcaption}
\usepackage{rotating}
\usepackage{pdflscape}

\usepackage{setspace,caption}
\usepackage[margin=1in]{geometry}

\usepackage{cancel}

\usepackage{calc}
\newlength{\depthofsumsign}
\DeclareMathOperator{\Cov}{Cov}

\DeclareMathOperator{\cum}{cum}
\DeclareMathOperator{\Var}{Var}

\newcommand{\ee}{\ensuremath{\mathrm{e}}}
\newcommand{\ii}{\ensuremath{\mathrm{i}}}

\newcommand{\IC}{\ensuremath{\mathbb{C}}}

\newcommand{\IH}{\ensuremath{\mathbb{H}}}

\newcommand{\IL}{\ensuremath{\mathbb{L}}}
\newcommand{\IN}{\ensuremath{\mathbb{N}}}
\newcommand{\IP}{\ensuremath{\mathbb{P}}}
\newcommand{\IR}{\ensuremath{\mathbb{R}}}
\newcommand{\IZ}{\ensuremath{\mathbb{Z}}}

\newcommand{\bfrakR}{\ensuremath{\boldsymbol{\beta}}}

\newcommand{\bX}{\ensuremath{\bm{x}}}

\newcommand{\ff}[2]{\mathfrak{f}_{#1, #2}}

\newcommand{\BB}[2]{\bm{B}_{#1, #2}}
\newcommand{\HH}[2]{\IH_{#1, #2}}
\newcommand{\LL}[2]{\IL_{#1, #2}}

\renewcommand{\emph}[1]{#1}

\newtheorem{theorem}{Theorem}
\newtheorem{assumption}{Assumption}
\newtheorem{proposition}{Proposition}

\newtheorem{lemma}{Lemma}

\title{Quantile Spectral Beta: A Tale of Tail Risks, Investment Horizons, and Asset Prices \thanks{We are grateful to the editor, and two anonymous referees for their useful comments and suggestions, which have greatly improved the paper. We appreciate helpful comments from Allan Timmermann, Tobias Kley, Michal Kejak, Ev\v zen Ko\v cenda, Luk\' a\v s V\' acha, and participants at various seminars, workshops and conferences. Support from the Czech Science Foundation under the 19-28231X (EXPRO) project is gratefully acknowledged. For estimation of quantile spectral betas, we provide package \texttt{QSbeta} in \textsf{R} available at \url{https://github.com/barunik/QSbeta}.}}

\author{%
Jozef {\sc Barun\'{i}k}$^{\rm a,b}$\thanks{Corresponding author, Tel. +420(776)259273, Email address: barunik@fsv.cuni.cz}, and
Mat\v ej {\sc Nevrla}$^{\rm a,b}$
\vspace{5mm} \\
 \small $^{\rm a}$ Institute of Economic Studies, Charles University, \vspace{-0.5mm}\\  \
 \small Opletalova 26, 110 00, Prague, Czech Republic \vspace{3mm} \\
 \small $^{\rm b}$ Department of Econometrics, IITA, The Czech Academy of Sciences, \vspace{-0.5mm}\\
  \small Pod Vodarenskou Vezi 4, 182 00, Prague, Czech Republic}

\begin{document}
\maketitle

\begin{abstract}
\noindent 

This paper investigates how two important sources of risk -- market tail risk and extreme market volatility risk -- are priced into the cross-section of asset returns across various investment horizons. To identify such risks, we propose a \textit{quantile spectral beta} representation of risk based on the decomposition of covariance between indicator functions that capture fluctuations over various frequencies. We study the asymptotic behavior of the proposed estimators of such risk. Empirically, we find that tail risk is a short-term phenomenon, whereas extreme volatility risk is priced by investors in the long term when pricing a cross-section of individual stocks. In addition, we study popular industry, size and value, profit, investment or book-to-market portfolios, as well as portfolios constructed from various asset classes, portfolios sorted on cash flow duration and other strategies. These results reveal that tail-dependent and horizon-specific risks are priced heterogeneously across datasets and are important sources of risk for investors. \\

\end{abstract}

\section{Introduction}
\label{sec:introduction}

The classical conclusion of the asset pricing literature states that the price of an asset should be equal to its expected discounted payoff. In the capital asset pricing model (CAPM) introduced by \cite{sharpe1964capital}, \cite{lintner1965security}, \cite{black1972capital}, we assume that the stochastic discount factor can be approximated by return on market portfolio; thus, expected excess returns can be fully described by their market betas based on covariance between asset return and market return. While early empirical evidence validated this prediction, decades of consequent research have called the ability of the traditional market beta to explain cross-sectional variation in returns into question. We aim to show that to understand the formation of expected returns, one has to look deeper into the features of asset returns that are crucial in terms of the preferences of a representative investor. We argue that two important, risk related features are tail events and frequency-specific (spectral) risk capturing behavior at different investment horizons. To characterize such general risks, we derive a novel \textit{quantile spectral} representation of beta that captures covariation between indicator functions capturing fluctuations of different parts of joint risky asset and market return distributions over various frequencies. Nesting the traditional beta, the new representation captures \textit{tail}-specific as well as horizon- or frequency-specific \textit{spectral} risks.

Intuitively, covariation stemming from (extremely) negative returns of risky assets and (extremely) negative returns of the market that are known as downside risk in the literature should be positively compensated. While early literature \citep{ang2006downside} empirically confirms the premium for bearing downside risk, \cite{levi2019symmetric} concludes that estimated downside betas do not provide superior predictions compared to standard betas. More recently, \cite{bollerslev2020realized} argue that we need to look at finer representations allowing combinations of positive and negative assets and market returns and suggest how such semibetas are priced.

The aim of this paper is to show that there is heterogeneity in the weights that investors assign to the risk for different investment horizons and different parts of the distribution of their future wealth. We argue that previous attempts have failed to fully account for more subtle implications arising from these heterogeneities. An asset drop that covaries with a drop in the market and, at the same time, is a low-frequency event with large persistence should be priced by investors differently than such extreme situations due to high-frequency, transitory events. While in the first situation investors will be pricing a persistent crash resulting in long-term fluctuations in the quantiles of the market's and risky asset's joint distribution, in the latter case the investor cares about the transitory crash resulting in short-run fluctuations. This essentially means that a covariance between the risky asset and discount factor will not only be different across all parts of the joint distribution but will also be different across various investment horizons. Intuitively, these co-occurrences of tail events will have either short-term or long-term effects on the marginal utility of investors. Looking at the beta representation that will capture such information empirically will also be informative for the rare disaster literature \citep{barro2006rare}.

Economists have long recognized that decisions under risk are more sensitive to changes in the probability of possible extreme events compared to the probability of a typical event. The expected utility might not reflect this behavior since it weighs the probability of outcomes linearly. Consequently, CAPM beta as an aggregate measure of risk may fail to explain the cross-section of asset returns. Several alternative notions have emerged in the literature. \cite{mao1970models} presents survey evidence showing that decision-makers tend to think of risk in terms of the possibility of outcomes below some target. For an expected utility-maximizing investor, \cite{bawa1977capital} has provided a theoretical rationale for using a lower partial moment as a measure of portfolio risk. Based on the rank-dependent expected utility due to \cite{yaari1987dual}, \cite{polkovnichenko2013probability} introduce utility with probability weights and derive the corresponding pricing kernel. As mentioned earlier, \cite{ang2006downside,lettau2014conditional} argue that downside risk -- the risk of negative returns -- is priced across asset classes and is not captured by CAPM betas. Furthermore, \cite{farago2017downside} extend the results using a general equilibrium model based on generalized disappointment aversion and show that downside risks in terms of market return and market volatility are priced in the cross-section of asset returns.\footnote{In addition, it is interesting to note that equity and variance risk premiums are also associated with compensation for jump tail risk \citep{bollerslev2011tails}. A more general exploration of the asymmetry of stock returns is provided by \cite{ghysels2016invest}, who propose a quantile-based measure of conditional asymmetry and show that stock returns from emerging markets are positively skewed. \cite{conrad2013ex} use option price data and find a relation between stock returns and their skewness. Another notable approach uses high-frequency data to define realized semivariance as a measure of downside risk \citep{barndorff2008measuring}. From a risk-measure standpoint, handling negative events, especially rare events, is a highly relevant theme in both practice and academia. The most prominent example is value-at-risk \citep{adrian2016covar,engle2004caviar}.}

The results described above lead us to question the role of expected utility maximizers in asset pricing. A recent strand of literature solves the problem by considering the quantile of utility instead of its expectation. This strand of literature complements the previously described empirical findings focusing on downside risk, as it highlights the notion of economic agents particularly averse to outcomes below some threshold compared to outcomes above this threshold. The concept of a quantile maximizer and its features was pioneered by \cite{manski1988ordinal}, and later axiomatized by \cite{rostek2010quantile}. Most recently, \cite{https://doi.org/10.3982/ECTA15146} developed a quantile optimizer model in a dynamic setting. A different approach to emphasizing investors’ aversion toward less favorable outcomes defines the theory based on Choquet expectations. This approach is based on a distortion function that alters the probability distribution of future outcomes by accentuating probabilities associated with the least desirable outcomes. This approach was utilized in finance, for example, by \cite{bassett2004pessimistic}.   

Whereas aggregating linearly weighted outcomes may not reflect the sensitivity of investors to tail risk, aggregating linearly weighted outcomes over various frequencies or economic cycles also may not reflect risk specific to different investment horizons. One may suspect that an investor cares differently about short-term and long-term risk according to their preferred investment horizon. Distinguishing between long-term and short-term dependence between economic variables has proven to be insightful since the introduction of cointegration \citep{engle1987co}. The frequency decomposition of risk thus uncovers another important feature of risk that cannot be captured solely by market beta, which captures risk averaged over all frequencies. This recent approach to asset pricing enables us to shed light on asset returns and investor behavior from a different point of view, highlighting heterogeneous preferences. Empirical justification is brought by \cite{boons2015horizon} and \cite{bandi2017horizon}, who show that exposure in long-term returns to innovations in macroeconomic growth and volatility of the matching half-life is significantly priced in a variety of asset classes. The results are based on the decomposition of time series into components of different persistence proposed by \cite{ortu2013long}. \cite{piccotti2016portfolio} further sets the portfolio optimization problem into the frequency domain using matching of the utility frequency structure and portfolio frequency structure, and \cite{chaudhuri2016spectral} present an approach to constructing a mean-variance-frequency optimal portfolio. This optimization yields the mean-variance optimal portfolio for a given frequency band, and thus it optimizes the portfolio for a given investment horizon.

From a theoretical point of view, preferences derived by \cite{epstein1989substitution} enable the study of frequency aspects of investor preferences and this has quickly become a standard in the asset pricing literature. With the important results of \cite{bansal2004risks}, long-run risk started to enter asset pricing discussions. \cite{dew2016asset} investigate frequency-specific prices of risk for various models and conclude that cycles longer than the business cycle are significantly priced in the market. Other papers utilize the frequency domain and Fourier transform to facilitate estimation procedures for parameters hard to estimate using conventional approaches. \cite{berkowitz2001generalized} generalizes band spectrum regression and enables the estimation of dynamic rational expectation models matching data only in particular ways, for example, forcing estimated residuals to be close to white noise. \cite{dew2016risky} proposes a spectral density estimator of the long-run standard deviation of consumption growth, which is a key component for determining risk premiums under Epstein-Zin preferences and shows superior performance compared to the previous approaches. \cite{crouzet2017multi} developed a model of a multifrequency trade set in the frequency domain and showed that restricting trading frequencies reduces price informativeness at those frequencies, reduces liquidity and increases return volatility. One of the rare exceptions that entertains the idea of combining horizon-specific risk with tail events is \cite{BARRO20211}, who show that most of the risk premium is attributable to rare event risk, but the long-run risk component contributes to fitting the Sharpe ratio as well.

The debate clearly indicates that the standard assumptions leading to classical asset pricing models do not correspond with reality. In this paper, we suggest that more general pricing models have to be defined and should take into consideration both the asymmetry of the dependence structure among the stock market and the relation of asymmetry to different investor behaviors at various investment horizons.

The main contribution of this paper is threefold. First, based on the frequency decomposition of covariance between indicator functions, we define the \textit{quantile spectral beta} of an asset capturing frequency-specific tail risks and corresponding ways of measuring the beta. The newly defined notion of a beta can be viewed as a disaggregation of a classical beta to a frequency- and tail-specific beta. With this notion, we describe how extreme market risks are priced in the cross-section of asset returns at various horizons. We define frequency-specific tail market risk that captures dependence between extremely low market and asset returns, as well as extreme market volatility risk that is characterized by dependence between extremely high increments of market volatility and extremely low asset returns. Second, we empirically motivate the emergence of such types of risks in the cross-section of asset returns. Third, we estimate models and document these types of risks on a wide number of popular datasets, including Fama-French industry, size and value, profit, investment and book-to-market portfolio, as well as portfolios constructed from various asset classes and sorted on cash flow durations.

The results of this paper suggest that tail risk is consistently priced in the cross-section of asset returns in the short term, while extreme market volatility risk is priced mainly in the long term. The result also holds when we control for popular factors, including moment-based factors that are designed to capture asymmetric features and popular downside risk models \citep{ang2006downside,lettau2014conditional,farago2017downside}. We also discuss how our new beta representation relates to other risk measures. Finally, we document that the final model capturing tail-specific risks across horizons significantly outperforms the other competing models that capture downside risks.

The rest of the paper is structured as follows. Section \ref{sec:qs_risk_motivation} motivates the importance of tail risks across horizons. Section \ref{sec:qs_risk} introduces the estimation of quantile spectral betas and discusses the asymptotic theory for the estimators, Section \ref{sec:pricing} defines the empirical models used for testing the significance of extreme risks, and Section \ref{sec:testing} conducts the empirical analysis on individual stocks as well as on various portfolios. Section \ref{sec:conclusion} then concludes. In the Appendix, we detail the main technical results regarding the quantile spectral betas, their relation to the rare disaster model, specifications of the competing measures of risk, and detailed results from the portfolio estimations. For estimation of quantile spectral betas, we provide package \texttt{QSbeta} in \textsf{R} available at \url{https://github.com/barunik/QSbeta}. Quantile spectral and cross-spectral densities as well as other quantities can be estimated using package \texttt{quantspec} in \textsf{R} available at \url{https://github.com/tobiaskley/quantspec} introduced by \cite{JSSv070i03}.


\section{Motivation: Why Should We Care About Tail Risks across Horizons}
\label{sec:qs_risk_motivation}

The empirical search for an explanation of why different assets earn different average returns centers around the use of return factor models arising from the Euler equation. With only the assumption of `no arbitrage', a stochastic discount factor $m_{t+1}$ exists, and under the expected utility maximization framework, for the $i$th excess return, $r_{i,t+1}$ satisfies $\mathbb{E}[m_{t+1}r_{i,t+1}] = 0$, hence
\begin{equation}
\label{eqn:eq1}
\mathbb{E}[r_{i,t+1}] = \frac{\mathbb{C}ov(m_{t+1}, r_{i,t+1})}{\mathbb{V}ar(m_{t+1})} \bigg(-\frac{\mathbb{V}ar(m_{t+1})}{\mathbb{E}[m_{t+1}]}\bigg) = \beta_{i}\lambda
\end{equation}
where loading $\beta_{i}$ can be interpreted as exposure to systematic risk factors and $\lambda$ as the risk price associated with factors. Equation \ref{eqn:eq1} assumes that the risk premium of an asset or a portfolio can be explained by its covariance with some reference economic or financial variable such as consumption growth or return on market portfolio. This simple pricing relation also assumes that independent common sources of systematic risk exist in the economy, and exposure to them can explain the cross-section of asset returns.\footnote{For example, this is the cornerstone of arbitrage pricing theory (APT) of \cite{ROSS1976341}.} This leads to the so-called factor fishing phenomenon, which tries to identify other risk factors in addition to the traditional market factors assumed by CAPM using a linear combination of factors that are assumed to have nonzero covariance with a risky asset, and to be independent of each other.

Covariance between the two variables of interest,
\begin{align} \label{eq:cov}
\gamma^k_{i,j} = \mathbb{C}ov \big(r_{j,t+k}, r_{i,t} \big) \equiv \mathbb{E}[(r_{j,t+k} - \bar{r}_{j}) (r_{i,t} - \bar{r}_i)],
\end{align}
which is central to the asset pricing literature, may not be sufficient in cases in which the investor cares about different parts of the distribution of her future wealth differently or in cases in which an investor cares about specific investment horizons. The empirical literature silently assumes that the risk factors aggregate information over the distribution of returns as well as investment horizons. Part of the literature tracing back to early work by \cite{roy1952safety,markowitz1952portfolio,hogan1974toward,bawa1977capital} argues that the reason we do not empirically find support for the above thinking is that the pricing relationship is fundamentally too simplistic. If investors are averse to volatility only when it leads to losses, not gains, the total variance as a relevant measure of risk should be disaggregated.

Later work by \cite{ang2006downside,lettau2014conditional,farago2017downside} shows that investors require an additional premium as compensation for exposures to disappointment-related risk factors called downside risk. Recently, \cite{lu2019bear} argued that bear risk capturing the left tail outcomes is even more important, and \cite{bollerslev2020realized} introduced betas based on semicovariances. In contrast to the promising results, \cite{levi2019symmetric} conclude that estimated downside betas do not provide superior predictions compared to standard aggregated betas, partially due to the difficulties of accurately determining downside betas from daily returns. With a similar argument of an overly simplistic pricing relation, another strand of the literature looks at frequency decomposition and explores the fact that risk factors of claims on the consumption risk should be frequency dependent since consumption has strong cyclical components \citep{bandi2018measuring, dew2016asset}. 

More recently, a new stream of literature led by \cite{https://doi.org/10.3982/ECTA15146} assumes agents have quantile preferences. In asset pricing, such an investor prefers future streams of quantiles of utilities leading to $q_{t, \tau} \big( m_{\tau, t+1} (1 + r_{i, t+1}) - 1 \big) = 0$. Assuming quantile preferences, our focus shifts from the search for the best proxy for a discount factor toward the capturing of the general dependence structures that reveal such flexible preferences. Measures we introduce in this paper allow us to identify risks associated with this type of preference.

Recognizing departures from overly simplistic assumptions in the data, we need to examine more general dependence measures since a simple covariance aggregating dependence across distributions as well as investment horizons will not be a sufficient measure of (in)dependence.

\begin{figure}
        \centering
        \caption{\footnotesize \textit{Dependence Structure between the Market and SMB and MOM Factor Portfolios}. Plots display covariance in the tail and across horizons defined by Eq. \ref{eq:ind_beta} that measures the general dependence between the market return and the SMB and MOM factors, respectively. Dashed lines represent 95\% confidence intervals under the null hypothesis that the two series are jointly normally distributed correlated random variables. Data are sampled with monthly frequency.}
        \includegraphics[width=\textwidth]{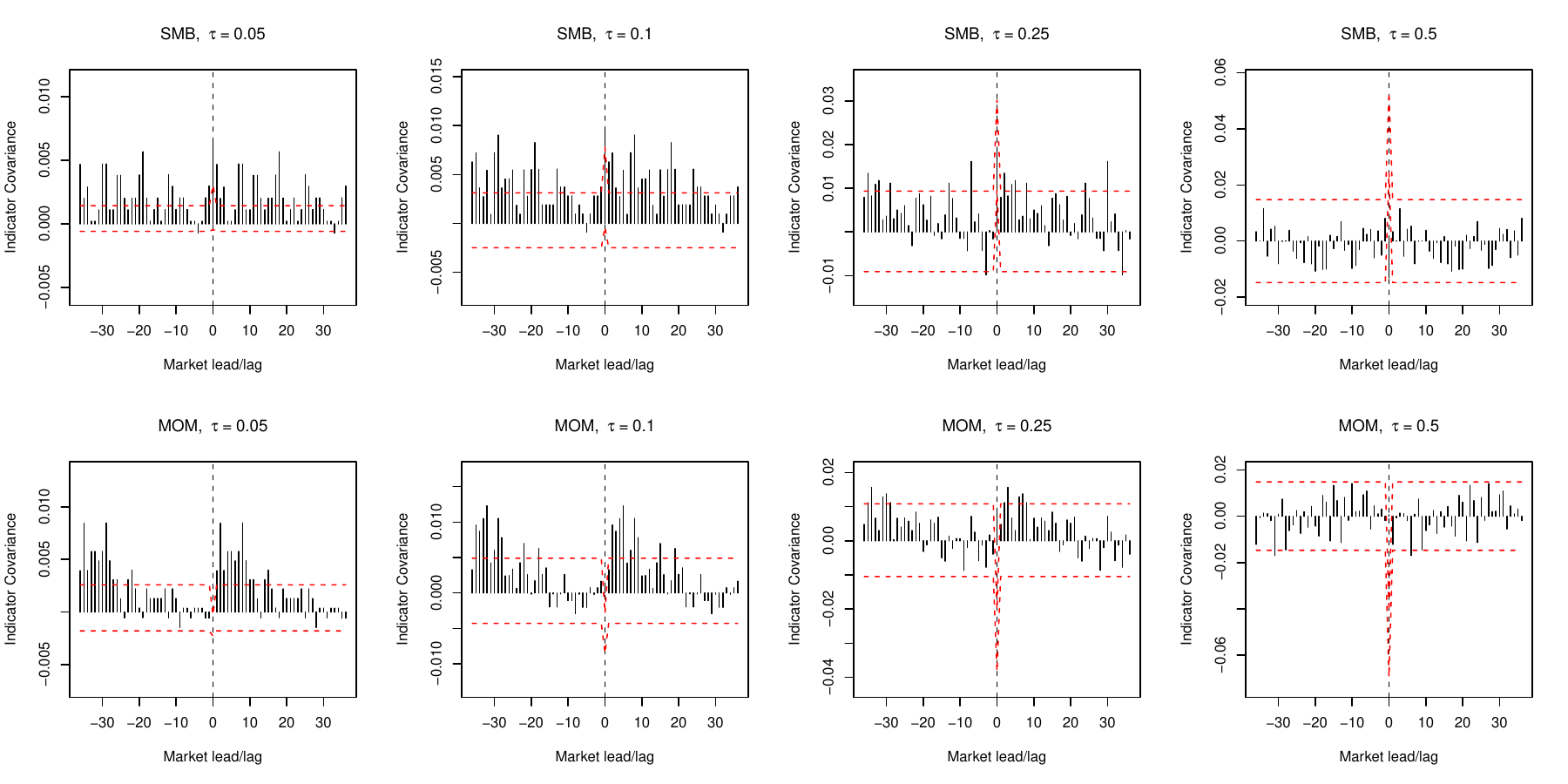}
    \label{fig:fac_dependence}
\end{figure}

To illustrate this discussion, we consider dependence between market returns and a popular small-minus-big portfolio (SMB) as well as momentum portfolio (MOM). While the literature assumes that these factors represent two independent sources of risk with contemporaneous correlation between them and the market being rather small, investigating the dependence in various parts of their joint distribution across different lags and leads reveals interesting relations. Instead of aggregate covariance between the market return and a factor portfolio, Figure \ref{fig:fac_dependence} depicts tail- and lead/lag-specific covariation for a threshold value given by $\tau$-quantile of the market return and a given lead/lag $k$ of the following form:
\begin{align}
\label{eq:ind_beta}
\mathbb{C}ov \big( I\{ r_{m,t-k} \leq q_{r_m}(\tau)) \}, I\{ r_{i,t} \leq q_{r_m}(\tau) \} \big),
\end{align}
where $r_{m,t}$ is the return of the market factor, $r_{i,t}$ is the return of either the SMB or the MOM portfolio, $I\{.\}$ is an indicator function and $q_{r_m}$ is the quantile function of the market return. This simple measure captures the probability of both returns being below some threshold value in some time interval given by lead/lag $k$. This can be seen from the fact that $\mathbb{C}ov \big( I\{ r_{m,t-k} \leq q_{r_m}(\tau)) \}, I\{ r_{i,t} \leq q_{r_m}(\tau) \} \big) =  Pr \big\{ r_{m,t-k} \leq q_{m}(\tau), r_{i,t} \leq q_{m}(\tau) \big\} - \tau \tau_i$. Therefore, this dependence essentially measures additional probability over the independence copula of both variables being below some threshold value.

Looking at the median dependence of market return on SMB or MOM portfolio returns (right column of plots for $\tau=0.5$), we observe that dependence can be fully characterized by rather weak contemporaneous covariation between the market and the SMB and MOM portfolio returns, since no significant relation exists at any lead or lag in the relationship.\footnote{Note that the dashed lines in the figure represent confidence intervals under the null hypothesis that the two series are jointly normally distributed correlated random variables.} Moving our attention toward the left tail of the joint distribution, more complicated dependence structures emerge. The departure from the joint Gaussian distribution is strongest in the left tail (left column of plots for $\tau=0.05$). The co-occurrences of large negative market returns with large negative SMB or MOM portfolio returns are significant and exist at various leads/lags. 

For example, if we look at the dependence between the market and SMB in the 5\% tail, we can observe that if the market is below this threshold, there is also a significant probability that the SMB portfolio will be below this threshold, with some delay. Similarly, the SMB downturn precedes the market downturn with significant probability.\footnote{A similar lead/lag investigation regarding business cycle indicators is performed in \cite{backus2010cyclical}.} Therefore, instead of arguing that the SMB factor proxies for an independent economic risk, the results suggest that the SMB portfolio captures more complicated market tail risk at some specific horizons. 

In other words, the left tail dependence shows that extreme market drop is correlated with extreme negative returns of SMB. This illustrates that large negative market returns are correlated with the situation in which large companies largely outperform small companies in the SMB portfolio. Hence, we document a joint probability of co-occurrence of the market extreme left tail event, and large companies outperform small companies, leading to an increase in default risk in the economy \citep{chan1985exploratory}. An important feature of the dependence not documented by earlier studies is its persistence structure shown by autocorrelations and the same strength for leading one another. At the same time, while momentum is negatively correlated with the market, the second row of Figure \ref{fig:fac_dependence} shows a significant lead-lag relationship of the momentum factor and stock market, pointing us to the intuition that extremely low market returns are cross-correlated with companies with low momentum outperforming those with high momentum.

Note that these observations are closely related to the literature on market frictions, price delays and aggregations and their asset pricing implications.\footnote{See, e.g., \cite{kamara_korajczyk_lou_sadka_2016, 10.1093/rfs/hhi023}} In that sense, we follow a similar vein of thought as \cite{bandi2018measuring}, with the important difference that we focus on the downside risk specifically.

This line of thinking may lead us to the conclusion that such general dependence structures can hardly be described by traditional contemporaneous correlation-based measures. The illustration suggests that there is no need for many factors to explain the average asset return, as carefully measured exposure to market risk can capture the risk investors care about. A natural way to summarize the dependence across these lead/lag relationships is to employ frequency analysis and precisely summarize this joint structure for specific horizons.

From an economic perspective, it is reasonable to assume that future marginal utility is affected by the realization of low quantile returns today, as this event may lead, for example, to bankruptcy or in other ways significantly shape the behavior of economic agents in the future. In other words, extreme market events can have either short-run or long-run effects on the marginal utility of investors. Previous studies, however, fail to fully account for horizon-specific information in tails, while one of the main reasons turns to the inability to measure such risks. Here, we propose robust methods for the measurement of such risks, and we argue that exploring the risk related to tail events as well as frequency-specific risk is crucial.

To see how tail-specific risks are priced across horizons by investors, we proceed as follows. First, we define a quantile risk measure based on the covariance between indicator functions, which has a natural economic interpretation in terms of probabilities. Second, we introduce its frequency decomposition and combine these two frameworks into the quantile spectral risk measure, which is the building block of our empirical model. This measure enables us to robustly test for the presence of extreme market risks over various horizons in asset prices. The aim is not to convince the reader that the functional form of the preferences precisely follows our model but to show that there is heterogeneity in the weights that investors assign to the risk for different investment horizons and different parts of the distribution of their future wealth. By estimating prices of risk for short- and long-term parts, we are able to identify the horizon that the investor cares most about. Moreover, by estimating prices of risk for various threshold values, we are able to identify the part of the joint distribution toward which the investor is the most risk averse.\footnote{Our investigation complements the work of \cite{10.1093/rfs/hhx012} and \cite{delikouras_kostakis_2019}. These studies investigate the position of the reference point of consumption growth and show that its correct location is crucial for fitting the model based on generalized disappointment aversion.} This is done by controlling for CAPM beta, and the influence of these new measures is measured as incremental information over simplifying assumptions that lead to the CAPM beta asset pricing models.

\section{Measuring the Tail Risks across Horizons: A Quantile Spectral Beta}
\label{sec:qs_risk}

Here, we formalize the discussion and provide more general measures that will provide a tool for inferring the discussed types of risks from data.

\subsection{Tail Risk}

Let us consider a bivariate, strictly stationary process $\bX_t = (m_t, r_t)'$ holding some reference economic or financial variable $m_t$ proxying risk and asset returns $r_t$. The marginal distribution functions of $m_{t}$ and $r_t$ will be denoted by $F_m$ and $F_r$ respectively, and by $q_{m}(\tau_m) := F_{m}^{-1}(\tau_m) := \inf\{ q \in \IR : \tau_m \leq F_{m}(q)\}$, and $q_{r}(\tau_r) := F_{r}^{-1}(\tau_r) := \inf\{ q \in \IR : \tau_r \leq F_{r}(q)\}$, where $\tau_{m},\tau_{r} \in [0,1]$ denote the corresponding quantile functions. 

Since we are interested in pricing extreme negative events, we want to measure dependence and risk in lower quantiles of the joint distribution that can be evaluated by quantile cross-covariance \citep{kley2014,barunik2019quantile}
\begin{align} \label{eq:cov_alt}
\gamma_k^{m,r}(\tau_m,\tau_r) \equiv \mathbb{C}ov \Big(I\{ m_{t+k} \leq q_m(\tau_m) \}, I\{ r_{t} \leq q_{r}(\tau_r) \} \Big),
\end{align}
$k \in \IZ$, and $I\{ A \}$ denotes the indicator function of event $A$. The measure is given by the covariance between two indicator functions and, together with $F_m$ and $F_r$, can fully describe the joint distribution of the pair of random variables $m_t$ and $r_{t}$, that is, provide a measure for their serial and cross-dependency structure. If the distribution functions of the variables are continuous, the quantity is essentially the difference between the copula of the pair $m_t$ and $r_{t}$ and the independent copula, i.e., $Pr \big\{ m_{t+k} \leq q_{m}(\tau_m), r_{t} \leq q_{r}(\tau_r) \big\} - \tau_m \tau_r$. Thus, covariance between indicators measures additional information from the copula over an independent copula about how likely it is that the series are jointly less than or equal to a given quantile of the variable $m_t$. It enables flexible measurement of both the cross-sectional structure and time-series structure of the pair of random variables.

Comparing these new quantities with their traditional counterparts, it can be observed that the covariance and means are essentially replaced by copulas and quantiles. A market beta associated with the tail risk can then be defined using Eq. \ref{eq:cov_alt}. This quantity would be similar to the tail risk measure of \cite{10.1093/rfs/hhz105}, which is also a function of the $\tau$ quantile threshold of consumption growth. The correlation between asset returns and consumption growth is then computed conditional on realizations of consumption growth below the threshold. It is also related to the negative semibetas of \cite{bollerslev2020realized}, which estimates the dependence between market return and asset return conditional on the co-occurrence of negative events for both market and asset.

\subsection{Tail Risks across Horizons: A Quantile Spectral Beta}

It is natural to further assume that economic agents care not only about different parts of the wealth distribution but also differently about long- and short-term investment horizons in terms of expected returns and associated risks. Investors may be interested in the long-term profitability of their portfolio and may not be concerned with short-term fluctuations. 
Frequency-dependent features of an asset return then play an important role for an investor. \cite{bandi2017horizon} argue that covariance between two returns can be decomposed into covariance between disaggregated components evolving over different time scales, and thus the risk on these components can vary. Hence, market beta can be decomposed into a linear combination of betas measuring dependence at various scales, i.e., dependence between fluctuations with various half-lives. Frequency-specific risk at a given time plays an important role in the determination of asset prices, and the price of risk in various frequency bands may differ, which means that the expected return can be decomposed into a linear combination of risks in various frequency bands.

A natural way to decompose covariance between two assets into dependencies over different horizons is in the frequency domain. A frequency domain counterpart of cross-covariance $\gamma_k$ is obtained as the Fourier transform of the cross-covariance functions $S_{m,r} (\omega) = \frac{1}{2 \pi} \sum_{k = -\infty}^{\infty} \gamma_k^{m,r}  e^{-\textrm{i} k \omega}$. Conversely, cross-covariance can be obtained from the inverse Fourier transform of its cross-spectrum as $\gamma_k^{m,r}  = \int_{-\pi}^{\pi} S_{m,r} (\omega) e^{\textrm{i} k \omega} d\omega$, where $S_{m,r} (\omega)$ is the cross-spectral density of random variables $m_{t}$ and $r_t$ and $\rm{i} = \sqrt{-1}$.

This representation of covariation allows us to decompose the covariance and variance into frequency components and disentangle the short-term dependence from the long-term dependence. Using a similar approach, \cite{bandi2017horizon} estimate the price of risk for different investment horizons and show that investors possess heterogeneous preferences over various economic cycles instead of looking only at averaged quantities over the whole frequency spectrum.

To uncover more general dependence structures, we propose to study the Fourier transform of the covariance of indicator functions $\gamma_k^{m,r}(\tau_m,\tau_r)$ instead. In this way, one can quantify the horizon-specific risk premium across the joint distribution. To define the new beta representation that will allow us to characterize such general risks, we use the so-called quantile cross-spectral densities introduced by \cite{barunik2019quantile} as a generalization of quantile spectral densities of~\cite{DetteEtAl2015}.

The cornerstone of this new beta representation lies in quantile cross-spectral density defined as
\begin{eqnarray}
\label{eq:qs_kernel}
f^{m,r}(\omega; \tau_m,\tau_r) &\equiv& \frac{1}{2 \pi} \sum_{k=-\infty}^{\infty} \gamma_k^{m,r} (\tau_m,\tau_r) e^{- \textrm{i} k \omega}\\
&\equiv& \frac{1}{2 \pi} \sum_{k=-\infty}^{\infty} \mathbb{C}ov \Big(I\{ m_{t+k} \leq q_m(\tau_m) \}, I\{ r_{t} \leq q_{r}(\tau_r) \} \Big) e^{- \textrm{i} k \omega}
\end{eqnarray}
with $\omega \in \IR$ and $\tau_m,\tau_r \in [0,1]$. A quantile cross-spectral density is obtained as a Fourier transform of covariances of indicator functions defined in Equation \ref{eq:cov_alt}, and will allow us to define beta that will capture the tail risks as well as spectral risks.

The \textit{quantile spectral} (QS) betas that characterize horizon- and tail-specific market risk at a given $\omega$, $\tau_m$ and $\tau_r$ are then defined as
\begin{align}
\label{eq:qs_beta}
\beta^{m,r}(\omega; \tau_m,\tau_r) \equiv \frac{f^{m,r}(\omega; \tau_m,\tau_r)}{f^{m,m}(\omega; \tau_m,\tau_m)},
\end{align}
and will be the key quantity in our analysis. To estimate the quantile spectral beta, we use the \emph{rank-based copula cross-periodogram} introduced by \cite{barunik2019quantile}
\begin{equation}
  \label{eqn:inr}
I_{n,R}^{m, r}(\omega; \tau_m, \tau_r):= \frac{1}{2\pi n}
d^{m}_{n,R}(\omega; \tau_m) d^{r}_{n,R}(-\omega; \tau_r),
\end{equation}
where $d_{n,R}^{m}(\omega; \tau_m) := \sum_{t=0}^{n-1} I\{\widehat F_{n,m}(m_t) \leq \tau_m\} \ee^{- \ii \omega t}$, and $d_{n,R}^{r}(\omega; \tau_r) := \sum_{t=0}^{n-1} I\{\widehat F_{n,r}(r_t) \leq \tau_r\} \textrm{e}^{- \ii \omega t}$ with $\widehat F_{n,m}(m_t)$ and $\widehat F_{n,r}(r_t)$ being empirical distribution functions of $m_t$ and $r_t$, respectively. A consistent estimator of the quantile cross-spectral density is then
\begin{equation}
  \label{eqn:DefRankEstimator}
  \widehat G^{m, r}_{n,R}(\omega; \tau_m, \tau_r) := \frac{2\pi}{n} \sum_{s=1}^{n-1} W_n\big( \omega - 2\pi s / n \big) I_{n,R}^{m, r}(2 \pi s / n, \tau_m, \tau_r),
\end{equation}
where $W_n$ denotes a sequence of weight functions, precisely to be defined in the next section studying the asymptotic properties of the proposed estimators. The estimator of the quantile spectral beta is then given by
\begin{equation}\label{def:Rhat}
  \widehat{\beta}^{m, r}_{n,R}(\omega; \tau_m, \tau_r)
:= \frac{\widehat G^{m, r}_{n,R}(\omega; \tau_m, \tau_r)}{\widehat G^{m, m}_{n,R}(\omega; \tau_m, \tau_m) }.
\end{equation}

Before we prove that $\widehat{\beta}^{m, r}_{n,R}(\omega; \tau_m, \tau_r)$ is a legitimate estimate of $\beta^{m,r}(\omega; \tau_m,\tau_r)$, we note that for serially uncorrelated variables (regardless of their joint or marginal distributions), the Fre\' chet/Hoeffding bounds give the limits that QS beta can attain in the case of a serially independent process as $\frac{\textrm{max} \{ \tau_m + \tau_r - 1, 0 \} - \tau_m \tau_r}{\tau_m(1-\tau_m)} \leq \beta^{m,r}(\omega; \tau_m,\tau_r) \leq \frac{\textrm{min}\{ \tau_m, \tau_r \} - \tau_m \tau_r}{\tau_m(1-\tau_m)}$.

\subsection{Asymptotic properties of the Quantile Spectral beta}
\label{sec:AsympQspec}
To derive the asymptotic properties of the quantile spectral beta, some assumptions need to be made. Recall (cf.~\citet{Brillinger1975}, p.~19) that the {$r$th order joint cumulant} $\cum(Z_1, \ldots, Z_r)$ of the random vector $(Z_1, \ldots, Z_r)$ is defined as
\[\cum(Z_1, \ldots, Z_r):= \sum_{\{\nu_1, \ldots, \nu_p\}} (-1)^{p-1} (p-1)! E \Big[ \prod_{j \in \nu_1} Z_j \Big] \cdots E \Big[ \prod_{j \in \nu_p} Z_j\Big],\]
with summation extending over all partitions $\{\nu_1, \ldots, \nu_p\}$, $p=1,\ldots,r$, of~$\{1,\ldots,r\}$.

\noindent Regarding the range of dependence of $\bX_t\in(m_t,r_t)'$, we make the following assumption:
\begin{assumption}\label{ass:exp_alpha_mix}
  The processes $(\bX_t)_{t \in \IZ}$ are strictly stationary and exponentially $\alpha$-mixing, that is, there exist constants $K < \infty$ and $\kappa \in (0,1)$, such that
\begin{equation}\label{eq:exp_alpha_mix}
\alpha(n) := \sup_{\substack{A \in \sigma(x_0, x_{-1}, \ldots)\\ B \in \sigma(x_n, x_{n+1}, \ldots)}} \big| \IP(A \cap B) - \IP(A) \IP(B) \big| \leq K \kappa^n, \quad n \in \IN.
\end{equation}
\end{assumption}
Note that the Assumption \ref{ass:exp_alpha_mix} is a bivariate extension of assumptions made in \cite{kley2014} and used in \cite{barunik2019quantile} to study quantile spectral quantities. It is important to observe that this assumption does not require the existence of any moments, which is in sharp contrast to classical assumptions, where moments up to the order of the respective cumulants must exist, and sets $A_j$ are not required to be general Borel sets, as in classical mixing assumptions. As noted in \cite{barunik2019quantile}, this assumption holds for a wide range of popular, linear and nonlinear, multivariate and univariate processes that are $\alpha$-mixing at an exponential rate, including traditional VARMA or vector-ARCH models.

To establish the consistency of the estimates, we further need to consider sequences of weights that asymptotically concentrate around multiples of $2\pi$.
\begin{assumption}\label{ass:W}
The weights are defined as
$
W_n(u) := \sum_{j=-\infty}^{\infty} b_n^{-1} W(b_n^{-1} [u + 2\pi j])
$, where $b_n > 0$, $n=1,2,\ldots$, is a sequence of scaling parameters satisfying~$b_n \rightarrow 0$ and $n b_n \rightarrow \infty$, as $n \rightarrow \infty$.
The weight function $W$ is real-valued, even has support $[-\pi,\pi]$, bound\-ed variation, and satisfies
$
\int_{-\pi}^{\pi} W(u) \text{d}u = 1.
$
\end{assumption}

The main result of this section will legitimize $\widehat\beta_{n,R}^{m,r}(\omega;\tau_m,\tau_r)$ as an estimator of the quantile spectral (QS) beta $\beta^{m,r}(\omega;\tau_m,\tau_r)$. The legitimacy of the estimates follows from the fact that the estimators converge weakly in the sense of \emph{Hoffman-J\o{}rgensen} (cf. Chapter 1 of~\cite{vanderVaartWellner1996}). We denote this mode of convergence by $\Rightarrow$. The estimators under consideration take values in the space of (elementwise) bounded functions $[0,1]^2 \rightarrow \IC^{d \times d}$, which we denote by $\ell_{\mathbb{C}^{d \times d}}^{\infty}([0,1]^2)$ \citep{kley2014}. While the results of empirical process theory are typically stated for spaces of real-valued, bounded functions, these results transfer immediately by identifying $\ell_{\mathbb{C}^{d \times d}}^{\infty}([0,1]^2)$ with $\ell^{\infty}([0,1]^2)^{2d^2}$.

Using Proposition \ref{prop:inr} in Appendix~\ref{app:main} and following \cite{kley2014} and \cite{barunik2019quantile}, we quantify uncertainty in estimating $f^{m,r}(\omega; \tau_m,\tau_r)$ by $\widehat G^{m, r}_{n,R}(\omega; \tau_m, \tau_r)$ asymptotically in the following theorem.

\begin{theorem} (\cite{barunik2019quantile})\label{thm:AsympDensityRankEstimator}
Let Assumptions~\ref{ass:exp_alpha_mix} and~\ref{ass:W} hold. Assume that the marginal distribution functions $F_m$ and $F_r$ are continuous and that constants $\kappa > 0$ and $k \in\IN$ exist, such that
$b_n = o(n^{-1/(2k+1)})$ and $b_n n^{1-\kappa} \rightarrow \infty$.
Then, for any fixed $\omega \in  \IR $,
\begin{equation}\label{eq:gntoh}
  \sqrt{n b_n} \Big(\widehat{G}^{m,r}_{n,R}(\omega; \tau_m, \tau_r) - f^{m,r}(\omega; \tau_m, \tau_r) - B_n^{m,r,(k)}(\omega; \tau_m, \tau_r)\Big)_{\tau_m,\tau_r \in [0,1]} \Rightarrow \IH^{m,r}(\omega; \cdot,\cdot),
\end{equation}
 where the bias is given by $B_n^{m,r,(k)}(\omega; \tau_m, \tau_r):=
  \sum_{\ell=2}^k \frac{b_n^{\ell}}{\ell!} \int_{ -\pi}^{\pi} v^{\ell} W(v) dv \frac{{\rm d}^{\ell}}{{\rm d}\omega^{\ell}}\mathfrak{f}^{m, r}(\omega; \tau_m, \tau_r)$. 
The process $\IH^{m,r}(\omega; \cdot,\cdot)$ is a centered, $\IC$-valued Gaussian process characterized by
\begin{multline} 
\Cov\big(\IH^{j_1, j_2}(\omega; u_1, v_1\big), \IH^{k_1, k_2}(\lambda; u_2, v_2)\big) \\ = 2\pi \Big(\int_{-\pi}^\pi W^2(\alpha){\rm d}\alpha \Big)
\Big( \mathfrak{f}^{j_1, k_1}(\omega; u_1, u_2) \mathfrak{f}^{j_2, k_2}(-\omega; v_1, v_2) \eta(\omega - \lambda) \\
+ \mathfrak{f}^{j_1, k_2}(\omega; u_1, v_2) \mathfrak{f}^{j_2, k_1}(-\omega; v_1, u_2)
\eta(\omega + \lambda) \Big),
\end{multline}
where $\eta(x) := I\{x = 0 (\mod 2\pi)\}$ [cf.~\cite[p.\,148]{Brillinger1975}] is the $2\pi$-periodic extension of Kronecker's delta function.
The family~$\{\IH(\omega; \, \cdot, \cdot),$ $ \omega \in  [0,\pi] \}$ is a collection of independent processes.
\end{theorem}

It is important to note that in sharp contrast to classical spectral analysis, where higher-order moments are required to obtain smoothness of the spectral density [cf.~\cite{Brillinger1975}, p.\,27], Assumption~\ref{ass:exp_alpha_mix} guarantees that the quantile cross-spectral density is an analytical function of $\omega$. Assume that $W$ is a kernel of order $p$; i.\,e., for some $p$, that satisfies $\int_{-\pi}^{\pi} v^j W(v) {\rm d}v = 0$, for all $j < p$, and $0 < \int_{-\pi}^{\pi} v^p W(v) {\rm d}v < \infty$; e.\,g., the Epanechnikov kernel is a kernel of order $p=2$. Then, the bias is of order $b_n^p$. As the variance is of order $(n b_n)^{-1}$, the mean squared error is minimal if $b_n \asymp n^{-1/(2p+1)}$. This optimal bandwidth fulfills the assumptions of Theorem~\ref{thm:AsympDensityRankEstimator}.
A detailed discussion of how Theorem~\ref{thm:AsympDensityRankEstimator} can be used to construct asymptotically valid confidence intervals can be found in \cite{barunik2019quantile}.

The independence of the limit $\{\IH(\omega; \, \cdot, \cdot),$ $ \omega \in [0,\pi] \}$ has two important implications. On the one hand, the weak convergence~(\ref{eq:gntoh}) holds jointly for any \emph{finite} fixed collection of frequencies~$\omega$. Furthermore, fixing $j_1, j_2$ and $\tau_1, \tau_2$, the CCR periodogram $\widehat G^{j_1, j_2}_{n,R}(\omega; \tau_1, \tau_2)$ and traditional smoothed cross-periodogram determined from the unobservable, bivariate time series
\begin{equation} 
\big(I\{F_{j_1}(X_{t,j_1}) \leq \tau_1\},I\{F_{j_1}(X_{t,j_2}) \leq \tau_2\}\big), \quad t = 0, \ldots, n-1,
\end{equation}
are asymptotically equivalent. Theorem~\ref{thm:AsympDensityRankEstimator} thus reveals that in the context of the estimation of the quantile cross-spectral density, the estimation of the marginal distribution has no impact on the limit distribution (cf. comment after Remark~3.5 in~\cite{kley2014}).

We are now ready to state the main result of this section.

\begin{theorem}\label{thm:AsympQSbeta}
Let Assumptions~\ref{ass:exp_alpha_mix} and~\ref{ass:W} hold. Assume that the marginal distribution functions $F_m$ and $F_r$ are continuous and that constants $\kappa > 0$ and $k \in\IN$ exist, such that
$b_n = o(n^{-1/(2k+1)})$ and $b_n n^{1-\kappa} \rightarrow \infty$. Assume that for some $\varepsilon \in (0,1/2)$, we have
$\inf_{\tau \in [\varepsilon, 1-\varepsilon]} \mathfrak{f}^{m,m}(\omega; \tau_m, \tau_m) > 0$, and $\inf_{\tau \in [\varepsilon, 1-\varepsilon]} \mathfrak{f}^{r, r}(\omega; \tau_r, \tau_r) > 0$. Then, for any fixed $\omega \in   \IR$,
\begin{equation}\label{eqn:convR}
\sqrt{n b_n} \Big( \widehat{\beta}_{n,R}^{m,r}(\omega; \tau_m, \tau_r) - \beta^{m,r}(\omega;\tau_m,\tau_r) - B_n^{m,r,(k)}(\omega; \tau_m, \tau_r) \Big)_{(\tau_m, \tau_r) \in [\varepsilon, 1-\varepsilon]^2}
\Rightarrow \frac{1}{\ff{m}{m}} \Big(\mathbb{H}_{m,m}- \frac{\ff{m}{r}}{\ff{m}{m}} \mathbb{H}_{m,r}\Big),
\end{equation}
where
\begin{equation} \label{def:biasR}
B_n^{m,r,(k)}(\omega; \tau_m, \tau_r) :=
\frac{1}{\ff{m}{m}} \Big( B_{m,m} - \frac{\ff{m}{r}}{\ff{m}{m}} B_{m,r} \Big)
\end{equation}
and we have written $\ff{a}{b}$ for the quantile cross-spectral density $\mathfrak{f}^{a, b}(\omega; \tau_a, \tau_b)$ as defined in \eqref{eq:qs_kernel},
$B_{a,b} := \sum_{\ell=2}^k \frac{b_n^{\ell}}{\ell!} \int_{ -\pi}^{\pi} v^{\ell} W(v) dv \frac{{\rm d}^{\ell}}{{\rm d}\omega^{\ell}}\mathfrak{f}^{a, b}(\omega; \tau_a, \tau_b)$, and $\HH{a}{b}$ for $\IH^{a, b}(\omega; \tau_a, \tau_b\big)$ defined as a centered, $\IC$-valued Gaussian process characterized by
\begin{multline} \label{eqn:CovH}
\Cov\big(\IH^{j_1, j_2}(\omega; u_1, v_1\big), \IH^{k_1, k_2}(\lambda; u_2, v_2)\big) \\ = 2\pi \Big(\int_{-\pi}^\pi W^2(\alpha){\rm d}\alpha \Big)
\Big( \mathfrak{f}^{j_1, k_1}(\omega; u_1, u_2) \mathfrak{f}^{j_2, k_2}(-\omega; v_1, v_2) \eta(\omega - \lambda) \\
+ \mathfrak{f}^{j_1, k_2}(\omega; u_1, v_2) \mathfrak{f}^{j_2, k_1}(-\omega; v_1, u_2)
\eta(\omega + \lambda) \Big),
\end{multline}
where $\eta(x):= I\{x = 0 (\mod 2\pi)\}$ [cf.~\cite[p.\,148]{Brillinger1975}] is the $2\pi$-periodic extension of Kronecker's delta function.
The family~$\{\IH(\omega; \, \cdot, \cdot),$ $ \omega \in [0,\pi] \}$ is a collection of independent processes.
\end{theorem}
\begin{proof}
  The proof is lengthy and technical, and it is therefore deferred to Appendix \ref{app:proofmain}.
\end{proof}
Convergence to a Gaussian process can be employed to obtain asymptotically valid pointwise confidence bands. A more detailed discussion on how to conduct inference is given in Appendix \ref{app:intervals}.

If $W$ is a kernel of order $p\geq 1$, we have that the bias is of order $b_n^p$. Thus, if we choose the mean square error minimizing bandwidth $b_n \asymp n^{-1/(2p + 1)}$, the bias will be of order $n^{-p/(2p + 1)}$.

Regarding the restriction $\varepsilon > 0$, note that the convergence~\eqref{eqn:convR} cannot hold if $(\tau_1, \tau_2)$ is on the border of the unit square, as the quantile coherency $\bfrakR(\omega;\tau_1,\tau_2)$ is not defined if $\tau_j \in \{0,1\}$, as this implies that $\Var(I\{F_{j}(X_{t,j}) \leq \tau_j\}) = 0$.

\section{Pricing Model for Extreme Risks across the Frequency Domain}
\label{sec:pricing}

Quantile spectral betas defined in the previous sections will be the cornerstone of our empirical models. We assume that QS betas for low threshold values will be significant determinants of risk priced heterogeneously across investment horizons. We will employ QS betas to study two kinds of risk related to the market return. First, we will investigate \textit{tail market risk} (TR), a risk representing dependence between extreme negative events of both market and asset returns at a given horizon. In case the stochastic discount factor is linear in factors and we consider the market return as a risk factor, we further look at the dependence between asset returns and market returns and the threshold values are based on quantiles of market returns.

It is useful to connect our notion of risks to a well-established rare disaster model of \cite{10.1257/mac.5.3.35}. QS betas between consumption growth and equity returns can be directly connected to permanent and transitory disasters that moreover unfold over multiple years or just one period. QS beta can be used to clearly distinguish between the dependence structures of these types that are otherwise invisible to investors. The detailed discussion with simulations is relegated to Appendix \ref{app:rare_disaster} due to the limited space of the paper.

Our notion of tail risk also relates to the downside risk of \cite{ang2006cross,lettau2014conditional}. While downside risk stems from covariation of asset returns and market return under some threshold, our notion stems from joint probability of the co-occurrence of extreme negative returns in both asset and market returns. This is more in line with the approach of semibetas \citep{bollerslev2020realized} but with an important feature of the persistence structure of such risks across investment horizons.

Second, we will examine \textit{extreme market volatility risk} (EVR), a risk capturing unpleasant situations in which extremely high levels of market volatility are linked with extremely low asset returns, again with respect to the investment horizon. We argue that both of these concepts capture important features of risk of an asset faced by the investor and thus should be priced in a cross-section of asset returns.

In each of the models defined in the paper, we control for CAPM beta as a baseline measure of risk. This ensures that if the QS betas are proven to be significant determinants of risk premium, they do not simply duplicate the information contained in the CAPM beta. Moreover, in the case of tail market risk, we define relative betas that explicitly capture the additional information over the CAPM beta only.

\subsection{Tail Market Risk}

For better interpretability, we construct a quantile spectral beta for a given frequency band corresponding to reasonable economic cycles. This definition is important since it allows us to define short-run or long-run bands covering corresponding frequencies and hence disaggregate beta based on the specific demands of a researcher.

We expect the dependence between market return and asset return during extreme negative joint events to be positively priced across assets. The stronger the relationship is, the higher the risk premium required by investors. In addition, we expect this risk to be priced heterogeneously across different investment horizons.

To capture the tail market risk measuring the probability of co-occurrence between (extreme) negative events of both market and asset returns at a given horizon, we define
\begin{align}
\label{eq:qs_beta_avg1}
\beta^{r_m,r_i}_{\text{TR}}(\Omega; \tau) \equiv  \sum_{\Omega \equiv [\omega_1, \omega_2)} \left(\frac{ \sum_{k=-\infty}^{\infty} \mathbb{C}ov \Big( I\{ r_{m,t+k} \leq q_{r_m}(\tau) \}, I\{ r_{i,t} \leq q_{r_m}(\tau) \} \Big) e^{- \textrm{i} k \omega}}{ \sum_{k=-\infty}^{\infty} \mathbb{C}ov \Big( I \{ r_{m,t+k} \leq q_{r_m}(\tau), I \{ r_{m,t} \leq q_{r_m}(\tau) \}\Big) e^{- \textrm{i} k \omega}}\right).
\end{align}
The numerator of Eq. (\ref{eq:qs_beta_avg1}) captures the probability of co-occurrence of the negative events at a given horizon, and the denominator captures information related to the probability of market tail events at a given horizon, which is related to the variation in market returns.

Similar to \cite{ang2006downside} and \cite{lettau2014conditional}, we define relative betas that capture additional information not contained in the classical CAPM beta. In this way, we can test the significance of tail market risk decomposed into the long- and short-term components to obtain their prices of risk separately. Because we want to quantify risk that is not captured by the CAPM beta, we propose to test the significance of tail market risk via differences in the QS beta and QS beta implied by the Gaussian white noise assumption. We call this \textit{relative} QS beta, and we compute it for a given frequency band $\Omega_j$ and given market $\tau$-quantile level as
\begin{equation}
\beta_{\text{rel}}^{r_m,r_i}(\Omega_j; \tau) \equiv \beta^{r_m,r_i}_{\text{TR}}(\Omega_{j}; \tau) - \beta_{\text{Gauss}}^{r_i} (\Omega_{j}, \tau),
\end{equation}
where $\beta_{\text{Gauss}}^{r_i} (\Omega_{j}, \tau) = \frac{C_{\text{Gauss}}(\tau, \tau_i; \rho) - \tau \tau_i}{\tau (1 - \tau)} $ with $C_{\text{Gauss}}$ being a Gaussian copula with correlation $\rho$ between market return and an asset's return.\footnote{This stems from the fact that quantile cross-spectral density corresponds to a difference of probabilities $Pr \big\{r_{i,t} \leq_{r_m}(\tau), r_{m,t} \leq q_{r_m}(\tau) \big\} - \tau \tau_i$, where $\tau$ and $\tau_i$ are probability levels under a Gaussian distribution, and $\tau_i$ is obtained as $\tau_i = F_{r_i} \{q_m(\tau)\}$.}

Assuming that all the relevant pricing information is contained in the CAPM beta, contemporaneous covariance between two time series should capture all the priced information. Moreover, if the series are jointly normally distributed and independent through time, the CAPM beta contains all the available information regarding the dependence. Hence, under the hypothesis that market and asset returns are correlated Gaussian noise, $\beta_{\text{rel}}^{r_m,r_i}(\Omega_j; \tau)$ will not carry any additional information, and CAPM characterizes the risks well. Note that $\beta_{\text{Gauss}}^{r_i} (\Omega_{j}, \tau)$ is constant across frequencies and depends only on the chosen quantile and correlation coefficient. On the other hand, if investors price information not captured by the CAPM beta, the QS beta estimated without any restriction may identify an additional dimension of risk not contained in the CAPM beta. More specifically, we can identify whether dependence in a specific part of the joint distribution and/or over a specific horizon is significantly priced.

If the CAPM beta captures all the risk information priced in the cross-section, the risk premium corresponding to the relative QS beta will be insignificant. Moreover, if the returns are Gaussian, the relative QS beta will be zero at all frequencies and quantiles.\footnote{Here, we briefly note that we set the threshold values in the covariance between indicators’ measure of dependence as a $\tau$ quantile of market return. In the case of TR betas, the thresholds for market and asset returns are the same and are given by the $\tau$ quantile of market return. In the case of EVR betas, the threshold for increments of market volatility is given by the $\tau$ quantile of the series of increments of market volatility, and the threshold for asset return is given by the $\tau$ quantile of market return. Note that one could flexibly choose the thresholds based on the best model fit specific to our datasets. For example, we may choose the threshold value to be asset specific by corresponding to the $\tau$ quantile of the asset return. We do not follow this approach because we do not explicitly care about dependence between quantiles in the cross-section. Rather, we care about dependence in extreme market situations.}

Our first model is hence a tail market risk (TR) model, which is defined as
\begin{equation}
\label{eq:tr_mod}
\mathbb{E}[r_{i,t+1}^e] = \sum_{j=1}^2 \beta_{\text{rel}}^{r_m,r_i}(\Omega_j; \tau) \lambda_{\text{\text{TR}}} (\Omega_j; \tau) + \beta^{r_m,r_i}_{\text{\text{CAPM}}} \lambda_{\text{CAPM}},
\end{equation}
where $r_{i,t+1}^e$ is the excess return of asset $i$,\footnote{Note that all the risk measures (in line with the literature) present in the paper are calculated using excess returns.} $\beta^{r_m,r_i}_{\text{\text{CAPM}}}$ is an aggregate CAPM beta, $\lambda_{\text{CAPM}}$ is the price of aggregate risk of the market captured by the classical beta, and $\lambda_{\text{TR}}(\Omega_{j}, \tau)$ is the price of tail risk (TR) for a given quantile and horizon (frequency band). We specify our models to include the disaggregation of risk into two horizons -- long and short. Long horizon is defined by corresponding frequencies of cycles of 3 years and longer, and short horizon by frequencies of cycles shorter than 3 years.\footnote{In Appendix \ref{app:diff_long}, we perform a robustness check by defining the horizons using 1.5 years as a threshold and the results do not qualitatively differ. Different specifications are available upon request.} The procedure for obtaining these betas is explained below.

The intuition behind the TR model defined in \ref{eq:tr_mod} is that the relative TR betas will be zero in the case of Gaussian data, and no association between tail risk and the risk premium should be documented since risk is perfectly described by variance. On the other hand, if the data distribution is not Gaussian, the relative TR betas will be significantly different from zero, and the significance of the estimated price of risk captures the pricing effect of the TR over the conventional measure of dependence based on the contemporaneous correlation. We explicitly wish to investigate whether the dependence information over the classical assumptions is a significant determinant of the excess returns, so it is not important whether the CAPM model is true or not.

This specification also relates to the models recently proposed in the literature.\footnote{\cite{barunik2019quantile} features a toy example of TR risk estimated on asset returns as well, but they do not investigate any asset pricing implications of the estimated risk.} Similar models of \cite{bandi2017horizon,bandi2018measuring} focus on the \textit{consumption} CAPM model and thus use consumption as their proxy for risk. In contrast to these attempts, we consider horizon-specific risk in \textit{in tails}.

From the TR perspective, the proposed model also relates to the model of \cite{bollerslev2020realized}, who investigate the pricing implications of the co-occurrence of the downside events of both market and asset returns. In contrast to our model, \cite{bollerslev2020realized} does not consider the horizon over which these risks unfold.

\subsection{Extreme Volatility Risk}

Assets with high sensitivities to innovations in aggregate volatility have low average returns \citep{ang2006cross}. We further focus on extreme events in volatility and investigate whether dependence between extreme market volatility and tail events of assets is priced across assets. Because time of high volatility within the economy is perceived as time with high uncertainty, investors are willing to pay more for the assets that yield high returns during these tumultuous periods and thus positively covary with innovations in market volatility. This drives the prices of these assets up and decreases expected returns. This notion is formally anchored in the intertemporal pricing model, such as the intertemporal CAPM model of \cite{merton1973intertemporal} or \cite{10.2307/2117530}. According to these models, market volatility is stochastic and causes changes in the investment opportunity set by changing the expected market returns or by changing the risk-return trade-off. Market volatility thus determines systematic risk and should determine the expected returns of individual assets or portfolios. Moreover, we assume that extreme events in market volatility play a significant role in the perception of systematic risk and that exposure to them affects the risk premium of assets.

In addition, decomposition of volatility into the short run and long run when determining asset premiums was proven to be useful \citep{adrian2008stock}. Moreover, \cite{bollerslev2016good} incorporated the notion of downside risk into the concept of volatility risk and showed that stocks with high negative realized semivariance yield higher returns. \cite{farago2017downside} examine downside volatility risk in their five-factor model and obtain a model with negative prices of risk of the volatility downside factor, yielding low returns for assets that positively covary with innovations of market volatility during disappointing events. Thus, we want to investigate which horizon and part of the joint distribution of market volatility and asset returns generate these findings.

We assume that assets that yield highly negative returns during times of large innovations of volatility are less desirable for investors, and thus, holding these assets should be rewarded by higher risk premiums. In addition, we assume that such risk will be horizon specific. To measure the extreme volatility risk, we define the beta that will capture the joint probability of co-occurrences of negative asset returns and the extreme increment of market volatility across horizons. Because of the nature of covariance between indicator functions, we work with negative market volatility innovations $-\Delta \sigma_t^2 = -(\sigma_t^2 - \sigma_{t-1}^2)$, where we estimate $\sigma_t$ with a popular GARCH(1,1). Then, the high volatility increments correspond to low quantiles of the distribution of the negative differences. If an asset positively covaries with increments of market volatility, the extreme volatility risk beta will be small, and vice versa. This is in contrast to most of the measures employed in similar analyses. We define the beta that captures extreme volatility risk across horizons as
\begin{align}
\label{eq:qs_beta_avg}
\beta^{r_i}_{\Delta \sigma^2}(\Omega; \tau) \equiv \sum_{\Omega \equiv [\omega_1, \omega_2)} \left( \frac{ \sum_{k=-\infty}^{\infty} \mathbb{C}ov \Big( I\{ -\Delta \sigma_{t+k}^2 \leq q_{-\Delta \sigma_{t}^2}(\tau) \}, I\{ r_{i,t} \leq q_{r_m}(\tau) \} \Big) e^{- \textrm{i} k \omega}}{ \sum_{k=-\infty}^{\infty} \mathbb{C}ov \Big( I \{ -\Delta \sigma_{t+k}^2 \leq q_{-\Delta \sigma_{t}^2}(\tau), I \{ -\Delta \sigma_{t}^2 \leq q_{-\Delta \sigma_{t}^2}(\tau) \}\Big) e^{- \textrm{i} k \omega}} \right)
\end{align}
Threshold values for asset returns are obtained in the same manner as for tail market risk and are derived from the distribution of the market returns, which means that $q_{r_m} (\tau)$ is used as an asset threshold value. For example, for model with $\tau = 0.05$, when computing extreme market volatility beta, as a threshold for negative innovations of market squared volatility, we use the 5\% quantile of its distribution (corresponding to the 95\% quantile of the original distribution), and the threshold for asset return is set to the 5\% quantile of the distribution of market returns.

Our second model, the extreme volatility risk (EVR) model, will test the significance of EVR betas and is defined as
\begin{align}
\label{eq:evr_mod}
\mathbb{E}[r_{i,t+1}^e] = \sum_{j=1}^2 \beta^{r_i}_{\Delta \sigma^2}(\Omega_j; \tau) \lambda_{\text{EV}} (\Omega_j; \tau) + \beta^{r_i}_{\text{CAPM}} \lambda_{\text{CAPM}},
\end{align}
where, as in the case of the TR model, we include the CAPM beta to control for the corresponding risk premium. In line with the results of the current literature (e.g.,   \cite{boons2015horizon}, \cite{doi:10.1111/jofi.12058}, or \cite{adrian2008stock}), we expect positive prices of risk corresponding to EVR betas. This is because EVR betas measure the dependence between extremely high increments of market volatility (i.e., low values of negative innovations of market volatility) and low values of asset returns. Therefore, if an asset yields low returns in times of high market volatility, investors will require high premiums to hold it. Note that our EVR model closely relates to the model of \cite{farago2017downside}, who introduces downside volatility betas without the frequency aspect of the risk.

Unlike the TR model, the EVR model does not take into consideration the Gaussianity of the data. The estimated price of EVR will directly measure the pricing implication of extreme dependence between market increments of volatility and asset returns.

\subsection{Full model}

Finally, to show the independence of the two horizon-specific tail risks, we also combine them into the third model that includes both tail market risk and extreme market volatility risk for both short- and long-run horizons, again controlling for a traditional CAPM beta. The model possesses the following form
\begin{align}
\label{eq:full_mod}
\mathbb{E}[r_{i,t+1}^e] = \sum_{j=1}^2 \beta_{rel}^{r_m,r_i}(\Omega_j; \tau) \lambda_{\text{TR}} (\Omega_j; \tau) + \sum_{j=1}^2 \beta^{r_i}_{\Delta \sigma^2}(\Omega_j; \tau) \lambda_{\text{EV}} (\Omega_j; \tau) + \beta^{r_m,r_i}_{\text{CAPM}} \lambda_{\text{CAPM}}.
\end{align}
We denote this model as the \textit{full model}. Assuming that TR and EVR are priced, using this model, we will investigate whether these risks are subsumed by each other or whether they describe independent dimensions of priced risk.

Throughout the paper, we focus on results for $\tau$ equal to 1\%, 5\%, 10\%, 15\%, 20\%, and 25\%. The choice of 1\%, 5\% and 10\% quantiles is natural and arises in many economic and finance applications. Most likely, the most prominent example is value-at-risk, which is a benchmark measure of risk widely used in practice and studied among academics. Remaining values of $\tau$, i.e., 15\%, 20\%, and 25\% capture general downside risk and thus more probable negative joint events.

\subsection{Estimation} 

To test our models, we use the standard \cite{fama1973risk} cross-sectional regressions. In the first stage, we estimate all required QS betas, relative QS betas, and CAPM betas for all assets. We define two nonoverlapping horizons: short and long. Horizon is specified by the corresponding frequency band. We specify the long horizon by frequencies with corresponding cycles of 3 years and longer, whereas short horizon indicate frequencies with corresponding cycles below 3 years.\footnote{For a robustness check using 1.5 years as a threshold value, see Appendix \ref{app:diff_long}}. QS betas for these horizons are obtained by averaging QS betas over corresponding frequency bands.

In the second stage, we use these betas as explanatory variables and regress average asset returns on them and obtain the model fit. We assess the significance of a given risk by the significance of its corresponding estimated price \footnote{As shown in \cite{shanken1992estimation}, if the betas are estimated over the whole period, the second-stage regression is $T$-consistent.}. In the case of the full model, we obtain the statistical inference on the estimated prices of risk by repeating cross-sectional regression at every time point, i.e., in every month $t=1,\ldots,T$, we estimate the model of the following form:
\begin{equation}
r_{i,t}^e = \sum_{j=1}^2 \widehat{\beta}_{rel}^{r_m,r_i}(\Omega_j; \tau) \lambda_{t,\text{TR}} (\Omega_j; \tau) + \sum_{j=1}^2 \widehat{\beta}^{r_i}_{\Delta \sigma^2}(\Omega_j; \tau) \lambda_{t,\text{EV}} (\Omega_j; \tau) + \widehat{\beta}^{r_m,r_i}_{\text{CAPM}} \lambda_{t,\text{CAPM}}.
\end{equation}

We obtain $T$ cross-sectional estimates of lambdas for each of the corresponding betas. Then, we estimate the prices of risk by time-series averages of the lambdas over the whole period
\begin{align}
\widehat{\lambda}_k(\Omega_j; \tau) = \frac{1}{T} \sum_{t=1}^T \widehat{\lambda}_{t,k}(\Omega_j; \tau), \quad j=1,2, \quad k=\text{TR}, \text{EVR}.
\end{align}
Standard errors and corresponding $t$-statistics are computed from $\sigma^2 \Big( \widehat{\lambda}_k(\Omega_j; \tau) \Big) = \frac{1}{T^2} \sum_{t=1}^T \Big(\hat{\lambda}_{t,k}\Omega_j; \tau) - \widehat{\lambda}_k(\Omega_j; \tau) \Big)^2$ for both horizons $j=\{1,2\}$ and risks $k=\{\text{TR}, \text{EVR}\}$.

The same estimation logic applies to other studied models. To take into account multiple hypothesis testing, we follow \cite{harvey2016and} and report $t$ -statistics of estimated parameters (below the actual estimates). The overall fit of the model is measured from the OLS regression of the average returns of the assets on their betas. Throughout the paper, we use the root mean squared pricing error (RMSPE) metric, which is a widely used metric for assessing model fit in the asset pricing literature, to assess the overall model performance.

As mentioned earlier, we estimate our models for various threshold values given by the $\tau$ quantile of market return. Furthermore, in Appendix \ref{sec:robust}, we compare our newly proposed measures with i) classical CAPM ii) downside risk model of \cite{ang2006downside} (DR1), iii) downside risk model of \cite{lettau2014conditional} (DR2), iv) 3-factor model of \cite{fama1993common}, v) GDA3 and GDA5 models of \cite{farago2017downside}, and vi) coskewness and cokurtosis measures. Details regarding the estimation of the risk measures of the competing models are summarized in Appendix \ref{app:comp_models}. All the models are estimated for comparison purposes without any restrictions in two stages, similar to our three- and five-factor models. Thus, GDA3 and GDA5 are, despite their theoretical background, estimated without setting any restriction to their coefficients and are also estimated in two stages.

\subsection{Size of the 2-Stage Estimation Procedure}

Naturally, there is a question of how the 2-stage procedure with estimates on frequency bands performs in typical (small) samples, which we encounter in finance. To give the reader a notion of these properties, we present a simulation exercise to investigate the statistical size of our testing approach. In each run, we simulate returns on either 300 or 30 assets to mirror the settings of our empirical investigation of individual stocks and portfolio returns. Each asset possesses a length of 720 observations using either the classical CAPM model or white noise as a data generating process. First, we simulate time series of returns on the market from the normal distribution $N(\mu, \sigma)$ with $\mu = 0.06/12$ and $\sigma = 0.2/\sqrt{12}$. Second, in the case of the CAPM model, we generate time series of asset returns by randomly drawing the CAPM beta from the normal distribution $N(\bar{\beta}, \sigma_{\beta})$, where $\bar{\beta} = 1$ and $\sigma_{\beta} = 0.5$, and then create the return as
\begin{equation}
R_{it} = \beta_i R_{mt} + \epsilon_{it}, \, i = 1, \ldots, N, \, t = 1, \ldots, T.
\end{equation}

In the case of the white noise model, we set all the CAPM betas equal to 0. In the third stage, for every stock, using the simulated data, we estimate their CAPM betas and QS betas (both TR and EVR) and regress the average returns on them using specifications of the TR model, EVR model and full model. We determine the number of cases where we incorrectly reject the null hypothesis that a given QS beta in a given model is a significant determinant of average returns. We set the significance level at $\alpha = 0.05$. Ideally, we would like to observe the rejection rates of approximately 5\%. The results are summarized in Table \ref{tab:test_size}, which shown that  the rejection rates typically correspond to the chosen significance level $\alpha$. This shows the validity of our approach; even for low values of $\tau$ and long horizons, there is no significant bias in the rejection rates.

\begin{table}[ht!]
\scriptsize
\centering
\caption{\textit{Size of the 2-Stage Estimation Procedure.} Here we report rejection rates of the 2-stage estimation procedure when the assets are generated using either the CAPM model or white noise. The significance level is set to $\alpha = 0.05$. The number of simulations is 500.}
\begin{tabular}{ccccccccccc}
 \toprule
	 	DGP &  \# of assets & $\tau$ & \multicolumn{2}{c}{Tail market risk} & \multicolumn{2}{c}{Extreme volatility risk} & \multicolumn{4}{c}{Full model}\\
		\cmidrule(r){1-3} \cmidrule(r){4-5} \cmidrule(r){6-7} \cmidrule(r){8-11}
 & & & $\lambda_{\text{long}}^{\text{TR}}$ & $\lambda_{\text{short}}^{\text{TR}}$  & $\lambda_{\text{long}}^{\text{EV}}$ & $\lambda_{\text{short}}^{\text{EV}}$ &  $\lambda^{\text{TR}}_{\text{\text{long}}}$ & $\lambda^{\text{TR}}_{\text{short}}$ & $\lambda^{\text{EV}}_{\text{long}}$ & $\lambda^{\text{EV}}_{\text{short}}$ \\ 
\cmidrule(r){4-5} \cmidrule(r){6-7} \cmidrule(r){8-11}
\multirow{10}{*}{\shortstack[c]{CAPM}} & \multirow{5}{*}{\shortstack[c]{$N=300$}} & 0.01 & 0.052 & 0.046  & 0.062 & 0.072 & 0.060 & 0.048 & 0.070 & 0.080 \\ 
& & 0.05 & 0.066 & 0.062 & 0.072 & 0.058 & 0.066 & 0.062 & 0.072 & 0.058 \\ 
& & 0.10 & 0.056 & 0.046 & 0.048 & 0.088 & 0.068 & 0.048 & 0.066 & 0.088 \\ 
& & 0.15 & 0.056 & 0.046 & 0.050 & 0.042 & 0.048 & 0.046 & 0.048 & 0.038 \\ 
& &  0.25 & 0.046 & 0.054 & 0.068 & 0.032 & 0.056 & 0.054 & 0.064 & 0.030 \\
  \cmidrule(r){2-11}
& \multirow{5}{*}{\shortstack[c]{$N=30$}} &  0.010 & 0.054 & 0.040 & 0.056 & 0.066 & 0.074 & 0.052 & 0.062 & 0.060 \\ 
& & 0.05 & 0.028 & 0.060 & 0.042 & 0.048 & 0.026 & 0.054 & 0.042 & 0.060 \\ 
& & 0.10 & 0.044 & 0.058 & 0.048 & 0.058 & 0.050 & 0.058 & 0.044 & 0.052 \\ 
& & 0.15 & 0.044 & 0.044 & 0.048 & 0.058 & 0.044 & 0.038 & 0.048 & 0.048 \\ 
& & 0.25 & 0.062 & 0.054 & 0.068 & 0.044 & 0.056 & 0.060 & 0.058 & 0.054 \\
	\midrule
\multirow{10}{*}{\shortstack[c]{White noise}} & \multirow{5}{*}{\shortstack[c]{$N=300$}} & 0.01 & 0.058 & 0.050 & 0.064 & 0.058 & 0.054 & 0.056 & 0.062 & 0.056 \\ 
 & & 0.05 & 0.040 & 0.064 & 0.068 & 0.040 & 0.036 & 0.064 & 0.064 & 0.042 \\ 
 & & 0.10 & 0.044 & 0.044 & 0.054 & 0.046 & 0.042 & 0.042 & 0.066 & 0.054 \\ 
 & & 0.15 & 0.044 & 0.042 & 0.060 & 0.054 & 0.046 & 0.046 & 0.072 & 0.050 \\ 
 & & 0.25 & 0.066 & 0.040 & 0.040 & 0.068 & 0.062 & 0.038 & 0.050 & 0.064 \\
  \cmidrule(r){2-11}
& \multirow{5}{*}{\shortstack[c]{$N=30$}} & 0.01 & 0.054 & 0.038 & 0.074 & 0.060 & 0.040 & 0.040 & 0.066 & 0.058 \\ 
 & & 0.05 & 0.050 & 0.060 & 0.036 & 0.038 & 0.048 & 0.064 & 0.038 & 0.040 \\ 
 & & 0.10 & 0.046 & 0.048 & 0.032 & 0.048 & 0.048 & 0.040 & 0.034 & 0.048 \\ 
 & & 0.15 & 0.052 & 0.048 & 0.042 & 0.060 & 0.048 & 0.040 & 0.038 & 0.052 \\ 
 & & 0.25 & 0.044 & 0.072 & 0.052 & 0.050 & 0.036 & 0.066 & 0.060 & 0.036 \\
	
     \bottomrule
\end{tabular}
\label{tab:test_size}
\end{table}


\section{Quantile Spectral Risk and the Cross-Sections of Expected Returns}
\label{sec:testing}

Here, we discuss how extreme risks are priced in the cross-section of asset returns across horizons. We focus on the results from the standard \cite{fama1973risk} cross-sectional predictive regressions of the three main models and use various cross-sections of asset returns. We show that the quantile spectral risks are priced heterogeneously across various asset classes. This provides a great opportunity for investors who prefer to avoid certain risks. By choosing a specific asset class in which a specific risk is not associated with a risk premium (i.e., assets with high exposure to this risk do not yield an extra premium and vice versa), investors can avoid this risk without paying extra money for it.

First, we investigate returns on individual stocks from the U.S. market. Next, we use standard Fama-French portfolios sorted on various characteristics. More specifically, we use 30 industry portfolios, 25 portfolios sorted on size and value and decile portfolios sorted either on operating profit, investment or book-to-market. Finally, we use three datasets previously introduced in the literature to illustrate some specific phenomena. First, we analyze the dataset of \cite{lettau2014conditional}, which contains portfolios constructed from various asset classes. Second, we analyze equity portfolios sorted by cash flow duration of \cite{WEBER2018486}. Third, we investigate data on investment strategies constructed across various asset classes from \cite{ilmanen2021factor}.

We report models estimated for various threshold values given by the $\tau$ quantile of the market return. We report models estimated for the 1\%, 5\%, 10\%, 15\%, 20\% and 25\% quantiles.\footnote{We had to rescale the data of \cite{lettau2014conditional} and \cite{WEBER2018486} to be comparable to the market return.} Throughout the paper, market return is computed using the value-weight average return on all CRSP stocks. As a risk-free rate, we use the Treasury bill rate from Ibbotson Associates.\footnote{All the data were obtained from \href{http://mba.tuck.dartmouth.edu/pages/faculty/ken.french/data_library.html}{Kenneth French's online data library}.}

\subsection{Individual Stocks}

We collect our data from the Center for Research in Securities Prices (CRSP) database on a monthly basis. The sample spans from July 1926 to December 2015; we select stocks with a long enough history to obtain precise estimates of our measures of risk. While the main results are presented with a sample of stocks with an available history of 60 years, to study the robustness of our results on a larger cross-section of data, we also report results based on stocks with a shorter history of 50 years. On the other hand, one can argue that the precision of the estimated measures of risk relies on the number of observations available in the tail; hence, we also report results based on stocks with 70 years of available history. We report estimation results in Table \ref{tab:coef_3factor}.

\begin{sidewaystable}[ph!]
\scriptsize
\centering
\caption{\textit{Individual stocks.} Prices of risk estimated on monthly stock data from the CRSP database sampled between July 1926 and December 2015. Models are estimated for various values of thresholds given by $\tau$. We employ three samples with varying numbers of minimum years. A long horizon is given by frequencies corresponding to a 3-year cycle and longer. Below the coefficients, we include Fama-MacBeth $t$-statistics.}
\begin{tabular}{cccccccccccccccc}
 \toprule
	 	& & \multicolumn{4}{c}{Tail market risk} & \multicolumn{4}{c}{Extreme volatility risk} & \multicolumn{6}{c}{Full model}\\
		\cmidrule(r){2-6} \cmidrule(r){7-10} \cmidrule(r){11-16}
  & $\tau$ & $\lambda_{\text{long}}^{\text{TR}}$ & $\lambda_{\text{short}}^{\text{TR}}$ & $\lambda^{\text{CAPM}}$ & RMSPE & $\lambda_{\text{long}}^{\text{EV}}$ & $\lambda_{\text{short}}^{\text{EV}}$ & $\lambda^{\text{CAPM}}$ & RMSPE & $\lambda^{\text{TR}}_{\text{\text{long}}}$ & $\lambda^{\text{TR}}_{\text{short}}$ & $\lambda^{\text{EV}}_{\text{long}}$ & $\lambda^{\text{EV}}_{\text{short}}$ & $\lambda^{\text{CAPM}}$ & RMSPE \\ 
\cmidrule(r){2-6} \cmidrule(r){7-10} \cmidrule(r){11-16}
\multirow{12}{*}{\shortstack[c]{70 years \\ (142 assets)}} & 0.01 & -0.059 & 0.657 & 0.754 & 26.683 & -0.086 & -0.275 & 0.960 & 28.542 & 0.125 & 0.620 & -0.210 & -0.168 & 0.834 & 26.387 \\ 
  & & -0.660 & 3.309 & 3.911 &  & -0.972 & -0.887 & 5.001 &  & 0.901 & 3.087 & -1.485 & -0.539 & 4.196 &  \\ 
 & 0.05 & 0.102 & 1.319 & 0.717 & 26.807 & 0.232 & 0.380 & 0.682 & 28.192 & 0.080 & 1.295 & 0.009 & 0.303 & 0.721 & 26.757 \\ 
  & & 0.500 & 3.520 & 3.871 &  & 1.394 & 0.753 & 3.028 &  & 0.221 & 3.534 & 0.030 & 0.595 & 2.978 &  \\ 
 & 0.1 & 0.368 & 1.203 & 0.739 & 27.121 & 0.474 & 0.472 & 0.558 & 27.212 & -0.064 & 1.037 & 0.422 & 0.311 & 0.555 & 26.555 \\ 
  & & 1.483 & 2.304 & 4.144 &  & 2.532 & 0.677 & 2.533 &  & -0.158 & 1.973 & 1.384 & 0.448 & 2.312 &  \\ 
 & 0.15 & 0.544 & 0.895 & 0.733 & 26.672 & 0.509 & 0.552 & 0.602 & 27.016 & 0.242 & 0.882 & 0.326 & 0.538 & 0.618 & 26.289 \\ 
  & & 2.327 & 1.511 & 4.075 &  & 2.783 & 0.778 & 2.903 &  & 0.730 & 1.487 & 1.223 & 0.752 & 2.875 &  \\ 
 & 0.2 & 0.665 & 0.279 & 0.784 & 26.995 & 0.702 & -0.272 & 0.605 & 25.796 & -0.041 & 0.848 & 0.693 & -0.094 & 0.586 & 25.431 \\ 
  & & 2.566 & 0.400 & 4.454 &  & 3.665 & -0.348 & 3.070 &  & -0.116 & 1.250 & 2.601 & -0.120 & 2.868 &  \\ 
 & 0.25 & 0.746 & -0.132 & 0.812 & 27.244 & 0.823 & -0.543 & 0.648 & 25.768 & 0.009 & 0.444 & 0.805 & -0.397 & 0.643 & 25.676 \\ 
  & & 2.805 & -0.181 & 4.662 &  & 3.816 & -0.678 & 3.366 &  & 0.026 & 0.616 & 2.810 & -0.498 & 3.284 &  \\ 
\cmidrule(r){2-6} \cmidrule(r){7-10} \cmidrule(r){11-16}

\multirow{12}{*}{\shortstack[c]{60 years \\ (267 assets)}} & 0.01 & -0.044 & 0.439 & 0.759 & 29.725 & -0.090 & 0.271 & 0.939 & 30.494 & 0.170 & 0.387 & -0.241 & 0.255 & 0.865 & 29.442 \\ 
  & & -0.633 & 2.693 & 4.126 &  & -1.293 & 1.007 & 5.144 &  & 1.431 & 2.391 & -2.010 & 0.939 & 4.659 &  \\ 
 & 0.05 & 0.189 & 1.219 & 0.660 & 28.674 & 0.243 & 0.614 & 0.645 & 29.945 & 0.115 & 1.271 & 0.027 & 0.653 & 0.661 & 28.600 \\ 
  & & 1.068 & 3.653 & 3.573 &  & 1.471 & 1.505 & 2.850 &  & 0.379 & 3.903 & 0.103 & 1.594 & 2.759 &  \\ 
 & 0.1 & 0.315 & 1.000 & 0.718 & 29.243 & 0.503 & 0.420 & 0.511 & 29.201 & -0.161 & 0.939 & 0.509 & 0.443 & 0.493 & 28.676 \\ 
  & & 1.388 & 2.450 & 4.022 &  & 2.471 & 0.830 & 2.281 &  & -0.489 & 2.282 & 1.737 & 0.875 & 2.058 &  \\ 
 & 0.15 & 0.500 & 0.779 & 0.709 & 29.257 & 0.441 & 0.550 & 0.603 & 29.608 & 0.297 & 0.819 & 0.233 & 0.443 & 0.626 & 29.099 \\ 
  & & 2.188 & 1.546 & 3.961 &  & 2.248 & 1.004 & 2.889 &  & 1.042 & 1.630 & 0.958 & 0.805 & 2.972 &  \\ 
 & 0.2 & 0.484 & 0.287 & 0.765 & 29.764 & 0.630 & -0.198 & 0.595 & 28.769 & -0.234 & 0.735 & 0.742 & -0.253 & 0.561 & 28.620 \\ 
  & & 2.002 & 0.537 & 4.359 &  & 3.153 & -0.338 & 3.024 &  & -0.824 & 1.438 & 3.007 & -0.430 & 2.808 &  \\ 
 & 0.25 & 0.487 & 0.242 & 0.785 & 30.045 & 0.619 & -0.463 & 0.661 & 29.389 & -0.053 & 0.574 & 0.629 & -0.443 & 0.657 & 29.341 \\ 
  & & 2.017 & 0.427 & 4.513 &  & 3.084 & -0.764 & 3.505 &  & -0.203 & 1.037 & 2.768 & -0.737 & 3.481 &  \\   
\cmidrule(r){2-6} \cmidrule(r){7-10} \cmidrule(r){11-16}
   
\multirow{12}{*}{\shortstack[c]{50 years \\ (528 assets)}} & 0.01 & -0.089 & 0.441 & 0.823 & 29.727 & -0.099 & 0.478 & 0.970 & 30.281 & 0.001 & 0.410 & -0.096 & 0.439 & 0.873 & 29.683 \\ 
  & & -1.655 & 2.953 & 4.631 &  & -1.783 & 2.508 & 5.420 &  & 0.009 & 2.787 & -1.077 & 2.337 & 4.816 &  \\ 
 & 0.05 & -0.022 & 1.185 & 0.760 & 29.233 & 0.061 & 0.649 & 0.800 & 30.059 & 0.019 & 1.268 & -0.059 & 0.780 & 0.781 & 29.039 \\ 
  & & -0.152 & 3.949 & 4.233 &  & 0.484 & 1.971 & 3.776 &  & 0.067 & 4.400 & -0.258 & 2.421 & 3.439 &  \\ 
 & 0.1 & 0.153 & 0.820 & 0.786 & 29.762 & 0.289 & 0.093 & 0.680 & 29.700 & -0.104 & 0.801 & 0.288 & 0.182 & 0.665 & 29.417 \\ 
  & & 0.794 & 2.450 & 4.436 &  & 1.785 & 0.223 & 3.224 &  & -0.333 & 2.450 & 1.167 & 0.436 & 2.971 &  \\ 
 & 0.15 & 0.348 & 0.862 & 0.767 & 29.600 & 0.148 & 0.023 & 0.783 & 30.112 & 0.476 & 0.830 & -0.149 & 0.101 & 0.813 & 29.570 \\ 
  & & 1.832 & 2.034 & 4.307 &  & 0.921 & 0.051 & 3.891 &  & 1.733 & 1.984 & -0.670 & 0.222 & 3.945 &  \\ 
 & 0.2 & 0.272 & 0.683 & 0.810 & 29.973 & 0.346 & -0.267 & 0.735 & 29.791 & -0.002 & 0.743 & 0.314 & -0.298 & 0.735 & 29.704 \\ 
  & & 1.410 & 1.437 & 4.605 &  & 2.011 & -0.566 & 3.789 &  & -0.009 & 1.605 & 1.374 & -0.622 & 3.695 &  \\ 
 & 0.25 & 0.257 & 0.822 & 0.821 & 30.035 & 0.322 & -0.051 & 0.765 & 30.004 & 0.034 & 0.946 & 0.281 & -0.081 & 0.764 & 29.885 \\ 
  & & 1.316 & 1.594 & 4.704 &  & 1.885 & -0.106 & 4.082 &  & 0.152 & 1.848 & 1.388 & -0.167 & 4.070 &  \\
   \bottomrule
\end{tabular}
\label{tab:coef_3factor}
\end{sidewaystable}

Models are estimated for different values of the threshold value given by the $\tau$ market quantile to capture the different probabilities of event co-occurrences. The results of the TR model show that the relative TR beta for short horizons is more significant for low values of $\tau$, corresponding to 0.01, 0.05 and 0.10, while for $\tau \ge 0.15$, the relative TR beta becomes significant for long horizons. This pattern is observed across all three samples, but it is weaker among stocks with a history of 50 years, especially regarding the prices of risk corresponding to the long relative TR betas. This result may be caused by the fact that long relative TR betas require a longer history of data to obtain precise estimates in comparison to the short TR betas.

Signs of the estimated prices of risk are intuitive. More extreme dependence between market and asset returns in both horizons leads to a higher risk premium, as we may expect. If an asset is likely to deliver poor performance when the market is in a downturn, this asset is not desirable from the point of view of an investor, and to decide to hold such asset, she would require a significant risk premium. From the magnitude of the coefficients, we infer that investors price tail risk in the short term more than in the long term. Moreover, it is important to note that these features are not subsumed by the CAPM beta, as we explicitly control for it in the model, and report TR betas relative to the CAPM beta, as discussed above.

Estimation results for the EVR model are captured in the middle panel of Table \ref{tab:coef_3factor}. In this case, parameters are not significant for low values of $\tau$, but starting with $\tau \ge 0.1$, long EVR becomes significantly priced in the cross-section. On the other hand, short-horizon EVR risk is not significantly priced for any values of $\tau$.

Significant prices of risk corresponding to long-horizon EVR betas for $\tau \ge 0.10$ possess intuitive positive signs, as we expected. The EVR betas capture dependence between extremely high increments of market volatility\footnote{Note that we work with negative increments of market volatility when we estimate the QS betas.} and extremely low asset returns, and the results are consistent with the current literature \citep{boons2015horizon,doi:10.1111/jofi.12058,adrian2008stock}. Moreover, these results are in line with the conclusions of long-run risk models. We observe few instances of unintuitive negative signs of prices of risk, but these coefficients are insignificant and observed mostly for low values of $\tau$, which may be caused by the measurement error for the corresponding betas. We may conclude that EVR betas, especially their long-term component, provide priced information regarding risk, which is moreover orthogonal to the information featured in the CAPM beta.

In terms of the RMSPE, the TR model delivers better results than the EVR model for low values of $\tau$, as short TR betas are significantly priced for these values of $\tau$. On the other hand, for higher values of $\tau$, the EVR model delivers improved values of RMSPE, as the long EVR betas for these $\tau$ values deliver a significant dimension of risk priced in the cross-section and TR betas possess higher explanatory power for lower values of $\tau$.

Moreover, we identify the fact that there is a complex interplay between the horizons and parts of the joint distribution priced in the cross-section. Extreme TR is mostly a short-run phenomenon, and TR associated with more probable joint events (higher values of $\tau$) is priced with respect to long-term dependence between the market and assets. On the other hand, EVR is not significantly priced in cases of extreme joint events, but as unpleasant events become more probable, the joint dependence between increments of market volatility and asset return in the long run becomes a significant determinant of risk premiums. In Table \ref{tab:coef_3factor_alt} in Appendix \ref{app:diff_long}, we present the results for 1.5-years being the threshold in the definition of the long horizon. The results are qualitatively very similar, and all the findings from the 3-year horizon hold for this case.

From the results above, we can conclude that tail market and extreme market volatility risks are priced in the cross-section of stock returns across different horizons. A natural question arises whether these risks capture different information or whether one measure can subsume the other. For this purpose, we test the full model, which contains both risks for a given $\tau$ level at the same time. Estimated parameters can be found in the right panel of Table \ref{tab:coef_3factor}. We observe results mostly consistent with the outcomes of the separate TR and EVR models. Significantly priced determinants of the risk are short-term TR for low values of $\tau$ and long-term EVR for the higher values of $\tau$, both priced across assets with expected positive signs. Tail risk is more significant for lower values of $\tau$, meaning that dependence between market return and asset return during extremely negative events is a significant determinant of the risk premium. On the other hand, long-term extreme volatility risk is significant for higher values of $\tau$ - approximately 0.2. This finding suggests that investors price downside dependence between asset returns and market volatility but focus on more probable market situations. We can deduce that the price of long-run risk mentioned by \cite{bansal2004risks} is hidden in this coefficient.

The main deviation of the full model from the results of the separate TR and EVR models is that the long TR betas for higher values of $\tau$ become insignificant, in contrast with the conclusions from the TR model. One potential explanation for this result is that only a small fraction of the market return fluctuations are due to its long-term component in comparison to the short-term component, and thus, the risk premium for this risk is only small. Another explanation is that the long-term aspect of the market tail risk may be fully captured by the extreme volatility risk, namely, the long TR betas are subsumed by the long EVR betas. This makes sense since variance is much more persistent than the market return (high portion of variance due to the long-term component) and thus investors fear the fluctuation in long-term variance much more than the variance in the short term.

In Appendix \ref{app:qs_features}, we use this data sample and show various features of the estimated QS betas. We present distributions of the estimated QS betas to give a notion of their estimated values. Next, we investigate the relation between QS betas and other risk measures previously proposed in the literature. Although the QS measures are correlated with some of the other variables discussed previously in the literature, they do not drive out the QS measures of risk. Moreover, these variables are, in most cases, subsumed by the variables from the full model. Our results are in agreement with recent results of \cite{bollerslev2020realized}, which show that the dependence characterized by the \textit{co-occurrence} of negative asset and negative market returns possesses the highest explanatory power on the formation of asset returns among all specifications of disaggregated conventional beta. Importantly, we explicitly show that the premium for this risk is generated by the dependence in the extreme left tail and by its short-term component. In addition, we extend the analysis to extreme volatility risk and show that investors focus on more probable joint negative outcomes that unfold over the long horizon.

\subsection{Other Portfolios}

Finally, we investigate the pricing implication across multiple datasets, including popular Fama-French portfolios sorted on various characteristics. We use 30 industry portfolios, 25 portfolios sorted by size and value and decile portfolios sorted by operating profit, investment or book-to-market portfolios of \cite{lettau2014conditional} constructed from various asset classes, equity portfolios sorted on cash flow duration of \cite{WEBER2018486} and finally investment strategies constructed across various asset classes from \cite{ilmanen2021factor}.

\begin{landscape}
\begin{figure}
        \centering
        \caption{\footnotesize \textit{Estimated t-statistics of quantile spectral risk for various portfolios. For each $\tau \in \{0.01,0.05,0.1,0.15,0.2,0.25\}$ quantile, we show Fama-MacBeth $t$-statistics of estimated prices of risk of the QS models. Vertical line Group 11 portfolios depicted by different symbols. The $t$-statistics of short-run (red color) and long-run (blue color) risks are shown for tail risk (TR) models by a solid line, extreme volatility risk (EVR) by a dashed line and the full model by dotted lines. Horizontal lines depict 1\% and 99\% quantiles of the standard normal distribution.}}
        \includegraphics[scale=0.45]{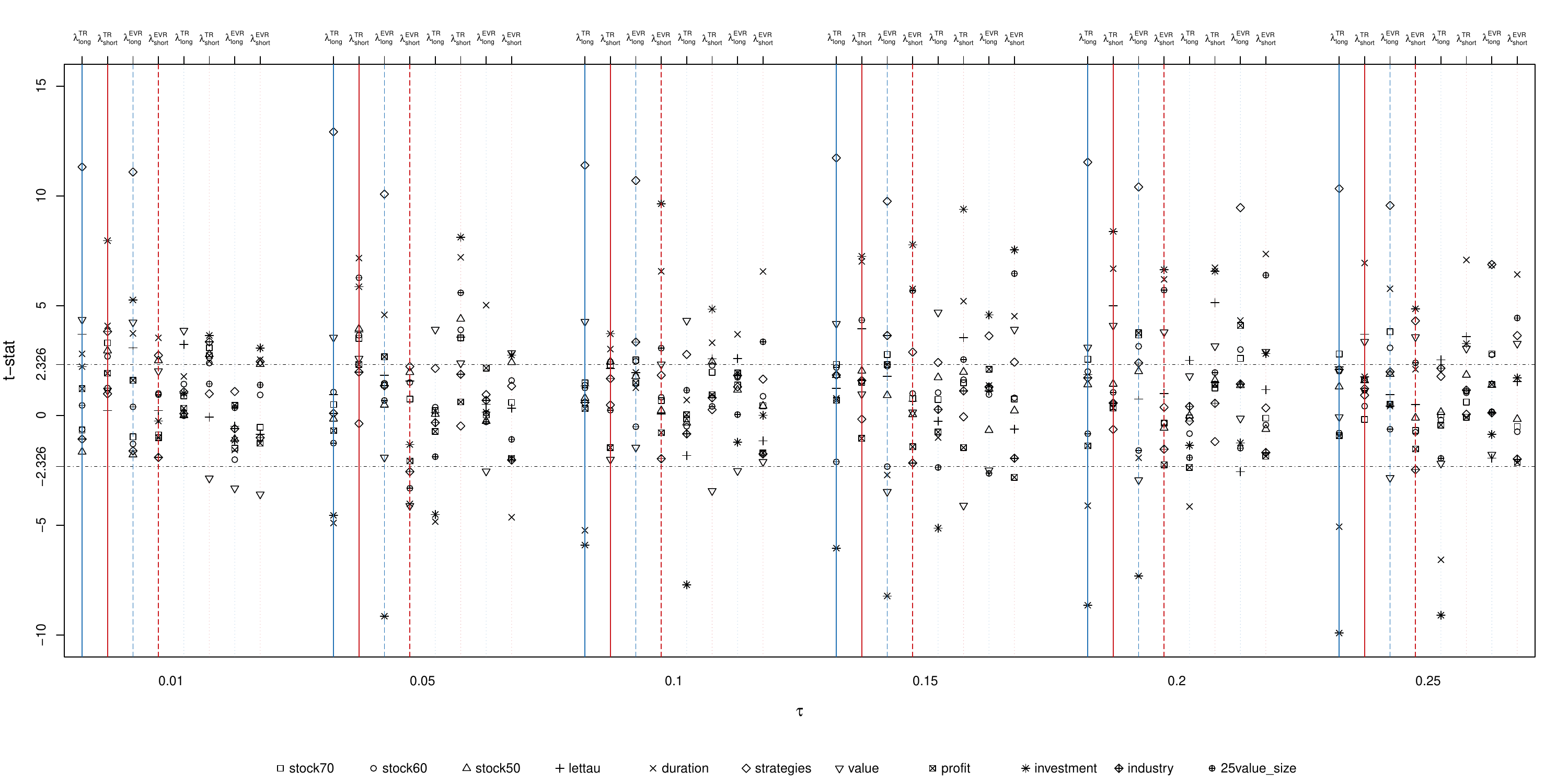}
    \label{fig:tstat_all}
\end{figure}
\end{landscape}

Figure \ref{fig:tstat_all} summarizes the estimation results for all these data. We report $t-$ statistics of estimated prices of QR risks over all portfolios and across tails, which gives a general overview of how tail- and horizon-specific risks are priced across a wide number of portfolios. Appendix \ref{app:portfolios_detailed} then provides a detailed summary of all results as well as a data description.

We conclude that a phenomenon of short-term tail risk (TR) is universally priced (although with varying magnitude) across most of the datasets. The results of EVR are slightly more mixed. In the case of individual stocks, it is mostly the long-term part of the EVR that is priced in the cross-section of the expected returns. The same is also true for the aggregated dataset of \cite{lettau2014conditional}. On the other hand, in the case of 25 portfolios sorted on size and value, the short-term part of EVR is priced. There are also datasets in which both components of the EVR risk are priced. These include equity portfolios sorted on cash flow duration of \cite{WEBER2018486} and investment strategies constructed using various asset classes of \cite{ilmanen2021factor}. This heterogeneity gives investors the opportunity to follow certain investment strategies according to their aversion to certain risk in a given horizon.


\section{Conclusion}
\label{sec:conclusion}

We introduce a novel approach for isolating the effects of various risk dimensions on the formation of expected returns. Until now, studies have focused either on exploring downside features of risk or on investigating its horizon-specific properties. We define novel measures that estimate risk in a specific part of the joint distribution over a specific horizon, and we show that extreme risks are priced in a cross-section of asset returns heterogeneously across horizons. Furthermore, we argue that it is important to distinguish between tail market risk and extreme volatility risk. Tail market risk is characterized by the dependence between a highly negative market and asset events. Extreme volatility risk is defined as the co-occurrence of extremely high increases in market volatility and highly negative asset returns. Negative events are derived from the distribution of market returns, and their respective quantiles are used to determine threshold values for computing quantile spectral betas.

To consistently estimate the models, data with a sufficiently long history must be employed. However, if these data are available, our measures of risk are able to outperform competing measures, and their performance is best for low threshold values, suggesting that investors require a risk premium for holding assets susceptible to extreme risks. Moreover, we show that the state-of-the-art downside risk measures do not capture the information contained in our newly proposed measures. Our results have important implications for asset pricing models. We show that only taking into account contemporaneous dependence averaged over the whole distribution when measuring risk exposure leads to the omitting of important information regarding the risk.

\setlength{\bibsep}{3pt}
\bibliographystyle{chicago}
\bibliography{bibliography}


\newpage
\clearpage
\appendix

\section{Technical Appendix}
\label{app:main}

In this section the proof of the results in Section~\ref{sec:AsympQspec} is given. Before we begin, note that by a trivial generalisation of Proposition~3.1 in \cite{kley2014} we have that Assumption~\ref{ass:exp_alpha_mix} implies that there exist constants $\rho\in(0,1)$ and $K < \infty$ such that, for arbitrary intervals $A_m,A_r \subset \mathbb{R}$, arbitrary times $t_m,t_r \in \IZ$,
\begin{equation}\label{eq:boundcum}
|\cum(I\{m_{t_m} \in A_m\},I\{r_{t_r} \in A_r\})| \leq K \rho^{|t_m-t_r|}.
\end{equation}
In addition, we will use the following lemma

\begin{lemma} (\cite{barunik2019quantile}) \label{lem:LipschitzLaplaceSD}
Under the assumptions of Theorem~\ref{thm:AsympDensityRankEstimator}, the derivative
\[\displaystyle{(\tau_m, \tau_r) \mapsto \frac{{\rm d}^k}{{\rm d}\omega^k}\mathfrak{f}^{m, r}(\omega; \tau_m, \tau_r)}\]
exists and satisfies, for any $k \in \IN_0$ and some constants $C,d$ that are independent of $a=(a_m,a_r),b=(b_m,b_r)$, but may depend on $k$,
\[
\sup_{\omega \in \IR } \Big|\frac{{\rm d}^k}{{\rm d}\omega^k}\mathfrak{f}^{m, r}(\omega; a_m, a_r)-\frac{{\rm d}^k}{{\rm d}\omega^k} \mathfrak{f}^{m, r}(\omega; b_m, b_r)\Big| \leq C \|a-b\|_1 (1+|\log\|a-b\|_1|)^D.
\]
\end{lemma}

Following proposition further provides asymptotic properties of  $I^{m, r}_{n,R}(\omega; \tau_m, \tau_r)$ 

\begin{proposition}\label{prop:inr} (\cite{barunik2019quantile})
Assume that $(\bX_t)_{t\in\IZ}$ is strictly stationary and satisfies Assumption~\ref{ass:exp_alpha_mix}. Further assume that the marginal distribution functions $F_m$, and $F_r$ are continuous. Then, for every fixed $\omega \neq 0 \mod 2\pi$,
\begin{equation}
\label{eq:IntoI}
\Big( I_{n, R}^{m,r}(\omega; \tau_m, \tau_r) \Big)_{ (\tau_m,\tau_r) \in [0,1]^2} \Rightarrow
\Big(\frac{1}{2\pi} \mathbb{D}^m(\omega;\tau_m) \mathbb{D}^r(-\omega;\tau_r) \Big)_{ (\tau_m,\tau_r) \in [0,1]^2},
\end{equation}
where $\mathbb{D}^m(\omega; \tau_m) $ and $\mathbb{D}^r(\omega; \tau_r) $, $\tau \in [0,1]$, $\omega \in \IR$ are centered, $\IC$-valued Gaussian processes with covariance structure of the following form
\[
\Cov(\mathbb{D}^{m}(\omega; \tau_m), \mathbb{D}^{r}(\omega; \tau_r))
= 2 \pi \mathfrak{f}^{m, r}(\omega; \tau_m, \tau_r).
\]
Moreover, the family~$\{\mathbb{D}^m(\omega; \, \cdot),\mathbb{D}^r(\omega; \, \cdot) \ : \ \omega \in [0,\pi] \}$ is a collection of independent processes. In particular, the weak convergence~(\ref{eq:IntoI}) holds jointly for any finite  fixed collection of frequencies~$\omega$.
\end{proposition}
For $\omega = 0 \mod 2\pi$ the asymptotic behaviour of the rank-based copula cross-periodogram is as follows: we have $d^j_{n,R}(0; \tau) = n \tau + o_p(n^{1/2})$, where the exact form of the remainder term depends on the number of ties in the process. Therefore, under the assumptions of Proposition~\ref{prop:inr}, we have $I_{n, R}^{m,r}(0; \tau_m, \tau_r) = n(2\pi)^{-1}\tau_m\tau_r 1 1' + o_p(1)$, where $1 := (1, 1)' \in \IR^2$.

\subsection{Proof of the Theorem~\ref{thm:AsympQSbeta}}
\label{app:proofmain}

\begin{proof}
By a Taylor expansion we have, for every $y, y_0 > 0$,

\[\frac{1}{y} - \frac{1}{y_0} = - \frac{1}{y_0^2} (y-y_0) + 2 \xi_{y,y_0}^{-3} (y-y_0)^2,\]
where $\xi_{y,y_0}$ is between $y$ and $y_0$. Let $R_n(y,y_0) := 2 \xi_{y,y_0}^{-3} (y-y_0)^2$, then
\begin{equation}
\label{eqn:expansion}
  \frac{x}{y} - \frac{x_0}{y_0}
  = \frac{x}{y} - \frac{x}{y_0} + \frac{x}{y_0} - \frac{x_0}{y_0}
  = \frac{1}{y_0} (y-y_0) - \frac{x_0}{y_0^2} (x - x_0) + r_n,
\end{equation}
where $r_n  = x R_n(y,y_0) + (x - x_0)^2 / y_0^2$

Write $\ff{a}{b}$ for $\mathfrak{f}^{a, b}(\omega; \tau_a, \tau_b)$, $G_{a,b}$ for $\widehat G_{n,R}^{a, b} (\omega; \tau_a, \tau_b)$, and $B_{a,b}$ for $B_n^{a,b,(k)}(\omega; \tau_a, \tau_b)$ and let
\begin{align*}
x   & := G_{a,b}      & y   & := G_{a,a} \\
x_0 & := \ff{a}{b} + B_{a,b}  &  y_0 & := \ff{a}{a} + B_{a,a}
\end{align*}
By Theorem \ref{thm:AsympDensityRankEstimator} differences $x-x_0$ and $y-y_0$ are in $O_p ((n b_n)^{-1/2})$, uniformly with respect to $\tau_m, \tau_r$. Under the assumption that $n b_n \rightarrow \infty$, as $n \rightarrow \infty$, this entails $G_{a,a} - B_{a,a} \rightarrow \ff{a}{a}$, in probability. For $\varepsilon \leq \tau_1, \tau_2 \leq 1-\varepsilon$, we have $\ff{a}{a} > 0$, such that, by the Continuous Mapping Theorem we have $(G_{a,a} - B_{a,a})^{-3} \rightarrow \ff{a}{a}^{-3}$, in probability. As $B_{a,a} = o(1)$, we have $y^{-3} - y_0^{-3} = o_p(1)$. Finally, due to
\[\xi_{y,y_0}^{-3} \leq y_n^{-3} \vee y_0^{-3} \leq (y_n^{-3} - y_0^{-3}) \vee 0 + y_0^{-3} = o_p(1) + O(1) = O_p(1),\]
we have that $R_n(y,y_0) = O_p((n b_n)^{-1})$. \\
So we have shown that 
\begin{multline*}
    \widehat{\beta}^{m, r}_{n,R}(\omega; \tau_m, \tau_r) - \frac{\ff{a}{b} + B_{a,b}}{\ff{a}{a} + B_{a,a}} 
    = \frac{1}{\ff{m}{m}} \Big([G_{m,m} - \ff{m}{m} - B_{m,m}]- \frac{\ff{m}{r}}{\ff{m}{m}} [G_{m,r} - \ff{m}{r} - B_{m,r}]\Big) + O_p \big(1/ (n b_n) \big),
\end{multline*}

with the $O_p$ holding uniformly with respect to $\tau_m,\tau_r$.Furthermore, note that setting
\begin{align*}
x   & := \ff{a}{b} + B_{a,b}     & y   & := \ff{a}{a} + B_{a,a}\\
x_0 & := \ff{a}{b}  &  y_0 & := \ff{a}{a}
\end{align*}
we have
\begin{equation*}
\begin{split}
\frac{\ff{a}{b} + B_{a,b}}{\ff{a}{a} + B_{a,a}}
- \frac{\ff{a}{b}}{\ff{a}{a}}
= \frac{1}{\ff{a}{a}} \Big( B_{a,a} - \frac{\ff{a}{b}}{\ff{a}{a}} B_{a,b} \Big)
+ O(|\BB{a}{a}|^2 + |\BB{a}{b}|^2).
\end{split}
\end{equation*}
By Lemma \ref{lem:LipschitzLaplaceSD} we have that 
\begin{equation*}
\sup_{\tau_m, \tau_r \in [\varepsilon, 1-\varepsilon]} \Big| \frac{{\rm d}^{\ell}}{{\rm d}\omega^{\ell}}\mathfrak{f}^{m, r}(\omega; \tau_m, \tau_r) \Big| \leq C_{\varepsilon,\ell}.
\end{equation*}
Therefore, $B_{m,r}$ satisfies
\begin{equation*}
\sup_{\tau_m, \tau_r \in [\varepsilon, 1-\varepsilon]} \Big| \sum_{\ell=2}^k \frac{b_n^{\ell}}{\ell!} \int_{ -\pi}^{\pi} v^{\ell} W(v) dv \frac{{\rm d}^{\ell}}{{\rm d}\omega^{\ell}}\mathfrak{f}^{m, r}(\omega; \tau_m, \tau_r) \Big| = o\big( (n b_n)^{-1/4} \big),
\end{equation*}
which implies that
\[|\BB{a}{a}|^2 + |\BB{a}{b}|^2 = o\big( (n b_n)^{-1/2} \big).\] 
Therefore,
\[\sqrt{n b_n} \Bigg( \widehat{\beta}^{m, r}_{n,R}(\omega; \tau_m, \tau_r) - \underbrace{\frac{\ff{a}{b}}{\ff{a}{a}}}_{=: \beta^{m,r}(\omega;\tau_m,\tau_r)} - \underbrace{\frac{1}{\ff{a}{a}} \Big( \BB{a}{a} - \frac{\ff{a}{b}}{\ff{a}{a}} \BB{a}{b} \Big)}_{=: B_n^{m,r,(k)}(\omega; \tau_m, \tau_r)} \Bigg)\]
and
\[\sqrt{n b_n} \frac{1}{\ff{m}{m}} \Big([G_{m,r} - \ff{m}{r} - B_{m,r}]- \frac{\ff{m}{r}}{\ff{m}{m}} [G_{m,m} - \ff{m}{m} - B_{m,m}]\Big)\]
are asymptotically equivalent in the sense that if one of the two converges weakly, then so does the other.
The assertion then follows by Theorem~\ref{thm:AsympDensityRankEstimator}, Slutzky's lemma and the Continuous Mapping Theorem.\hfill

\end{proof}

\subsection{Construction of pointwise confidence bands for Quantile Spectral Beta}
\label{app:intervals}

Following \cite{barunik2019quantile} and Theorem \ref{thm:AsympQSbeta}, we construct pointwise asymptotic $(1-\alpha)$ level confidence bands for the real and imaginary parts of $\beta_{n,R}^{m,r}(\omega;\tau_m,\tau_r)$ as follows:

\[C^{(2)}_{{\rm r},n}(\omega_{kn}; \tau_m, \tau_r) := \Re \widehat{\beta}^{m, r}_{n,R}(\omega_{kn}; \tau_m, \tau_r) \pm \Re \sigma_{(2)}^{m, r}(\omega_{kn}; \tau_m, \tau_r) \Phi^{-1}(1-\alpha/2),\]
for the real part, and 
\[C^{(2)}_{{\rm i},n}(\omega_{kn}; \tau_, \tau_r) := \Im \widehat{\beta}^{m, r}_{n,R}(\omega_{kn}; \tau_m, \tau_r) \pm \Im \sigma_{(2)}^{m, r}(\omega_{kn}; \tau_m, \tau_r) \Phi^{-1}(1-\alpha/2),\]
for the imaginary part of the quantile spectral beta. Here, $\Phi$ stands for the cdf of the standard normal distribution,
\begin{equation*}
\big( \Re \sigma_{(2)}^{m, r}(\omega_{kn}; \tau_m, \tau_r) \big)^2
:= 0 \vee \begin{cases}
      0 & \text{if $m = r$} \\
        & \text{\quad and $\tau_m = \tau_r$,} \\
      \frac{1}{2} \big(\mathbb{C}ov( \LL{m}{r}, \LL{m}{r} ) + \Re \mathbb{C}ov( \LL{m}{r}, \LL{r}{m} ) \big) & \text{otherwise},
\end{cases}
\end{equation*}
and
\begin{equation*}
\big( \Im \sigma_{(2)}^{m, r}(\omega_{kn}; \tau_m, \tau_r) \big)^2
:= 0 \vee \begin{cases}
      0 & \text{if $m = r$}\\
        & \text{\quad  and $\tau_m = \tau_r$,}\\
      \frac{1}{2} \big(\mathbb{C}ov( \LL{m}{r}, \LL{m}{r} ) - \Re \mathbb{C}ov( \LL{m}{r}, \LL{r}{m} ) \big) & \text{otherwise}.
\end{cases}
\end{equation*}
where $\LL{a}{b}=\frac{1}{\ff{a}{a}} \Big(\mathbb{H}_{a,a}- \frac{\ff{a}{b}}{\ff{a}{a}} \mathbb{H}_{a,b}\Big)$. The definition of $\sigma_{(2)}^{m, r}(\omega_{kn}; \tau_m, \tau_r)$ is motivated by noting that for any complex-valued random variable $Z$, with complex conjugate $\bar{Z}$,
\begin{equation}\label{eqn:rem:CI:f:2}
\Var(\Re Z) = \frac{1}{2} \big( \Var(Z) + \Re \mathbb{C}ov(Z, \bar{Z})\big);\ \
\Var(\Im Z) = \frac{1}{2} \big( \Var(Z) - \Re \mathbb{C}ov(Z, \bar{Z})\big),
\end{equation}
and we have $\overline{\LL{m}{r}} = \LL{r}{m}$. Furthermore, note that $\widehat{\beta}^{m, r}_{n,R}(\omega_{kn}; \tau_m, \tau_r) = 1$, if $m = r$ and $\tau_m = \tau_r$. We have used $\mathbb{C}ov( \LL{a}{b}, \LL{c}{d} )$ to denote an estimator for
\[\mathbb{C}ov\big(\IL^{a, b}(\omega_{kn}; \tau_a, \tau_b \big), \IL^{c, d}(\omega_{kn}; \tau_c, \tau_d )\big).\]
Recalling the definition of the limit process in Theorem~\ref{thm:AsympQSbeta} we derive the following expression:
\begin{equation*}
\begin{split}
  \mathbb{C}ov( \LL{a}{b}, \LL{c}{d} ) &= \frac{1}{\ff{a}{a} \ff{c}{c}}
  \mathbb{C}ov \Big( \HH{a}{a} - \frac{\ff{a}{b}}{\ff{a}{a}} \HH{a}{b},
  \HH{c}{c} - \frac{\ff{c}{d}}{\ff{c}{c}} \HH{c}{d} \Big) \\
  & = \frac{\mathbb{C}ov( \HH{a}{a}, \HH{c}{c})}{\ff{a}{a} \ff{c}{c}} 
    - \frac{\overline{\ff{c}{d}} \mathbb{C}ov( \HH{a}{a}, \HH{c}{d})}{\ff{a}{a} \ff{c}{c}^2 } \\
  & -  \frac{\ff{a}{b} \mathbb{C}ov( \HH{a}{b}, \HH{c}{c})}{\ff{a}{a}^2 \ff{c}{c} }
    +\frac{\ff{a}{b}\overline{\ff{c}{d}} \mathbb{C}ov( \HH{a}{b}, \HH{c}{d})}{\ff{a}{a}^2 \ff{c}{c}^2},
  \end{split}
\end{equation*}
where we have written $\ff{i}{j}$ for the quantile spectral density $\mathfrak{f}^{i, j}(\omega_{kn}; \tau_i, \tau_j)$, and $\HH{i}{j}$ for the limit distribution $\IH^{i, j}(\omega_{kn}; \tau_i, \tau_j\big)$ for any $i,j = a,b,c,d$.

Thus, considering the special case where $a = c = m$ and $b = d = r$, we have
\begin{equation*}
\begin{split}
   \mathbb{C}ov( \LL{m}{r}, \LL{m}{r} ) &= \frac{1}{\ff{m}{m}^2}\mathbb{C}ov( \HH{m}{m}, \HH{m}{m})
    - \frac{\ff{r}{m}}{\ff{m}{m}^3}\mathbb{C}ov( \HH{m}{m}, \HH{m}{r}) \\
    &- \frac{\ff{m}{r}}{\ff{m}{m}^3}\mathbb{C}ov( \HH{m}{r}, \HH{m}{m})
    + \frac{|\ff{m}{r}|^2}{\ff{m}{m}^4}\mathbb{C}ov( \HH{m}{r}, \HH{m}{r}).
  \end{split}
\end{equation*}
and for the special case where $a = d = m$ and $c = b = r$ we have 
\begin{equation*}
\begin{split}
   \mathbb{C}ov( \LL{m}{r}, \LL{r}{m} ) &= \frac{1}{\ff{m}{m}\ff{r}{r}}\mathbb{C}ov( \HH{m}{m}, \HH{r}{r})
    - \frac{\ff{m}{r}}{\ff{m}{m}\ff{r}{r}^2}\mathbb{C}ov( \HH{m}{m}, \HH{r}{m}) \\
    &- \frac{\ff{m}{r}}{\ff{m}{m}^2\ff{r}{r}}\mathbb{C}ov( \HH{m}{r}, \HH{r}{r})
    + \frac{\ff{m}{r}^2}{\ff{m}{m}^2\ff{r}{r}^2}\mathbb{C}ov( \HH{m}{r}, \HH{r}{m}).
  \end{split}
\end{equation*}
Finally, we substitute consistent estimators for the unknown quantities. To do so we abuse notation using $\ff{a}{b}$ to denote $\tilde{G}^{a, b}_{n,R}(\omega_{kn}; \tau_a, \tau_b)$ and motivated by Theorem~7.4.3 in~\cite{Brillinger1975}, we use 
\begin{multline}\label{eqn:rem:CI:f:1}
\Big(\frac{2\pi}{n \cdot W_n^k}\Big)
\times \Bigg[ \sum_{s=1}^{n-1} W_n\big(2\pi (k -  s) / n \big) W_n\big(2\pi (k -  s) / n \big) \tilde G^{a, c}_{n,R}(\tau_a, \tau_c; 2\pi s / n) \tilde G^{b, d}_{n,R}(\tau_b, \tau_d; - 2\pi s / n) \\
+ \sum_{s=1}^{n-1} W_n\big(2\pi (k -  s) / n \big) W_n\big(2\pi (k +  s) / n \big) \tilde G^{a, d}_{n,R}(\tau_a, \tau_d; 2\pi s / n)  \tilde G^{b, c}_{n,R}(\tau_b, \tau_c; -2\pi s / n)   \Bigg]
\end{multline}
to estimate $\Cov( \HH{a}{b}, \HH{c}{d})$.

\newpage
\clearpage
\section{Rare Disaster Risk Model and QS Betas}
\label{app:rare_disaster}

We show how the QS betas relate to the asset pricing model of \cite{10.1257/mac.5.3.35}. Their extension of disaster risk model originally proposed by \cite{RIETZ1988117} and \cite{barro2006rare} enables disasters to unfold over multiple periods and partially recover after the disaster. We argue that the QS betas can capture the complex joint dynamics between consumption growth and equity return. To do that, we simulate consumption growth and solve for equity return from three specification of the rare disaster model: 1) Model in which a disaster unfolds over multiple periods, and then a partial recovery occurs. 2) Model with unfolding disaster over multiple periods, but the disaster is permanent. 3) Model with one period disaster which is permanent. We assume preferences of \cite{epstein1989substitution} and \cite{10.2307/2937817} and follow \cite{10.1257/mac.5.3.35} in the estimation procedure using their dataset, solution procedure and values of preference parameters.\footnote{The code supplementing \cite{10.1257/mac.5.3.35} can be downloaded from \url{https://eml.berkeley.edu/~enakamura/papers.html}} Namely, we set the CRRA, $\gamma = 6.5$, the IES, $\psi = 2$ and the discount factor, $\beta = \textrm{exp}(-0.034)$.

\begin{figure}
        \centering
        \caption{\footnotesize \textit{QS Betas between Consumption Growth and Equity Return.}. First row depicts typical disasters for various specifications of rare disaster risk model as specified in \cite{10.1257/mac.5.3.35}. Second row captures QS betas and their 90\% confidence intervals for those specifications. For each specification, QS betas are estimated using 100 simulations of consumption growth series and equity return of length 50,000. Models and parameter values follow \cite{ 10.1257/mac.5.3.35}.}
        \includegraphics[width=\textwidth]{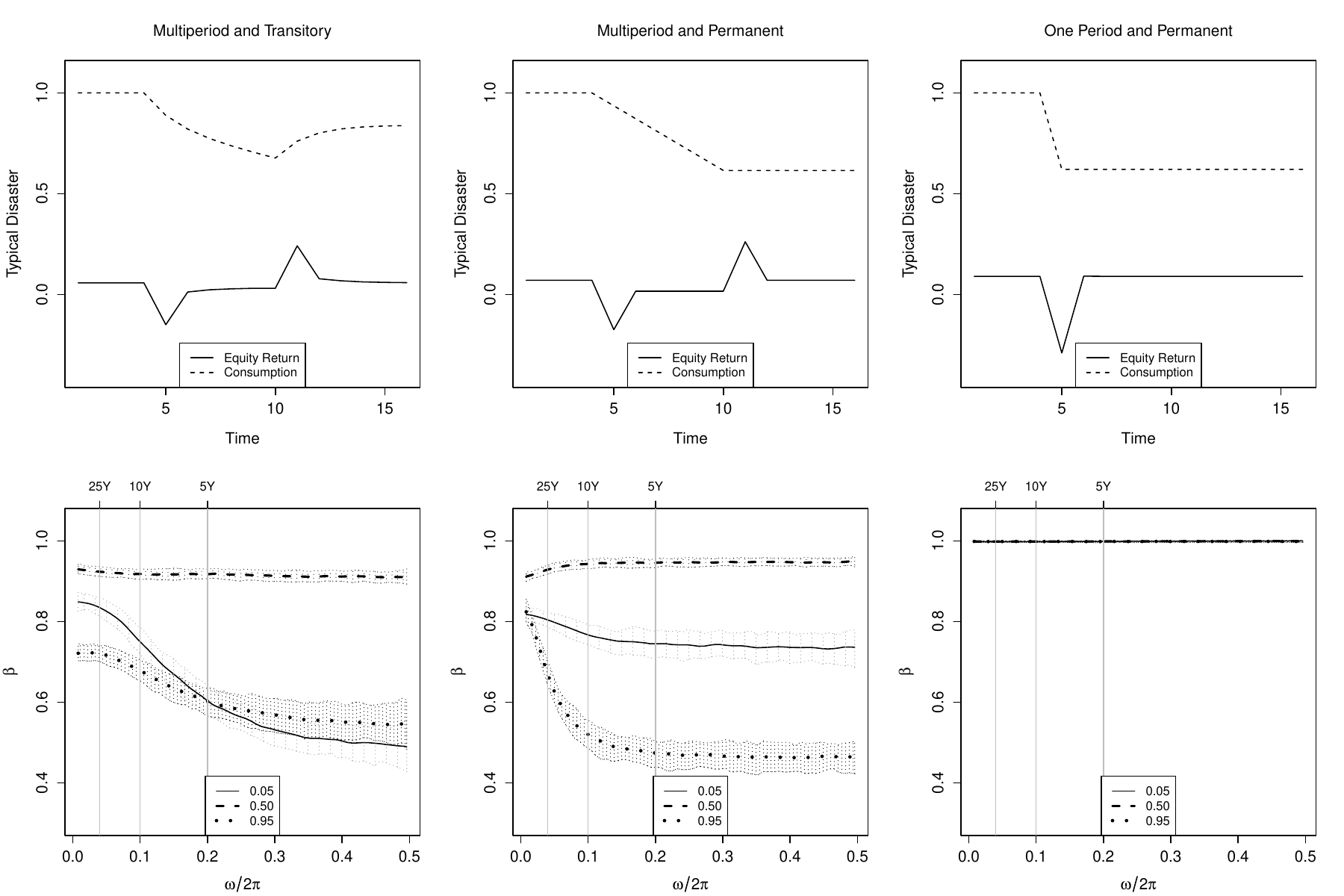}
    \label{fig:rare_disaster}
\end{figure}

Figure \ref{fig:rare_disaster} presents the main results. The first row of the figure contains courses of typical disasters with respect to the detrended consumption and equity return (return on unleveraged consumption claim). We observe that at the onset of the disaster (first drop of the consumption), there is a visible contemporaneous drop at equity return, as well. In case of unfolding disasters, after the end of the disaster period, there is a noticeable positive jump in the equity return. The lower panel of the figure contains QS betas and their 90\% confidence intervals simulated from the respective models. Each model is simulated 100 times and each simulation produces a time series of length 50,000 years (we simulate yearly observations).

We can see that the dependence in the median (given by the line corresponding to $\tau = 0.50$) does not dramatically differ across the specifications and is constant over horizons. This implies that using a simple covariance based measure, we cannot distinguish between joint dynamics across different specifications. The most important part of the joint structure contain the tails of the joint distribution over specific horizons. We may think of one period and permanent specification as a benchmark specification. In this case, on average, the extreme events occur contemporaneously and thus the beta across horizons is flat. If we look at the cases with unfolding disasters, the QS betas for the left tail due to the persistency of the disaster posses its peak at the longer horizons. For the case of multiperiod and transitory disaster, the QS betas for the upper tail are very similar to the QS betas for the lower tail, because after the end of the disaster, consumption partially recovers over multiple periods, which mirrors the joint dynamics at the onset of the disaster. On the other hand, in case of multiperiod permanent disaster, at the end of the disaster, there is a positive jump in equity return, but there is no recovery in the consumption. This makes the QS betas peaking at the longer horizons, as there is typically no contemporaneous positive jump in consumption growth and equity return at the end of the disaster.

\newpage
\clearpage
\section{Features of QS Betas}
\label{app:qs_features}

\subsection{Summary Statistics about Quantile Spectral Betas}

\begin{table}[ht!]
\scriptsize
\centering
\caption{\textit{Descriptive Statistics.} Table summarizes basic descriptive statistics and correlation structure for all betas from our Full model for the two choices of the quantile levels. Betas are computed using CRSP database sampled between July 1926 and December 2015. Presented results are computed on our largest sample, i.e., using stocks with at least 50 years of history. Long horizon is given by frequencies corresponding to 3-year cycle and longer.}
\begin{tabular}{ccccccccccc}
  \toprule
  		& \multicolumn{5}{c}{$\tau = 0.05$} & \multicolumn{5}{c}{$\tau = 0.10$}\\
		\cmidrule(r){2-6} \cmidrule(r){7-11}
  & $\beta^{CAPM}$ & $\beta_{long}^{rel}$ & $\beta_{short}^{rel}$ & $\beta_{long}^{EVR}$ & $\beta_{short}^{EVR}$ & $\beta^{CAPM}$ & $\beta_{long}^{rel}$ & $\beta_{short}^{rel}$ & $\beta_{long}^{EVR}$ & $\beta_{short}^{EVR}$ \\ 
\cmidrule(r){2-6} \cmidrule(r){7-11}
Mean & 1.068 & 0.310 & 0.098 & 0.726 & 0.016 & 1.068 & 0.197 & 0.051 & 0.632 & 0.015 \\ 
Median &   1.084 & 0.324 & 0.096 & 0.715 & 0.016 & 1.084 & 0.191 & 0.048 & 0.634 & 0.016 \\ 
 St. Dev. & 0.372 & 0.208 & 0.083 & 0.296 & 0.065 & 0.372 & 0.164 & 0.064 & 0.212 & 0.051 \\ 
  \cmidrule(r){2-6} \cmidrule(r){7-11}
$\beta^{CAPM}$ & 1.000 & 0.234 & -0.188 & 0.595 & 0.041 & 1.000 & -0.040 & -0.100 & 0.435 & 0.066 \\ 
$\beta_{long}^{rel}$ & 0.234 & 1.000 & 0.147 & 0.688 & 0.032 & -0.040 & 1.000 & 0.275 & 0.595 & 0.055 \\ 
$\beta_{short}^{rel}$ & -0.188 & 0.147 & 1.000 & -0.062 & -0.053 & -0.100 & 0.275 & 1.000 & 0.104 & -0.073 \\ 
$\beta_{long}^{EV}$ & 0.595 & 0.688 & -0.062 & 1.000 & 0.025 & 0.435 & 0.595 & 0.104 & 1.000 & 0.112 \\ 
$\beta_{short}^{EV}$ & 0.041 & 0.032 & -0.053 & 0.025 & 1.000 & 0.066 & 0.055 & -0.073 & 0.112 & 1.000 \\ 
   \bottomrule
\end{tabular}
\label{tab:desc}
\end{table}

We are interested to see what distributions of estimated quantile spectral betas reveal, and so we display the unconditional distribution of the estimated betas used in the TR, EVR and Full models. Table \ref{tab:desc} summarizes descriptive statistics for all estimated betas. We focus on two values of $\tau$ - 0.05, and 0.10, and present cross-sectional means, medians and standard deviations of the estimated parameters in the top panel. We observe that all the betas are on average positive. This is particularly interesting for relative TR betas, which means that, roughly speaking, average stock posses higher tail dependence with market than suggested by the simple covariance based measures. Bottom panel of Table \ref{tab:desc} presents correlation structure of TR, EVR and CAPM betas. We observe higher values of correlation between long-term betas, and also between long-term EVR and CAPM betas. Nevertheless, all these correlation are far below 1, which suggests that all the variables may posses different and potentially important information regarding the risk associated with the assets. Another interesting observations is that the relative TR betas, both long- and short-term, are almost uncorrelated with the CAPM betas, which is exactly what we want to see given their definition.

\begin{figure}[ht!]
        \centering
        \caption{\textit{Distribution of TR and EVR betas at different tails}. Plots displays kernel density estimates of the unconditional distribution of the short-term and long-term TR and EVR betas. Presented results are computed on our largest cross-section, i.e., using stocks with at least 50 years of history. Long horizon is given by frequencies corresponding to 3-year cycle and longer.}
        \includegraphics[scale = 0.45]{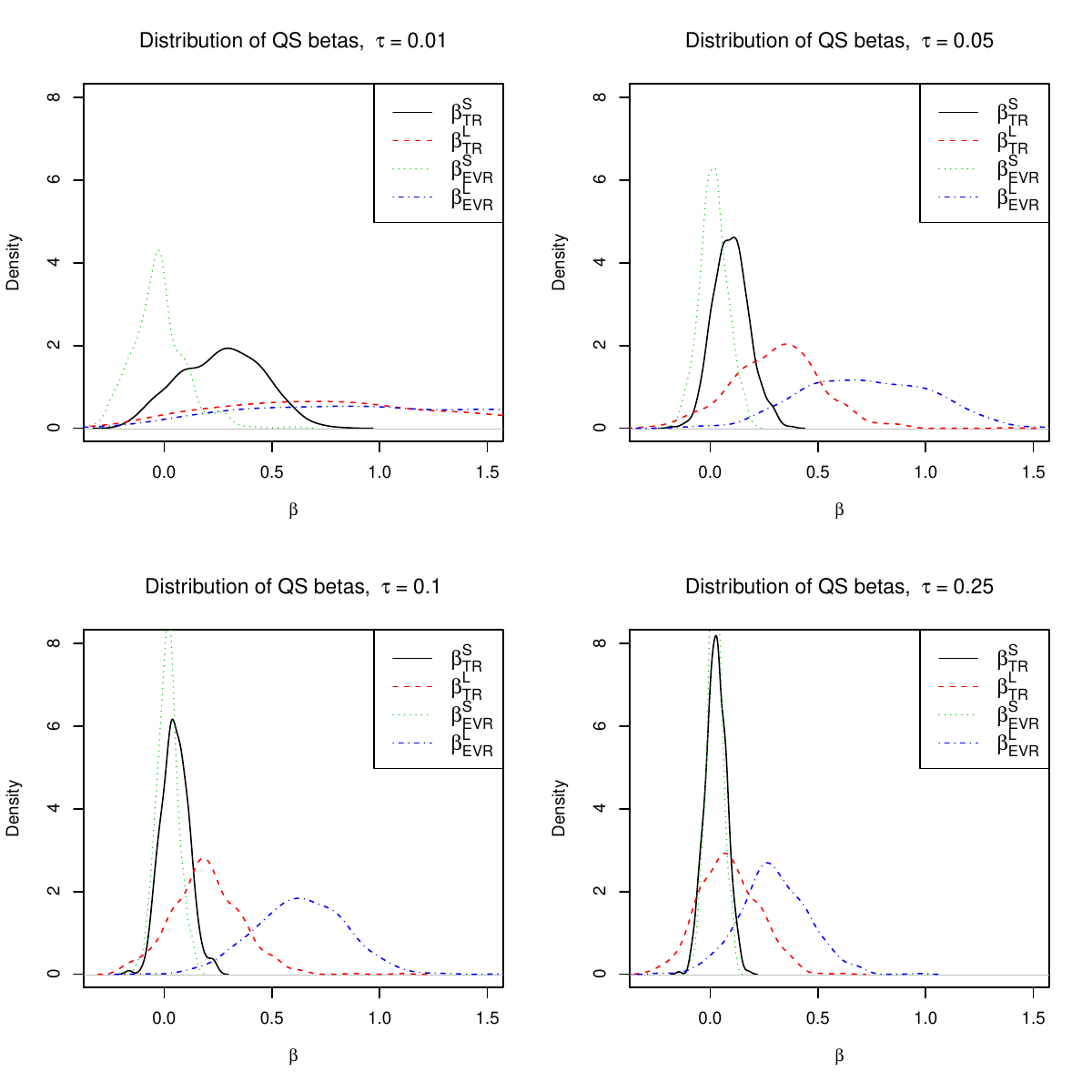}
    \label{fig:distribution}
\end{figure}

To further visualize the distributional features, Figure \ref{fig:distribution} presents unconditional distributions of the betas for four different threshold value for quantile levels. We observe the highest dispersion of  betas for the lowest values of $\tau$ corresponding the the most extreme case. As we move to higher values of $\tau$, the distributions exhibit less and less variance. Moreover, the distribution of long-term betas is wider than the distribution of the short-term betas for the respective risks.

\subsection{Robustness Checks: Tail Risk across Horizons and Other Risk Factors}
\label{sec:robust}

Large number of other risk factors and firm characteristics have been documented by the literature as significant drivers of the cross-sectional variation in equity returns \citep{harvey2016and}. While we do not attempt to include the whole exhaustive set of all controls, we would like to see if our newly defined risk factors are not subsumed by a subset of prominent variables, as well as variables related to the tails and moments of the return distribution. Hence we naturally focus on the downside measures and we use downside risks proposed by \cite{ang2006downside}, downside risk beta specification of \cite{lettau2014conditional} as well as recently proposed five factor generalized disappointment aversion (GDA5) model by \cite{farago2017downside}. Further we use coskewness and cokurtotis measures, as well as size, book-to-market and momentum factors used by \cite{fama1993common}.

To investigate whether our newly proposed measures of risk can be driven out by other determinants of risk proposed earlier in the literature, we include these risks as control variables in the previous regressions. First, we focus on the GDA5 model proposed by \cite{farago2017downside} as these are the risks most closely related to ours. It contains two measures of tail market risk as well as two measures of extreme volatility risk, but focuses on various specifications of downside dependence and does not take into consideration frequency aspect of the risks. Based on these competing measures, we compare risk measures associated with market return, and market volatility increments separately. The aim of this analysis is to decide which measures of risk better capture the notion of extreme risks associated with risk premium. The detailed specification of the corresponding betas can be found in Appendix \ref{app:comp_models}. 

Table \ref{tab:coef_hr_gda} reports the risk premium of our quantile spectral risks controlled for the GDA5 risks. In case of tail market risk presented in left panel, we see that GDA5 measures of risk ($\lambda_D$ and $\lambda_{WD}$) do not drive out our measures for any value of $\tau$ and remain insignificant when we include our TR measures. Moreover, the pattern of prices of risk corresponding to TR betas remain the same as in the TR and Full model specifications. This clearly suggests that our measures captures the asymmetric features of risk priced in the cross-section of assets. 

In case of extreme volatility risk, we see from the right panel of Table \ref{tab:coef_hr_gda} that the situation is similar. Especially, the price of risk for long-term EVR betas stays significantly strong for higher values of quantile. In addition, short-term EVR betas emerge as a significant predictors for the lower values of $\tau$. On the other hand, GDA5 measures of volatility risk remain insignificant in all of the cases. All the results suggest that our model brings an improvement in terms of identifying form of asymmetric risk which is priced in the cross-section of asset returns.

\begin{sidewaystable}[ph!]
\scriptsize
\centering
\caption{\textit{Estimated Coefficients of the TR, EVR and Full Models controlled for GDA5 Measures}. Table reports coefficients and their $t$-statistics from the horse race estimations. Displayed are prices of risk of three-factor models also including the GDA5 measures for corresponding risks. We use CRSP database between July 1926 and December 2015. Models are estimated for various values of thresholds given by $\tau$. We employ 3 samples with varying number of minimum years. Long horizon is given by frequencies corresponding to 3-year cycle and longer. Below the coefficients, we include Fama-MacBeth $t$-statistics.}
\begin{tabular}{cccccccccccccc}
  \toprule
		& \multicolumn{6}{c}{Tail market risk} & \multicolumn{6}{c}{Extreme volatility risk}\\
		\cmidrule(r){2-8} \cmidrule(r){9-14}
 & $\tau$ & $\lambda_D$ & $\lambda_{WD}$ & $\lambda_{\text{long}}^{\text{TR}}$ & $\lambda_{\text{short}}^{\text{TR}}$ & $\lambda^{\text{CAPM}}$ & RMSPE & $\lambda_X$ & $\lambda_{XD}$ & $\lambda^{\text{EV}}_{\text{long}}$ & $\lambda^{\text{EV}}_{\text{short}}$ & $\lambda^{\text{CAPM}}$ & RMSPE \\ 
  \midrule
 
\multirow{12}{*}{\shortstack[c]{70 years \\ (142 assets)}}  & 0.01 & -0.027 & 0.118 & -0.034 & 0.628 & 0.760 & 26.377 & 0.848 & 0.775 & -0.096 & 0.773 & 0.735 & 28.313 \\ 
 &  & -0.987 & 0.597 & -0.364 & 3.231 & 3.879 &  & 1.326 & 1.450 & -1.155 & 3.892 & 3.781 &  \\ 
 & 0.05 & -0.024 & 0.207 & 0.050 & 1.149 & 0.779 & 26.434 & 0.108 & 0.285 & 0.122 & 1.294 & 0.704 & 27.698 \\ 
  & & -0.908 & 1.037 & 0.234 & 3.066 & 4.153 &  & 0.169 & 0.526 & 0.612 & 3.431 & 3.805 &  \\ 
 & 0.1 & -0.017 & 0.253 & 0.305 & 0.795 & 0.799 & 26.688 & 0.157 & 0.343 & 0.388 & 1.220 & 0.724 & 26.560 \\ 
 &  & -0.622 & 1.279 & 1.243 & 1.488 & 4.436 &  & 0.246 & 0.633 & 1.580 & 2.344 & 4.066 &  \\ 
 & 0.15 & -0.010 & 0.214 & 0.449 & 0.746 & 0.779 & 26.326 & 0.219 & 0.367 & 0.572 & 0.786 & 0.720 & 26.627 \\ 
 &  & -0.357 & 1.090 & 1.992 & 1.210 & 4.297 &  & 0.344 & 0.678 & 2.422 & 1.325 & 4.018 &  \\ 
 & 0.2 & -0.010 & 0.244 & 0.526 & 0.301 & 0.824 & 26.532 & 0.302 & 0.492 & 0.723 & 0.176 & 0.766 & 25.345 \\ 
 &  & -0.324 & 1.258 & 2.115 & 0.401 & 4.630 &  & 0.473 & 0.900 & 2.732 & 0.257 & 4.374 &  \\ 
 & 0.25 & -0.018 & 0.268 & 0.587 & -0.064 & 0.856 & 26.693 & 0.377 & 0.554 & 0.850 & -0.332 & 0.793 & 25.364 \\ 
 &  & -0.587 & 1.423 & 2.300 & -0.086 & 4.878 &  & 0.587 & 1.007 & 3.089 & -0.468 & 4.576 &  \\
  
  \cmidrule(r){2-8} \cmidrule(r){9-14}
  
\multirow{12}{*}{\shortstack[c]{60 years \\ (267 assets)}}  & 0.01 & -0.014 & 0.141 & -0.011 & 0.338 & 0.765 & 29.578 & 0.367 & 0.032 & -0.060 & 0.495 & 0.753 & 30.288 \\ 
  & & -0.584 & 0.809 & -0.151 & 2.269 & 4.168 &  & 0.631 & 0.068 & -0.950 & 3.104 & 4.069 &  \\ 
 & 0.05 & -0.009 & 0.110 & 0.187 & 1.094 & 0.681 & 28.606 & -0.050 & -0.139 & 0.166 & 1.218 & 0.663 & 29.764 \\ 
  & & -0.364 & 0.627 & 1.030 & 3.437 & 3.719 &  & -0.086 & -0.297 & 0.949 & 3.679 & 3.608 &  \\ 
 & 0.1 & 0.000 & 0.164 & 0.283 & 0.850 & 0.740 & 29.013 & -0.044 & -0.192 & 0.284 & 1.004 & 0.721 & 28.928 \\ 
 &  & 0.002 & 0.946 & 1.277 & 2.166 & 4.166 &  & -0.076 & -0.408 & 1.253 & 2.453 & 4.071 &  \\ 
 & 0.15 & 0.001 & 0.182 & 0.464 & 0.627 & 0.733 & 28.962 & -0.076 & -0.234 & 0.479 & 0.774 & 0.713 & 29.459 \\ 
 &  & 0.022 & 1.067 & 2.097 & 1.279 & 4.108 &  & -0.131 & -0.500 & 2.104 & 1.570 & 4.020 &  \\ 
 & 0.2 & -0.007 & 0.244 & 0.449 & 0.092 & 0.796 & 29.317 & -0.032 & -0.147 & 0.473 & 0.240 & 0.765 & 28.616 \\ 
 &  & -0.259 & 1.414 & 1.906 & 0.173 & 4.541 &  & -0.056 & -0.314 & 1.966 & 0.458 & 4.401 &  \\ 
 & 0.25 & -0.005 & 0.237 & 0.425 & 0.199 & 0.813 & 29.566 & -0.028 & -0.143 & 0.477 & 0.182 & 0.784 & 29.268 \\ 
 &  & -0.205 & 1.399 & 1.814 & 0.367 & 4.691 &  & -0.049 & -0.305 & 1.995 & 0.337 & 4.557 &  \\
 
 \cmidrule(r){2-8} \cmidrule(r){9-14}
 
\multirow{12}{*}{\shortstack[c]{50 years \\ (528 assets)}} &  0.01 & -0.028 & 0.114 & -0.054 & 0.396 & 0.823 & 29.452 & 0.254 & -0.016 & -0.108 & 0.489 & 0.822 & 29.913 \\ 
 &  & -1.399 & 0.799 & -0.968 & 3.126 & 4.643 &  & 0.465 & -0.037 & -2.154 & 3.272 & 4.611 &  \\ 
 & 0.05 & -0.028 & 0.118 & 0.009 & 1.097 & 0.778 & 29.023 & 0.016 & -0.098 & -0.049 & 1.191 & 0.762 & 29.791 \\ 
 &  & -1.389 & 0.781 & 0.061 & 4.134 & 4.373 &  & 0.030 & -0.229 & -0.359 & 4.120 & 4.279 &  \\ 
 & 0.1 & -0.026 & 0.204 & 0.161 & 0.503 & 0.826 & 29.567 & -0.056 & -0.165 & 0.120 & 0.843 & 0.788 & 29.443 \\ 
 &  & -1.261 & 1.338 & 0.821 & 1.614 & 4.726 &  & -0.104 & -0.390 & 0.652 & 2.559 & 4.499 &  \\ 
 & 0.15 & -0.025 & 0.224 & 0.347 & 0.452 & 0.809 & 29.393 & -0.122 & -0.219 & 0.331 & 0.868 & 0.769 & 29.895 \\ 
 &  & -1.172 & 1.490 & 1.827 & 1.077 & 4.610 &  & -0.227 & -0.518 & 1.772 & 2.131 & 4.378 &  \\ 
 & 0.2 & -0.031 & 0.268 & 0.279 & 0.153 & 0.860 & 29.664 & -0.107 & -0.203 & 0.260 & 0.679 & 0.811 & 29.577 \\ 
 &  & -1.357 & 1.786 & 1.450 & 0.307 & 4.945 &  & -0.200 & -0.476 & 1.350 & 1.516 & 4.685 &  \\ 
 & 0.25 & -0.024 & 0.246 & 0.250 & 0.478 & 0.863 & 29.723 & -0.110 & -0.203 & 0.251 & 0.804 & 0.822 & 29.759 \\ 
 &  & -1.125 & 1.673 & 1.287 & 0.922 & 5.015 &  & -0.204 & -0.472 & 1.284 & 1.672 & 4.795 &  \\
 
   \bottomrule
\end{tabular}
	\label{tab:coef_hr_gda}
\end{sidewaystable}

\begin{figure}
        \centering
        \caption{\footnotesize \textit{Correlations with Other Risk Measures}. Plots display correlations between the QS betas and various other risk measures widely used in the asset pricing literature using CRSP database between July 1926 and December 2015. Presented results are computed on our largest sample, i.e., using stocks with at least 50 years of history. Long horizon is given by frequencies corresponding to 3-year cycle and longer.}
        \includegraphics[scale = 0.45]{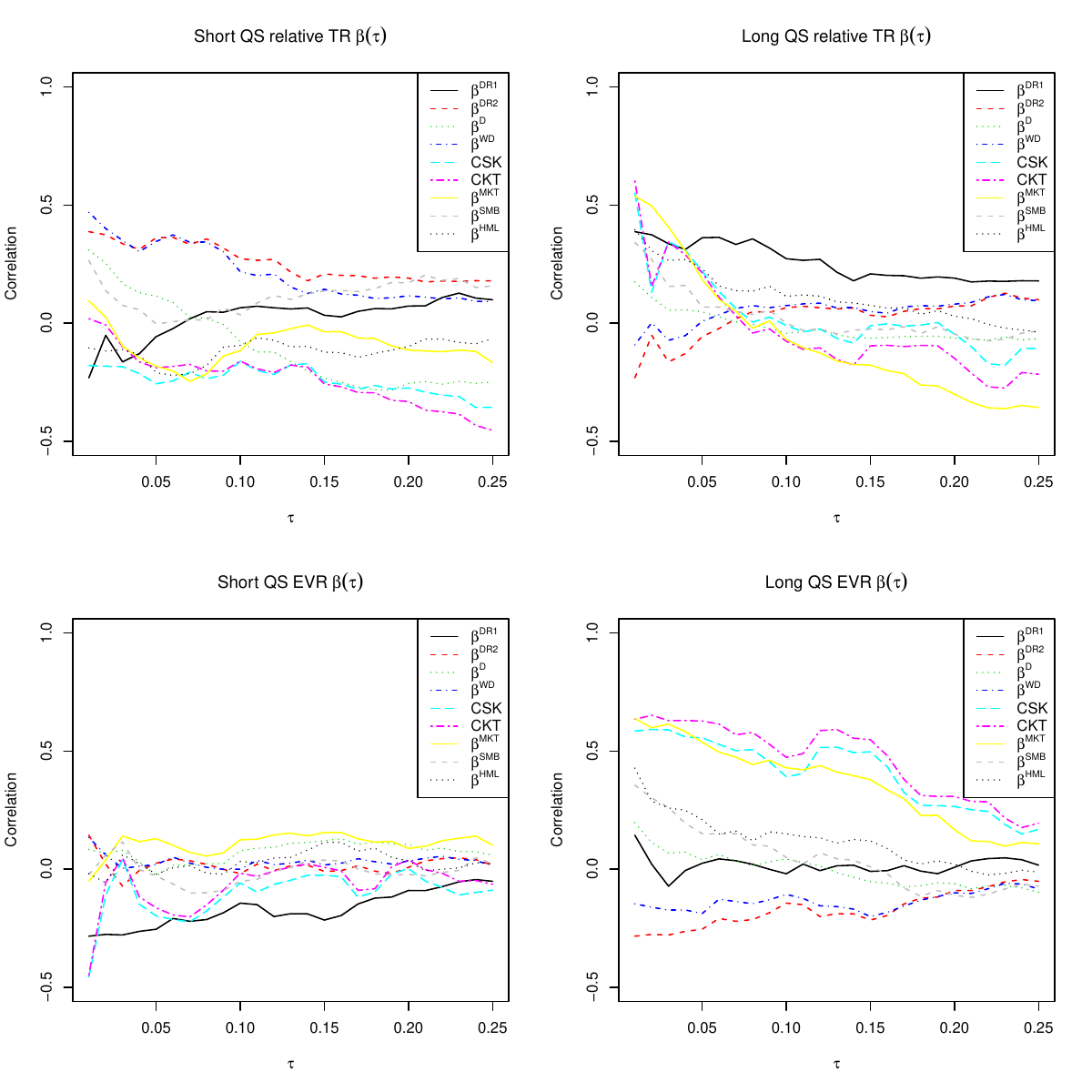}
    \label{fig:correlations}
\end{figure}

From these results, we can infer that our QS measures may potentially provide an additional information not captured by other risk measures. To further investigate this hypothesis, we present correlation structure of our QS measures with all other highly discussed asset pricing risk measures in Figure \ref{fig:correlations}. Details regarding their specifications are contained in the Applendix \ref{app:comp_models}. We plot dependence between them and the QS measures with respect to the value of quantile of the threshold value. Generally, our measures posses the highest correlation with coskewness and cokurtosis and market beta (computed using FF3 specification) in the extreme left tail and long horizon, while they show high correlation with downside risk measures in extreme left tail at short horizon. This suggests that downside risk measures capture short-term risk while moment-based risk measures are more related to the extreme volatility in the long-term. Although the correlations in few cases exceed 0.5 in absolute value, all the values are well below 1 suggesting potentially important additional information regarding the risk.

Next, we check whether these measures can drive out our QS measures in the cross-sectional estimation. Table \ref{tab:coef_hr_csk} reports the results of risk prices controlled for  coskewness and cokurtosis risks. We first include coskewness into our Full model and check whether it can drive out our risk measures. We can see that although the coskewness is significant, it does not drive out our QS measures, which follow the same pattern as in the case of previous specifications of the models. Table \ref{tab:coef_hr_csk} also reports in the right panel horse race regression including cokurtosis. We observe that cokurtosis does not bring any new explanatory information when included in our full model, as the corresponding estimated coefficients for cokurtosis are insignificant for all specifications.

\begin{sidewaystable}[ph!]
\scriptsize
\centering
\caption{\textit{Estimated Coefficients of the TR, EVR and Full Models Controlled for Coskewness and Cokurtosis.}  Displayed are prices of risk of full models also including either coskewness or cokurtosis. We use CRSP database between July 1926 and December 2015. Models are estimated for various values of thresholds given by $\tau$. We employ 3 samples with varying number of minimum years. Long horizon is given by frequencies corresponding to 3-year cycle and longer. Below the coefficients, we include Fama-MacBeth $t$-statistics.}
\begin{tabular}{cccccccccccccccc}
  \toprule
		& \multicolumn{7}{c}{Coskewness} & \multicolumn{7}{c}{Cokurtosis}\\
		\cmidrule(r){2-9} \cmidrule(r){10-16}
 & $\tau$ & $\lambda^{CSK}$ &$\lambda^{\text{TR}}_{\text{\text{long}}}$ & $\lambda^{\text{TR}}_{\text{short}}$ & $\lambda^{\text{EV}}_{\text{long}}$ & $\lambda^{\text{EV}}_{\text{short}}$ & $\lambda^{\text{CAPM}}$ & RMSPE & $\lambda^{CKT}$ &$\lambda^{\text{TR}}_{\text{\text{long}}}$ & $\lambda^{\text{TR}}_{\text{short}}$ & $\lambda^{\text{EV}}_{\text{long}}$ & $\lambda^{\text{EV}}_{\text{short}}$ & $\lambda^{\text{CAPM}}$ & RMSPE \\ 
  \cmidrule(r){2-9} \cmidrule(r){10-16}
  
 \multirow{12}{*}{\shortstack[c]{70 years \\ (142 assets)}}  & 0.01 & -0.255 & 0.122 & 0.493 & -0.122 & -0.320 & 0.790 & 25.630 & -0.020 & 0.113 & 0.676 & -0.197 & -0.241 & 0.928 & 26.332 \\ 
 &  & -1.321 & 0.879 & 2.420 & -0.854 & -1.021 & 4.152 &  & -0.942 & 0.820 & 3.374 & -1.390 & -0.777 & 4.217 &  \\ 
 & 0.05 & -0.316 & 0.016 & 0.983 & 0.093 & -0.179 & 0.747 & 25.911 & -0.009 & 0.017 & 1.288 & 0.094 & 0.236 & 0.725 & 26.756 \\ 
 &  & -1.701 & 0.044 & 2.674 & 0.327 & -0.343 & 3.279 &  & -0.452 & 0.049 & 3.492 & 0.330 & 0.467 & 2.874 &  \\ 
 & 0.1 & -0.345 & -0.147 & 0.633 & 0.470 & -0.453 & 0.610 & 25.532 & -0.022 & -0.294 & 0.952 & 0.642 & 0.165 & 0.575 & 26.478 \\ 
 &  & -1.833 & -0.368 & 1.183 & 1.582 & -0.642 & 2.684 &  & -0.988 & -0.709 & 1.817 & 1.921 & 0.242 & 2.310 &  \\ 
 & 0.15 & -0.344 & -0.028 & 0.301 & 0.584 & -0.205 & 0.616 & 25.214 & -0.021 & 0.106 & 0.647 & 0.541 & 0.425 & 0.663 & 26.262 \\ 
 &  & -1.644 & -0.084 & 0.475 & 2.165 & -0.280 & 3.040 &  & -0.903 & 0.319 & 1.062 & 1.926 & 0.595 & 2.721 &  \\ 
 & 0.2 & -0.281 & -0.152 & 0.327 & 0.743 & -0.395 & 0.630 & 24.616 & -0.023 & -0.135 & 0.509 & 0.845 & -0.052 & 0.668 & 25.353 \\ 
 &  & -1.429 & -0.443 & 0.466 & 2.925 & -0.510 & 3.234 &  & -0.993 & -0.389 & 0.720 & 3.111 & -0.067 & 2.735 &  \\ 
 & 0.25 & -0.279 & 0.001 & -0.310 & 0.754 & -0.761 & 0.709 & 24.895 & -0.017 & -0.005 & 0.021 & 0.896 & -0.477 & 0.721 & 25.653 \\ 
 &  & -1.417 & 0.004 & -0.402 & 2.835 & -0.960 & 3.752 &  & -0.738 & -0.015 & 0.027 & 3.089 & -0.607 & 2.921 &  \\
  
  \cmidrule(r){2-9} \cmidrule(r){10-16}
  
\multirow{12}{*}{\shortstack[c]{60 years \\ (267 assets)}}  & 0.01 & -0.342 & 0.159 & 0.218 & -0.091 & 0.021 & 0.776 & 28.728 & -0.011 & 0.169 & 0.409 & -0.229 & 0.222 & 0.908 & 29.442 \\ 
 &  & -1.988 & 1.343 & 1.308 & -0.751 & 0.074 & 4.427 &  & -0.531 & 1.424 & 2.462 & -1.857 & 0.798 & 4.486 &  \\ 
 & 0.05 & -0.340 & -0.071 & 0.767 & 0.252 & 0.144 & 0.636 & 28.063 & -0.014 & 0.026 & 1.225 & 0.152 & 0.519 & 0.672 & 28.600 \\ 
 &  & -2.095 & -0.233 & 2.466 & 0.950 & 0.338 & 2.772 &  & -0.749 & 0.091 & 3.818 & 0.629 & 1.236 & 2.697 &  \\ 
 & 0.1 & -0.377 & -0.385 & 0.559 & 0.597 & -0.272 & 0.545 & 27.857 & -0.034 & -0.499 & 0.805 & 0.812 & 0.163 & 0.544 & 28.572 \\ 
 &  & -2.406 & -1.164 & 1.441 & 2.017 & -0.522 & 2.338 &  & -1.632 & -1.518 & 2.029 & 2.625 & 0.325 & 2.188 &  \\ 
 & 0.15 & -0.368 & -0.026 & 0.067 & 0.505 & -0.079 & 0.624 & 28.124 & -0.018 & 0.176 & 0.618 & 0.397 & 0.281 & 0.668 & 29.093 \\ 
 &  & -2.246 & -0.092 & 0.139 & 1.963 & -0.139 & 3.121 &  & -0.860 & 0.645 & 1.259 & 1.707 & 0.514 & 2.848 &  \\ 
 & 0.2 & -0.350 & -0.266 & -0.058 & 0.715 & -0.562 & 0.639 & 27.744 & -0.025 & -0.321 & 0.360 & 0.876 & -0.254 & 0.661 & 28.553 \\ 
 &  & -2.185 & -0.941 & -0.114 & 2.973 & -0.954 & 3.313 &  & -1.211 & -1.161 & 0.701 & 3.568 & -0.433 & 2.815 &  \\ 
 & 0.25 & -0.361 & -0.034 & -0.471 & 0.553 & -1.001 & 0.747 & 28.311 & -0.014 & -0.065 & 0.252 & 0.681 & -0.524 & 0.725 & 29.336 \\ 
 &  & -2.241 & -0.132 & -0.845 & 2.564 & -1.665 & 4.073 &  & -0.691 & -0.252 & 0.448 & 2.990 & -0.889 & 3.127 &  \\
 
 \cmidrule(r){2-9} \cmidrule(r){10-16}
 
 \multirow{12}{*}{\shortstack[c]{50 years \\ (528 assets)}}  & 0.01 & -0.406 & 0.046 & 0.184 & -0.014 & 0.116 & 0.820 & 29.347 & -0.020 & -0.003 & 0.438 & -0.070 & 0.365 & 0.940 & 29.666 \\ 
 &  & -2.735 & 0.522 & 1.259 & -0.156 & 0.599 & 4.720 &  & -0.973 & -0.035 & 2.858 & -0.757 & 1.906 & 4.605 &  \\ 
 & 0.05 & -0.398 & -0.112 & 0.754 & 0.159 & 0.273 & 0.746 & 28.705 & -0.026 & -0.100 & 1.215 & 0.123 & 0.569 & 0.815 & 29.028 \\ 
 &  & -2.693 & -0.403 & 2.862 & 0.686 & 0.837 & 3.383 &  & -1.404 & -0.396 & 4.236 & 0.602 & 1.721 & 3.413 &  \\ 
 & 0.1 & -0.447 & -0.292 & 0.320 & 0.417 & -0.181 & 0.682 & 28.936 & -0.039 & -0.412 & 0.656 & 0.603 & 0.009 & 0.716 & 29.371 \\ 
 &  & -3.130 & -0.933 & 1.042 & 1.660 & -0.433 & 3.115 &  & -1.955 & -1.420 & 2.093 & 2.378 & 0.022 & 3.058 &  \\ 
 & 0.15 & -0.419 & 0.169 & 0.139 & 0.143 & -0.323 & 0.793 & 29.185 & -0.018 & 0.362 & 0.660 & 0.008 & -0.067 & 0.852 & 29.531 \\ 
 &  & -2.872 & 0.619 & 0.363 & 0.620 & -0.697 & 4.066 &  & -0.920 & 1.416 & 1.691 & 0.036 & -0.148 & 3.734 &  \\ 
 & 0.2 & -0.412 & -0.077 & -0.056 & 0.365 & -0.522 & 0.776 & 29.196 & -0.030 & -0.120 & 0.332 & 0.498 & -0.382 & 0.828 & 29.700 \\ 
 &  & -2.871 & -0.309 & -0.129 & 1.636 & -1.091 & 4.030 &  & -1.516 & -0.509 & 0.772 & 2.256 & -0.798 & 3.596 &  \\ 
 & 0.25 & -0.414 & 0.043 & -0.123 & 0.263 & -0.607 & 0.828 & 29.387 & -0.023 & 0.020 & 0.475 & 0.367 & -0.263 & 0.854 & 29.880 \\ 
  & & -2.905 & 0.191 & -0.259 & 1.361 & -1.262 & 4.514 &  & -1.142 & 0.089 & 1.016 & 1.872 & -0.547 & 3.746 & \\
 
   \bottomrule
\end{tabular}
	\label{tab:coef_hr_csk}
\end{sidewaystable}

In addition, Table \ref{tab:coef_hr_dr} reports the results controlled for the two specification of relative downside betas. In the left panel, we report results with downside risk specification of \cite{ang2006downside}. We observe that the downside risk beta does not capture any additional important dimension of risk when included in our full model specification. The same is true for the downside risk model of \cite{lettau2014conditional}, which is captured in the right panel.

\begin{sidewaystable}[ph!]
\scriptsize
\centering
\caption{\textit{Estimated Coefficients of the TR, EVR and Full Models controlled for Downside Risk Betas.} Displayed are prices of risk of full models also including either downside risk beta of \cite{ang2006downside} or downside risk beta specification of \cite{lettau2014conditional}. We use CRSP database between July 1926 and December 2015. Models are estimated for various values of thresholds given by $\tau$. We employ 3 samples with varying number of minimum years. Long horizon is given by frequencies corresponding to 3-year cycle and longer. Below the coefficients, we include Fama-MacBeth $t$-statistics.}
\begin{tabular}{cccccccccccccccc}
  \toprule
		& \multicolumn{7}{c}{DR beta of  \cite{ang2006downside}} & \multicolumn{7}{c}{DR of \cite{lettau2014conditional}}\\
		\cmidrule(r){2-9} \cmidrule(r){10-16}
 & $\tau$ & $\lambda^{DR1}$ & $\lambda^{\text{TR}}_{\text{\text{long}}}$ & $\lambda^{\text{TR}}_{\text{short}}$ & $\lambda^{\text{EV}}_{\text{long}}$ & $\lambda^{\text{EV}}_{\text{short}}$ & $\lambda^{\text{CAPM}}$ & RMSPE & $\lambda^{DR2}$ &$\lambda^{\text{TR}}_{\text{\text{long}}}$ & $\lambda^{\text{TR}}_{\text{short}}$ & $\lambda^{\text{EV}}_{\text{long}}$ & $\lambda^{\text{EV}}_{\text{short}}$ & $\lambda^{\text{CAPM}}$ & RMSPE \\ 
  \cmidrule(r){2-9} \cmidrule(r){10-16}
\multirow{12}{*}{\shortstack[c]{70 years \\ (142 assets)}}  &  0.01 & -0.017 & 0.129 & 0.629 & -0.210 & -0.194 & 0.826 & 26.387 & -0.017 & 0.129 & 0.629 & -0.210 & -0.194 & 0.826 & 26.387 \\ 
  & & -0.044 & 0.922 & 3.240 & -1.484 & -0.631 & 4.104 &  & -0.044 & 0.922 & 3.240 & -1.484 & -0.631 & 4.104 &  \\ 
 & 0.05 & 0.149 & 0.042 & 1.250 & -0.005 & 0.285 & 0.754 & 26.736 & 0.149 & 0.042 & 1.250 & -0.005 & 0.285 & 0.754 & 26.736 \\ 
 &  & 0.410 & 0.119 & 3.448 & -0.017 & 0.559 & 3.216 &  & 0.410 & 0.119 & 3.448 & -0.017 & 0.559 & 3.216 &  \\ 
 & 0.1 & 0.082 & -0.086 & 0.997 & 0.412 & 0.212 & 0.573 & 26.548 & 0.082 & -0.086 & 0.997 & 0.412 & 0.212 & 0.573 & 26.548 \\ 
 &  & 0.235 & -0.216 & 1.895 & 1.414 & 0.306 & 2.451 &  & 0.235 & -0.216 & 1.895 & 1.414 & 0.306 & 2.451 &  \\ 
 & 0.15 & 0.128 & 0.211 & 0.864 & 0.325 & 0.367 & 0.636 & 26.259 & 0.128 & 0.211 & 0.864 & 0.325 & 0.367 & 0.636 & 26.259 \\ 
 &  & 0.363 & 0.638 & 1.433 & 1.272 & 0.512 & 3.036 &  & 0.363 & 0.638 & 1.433 & 1.272 & 0.512 & 3.036 &  \\ 
 & 0.2 & 0.121 & -0.098 & 0.840 & 0.717 & -0.207 & 0.594 & 25.397 & 0.121 & -0.098 & 0.840 & 0.717 & -0.207 & 0.594 & 25.397 \\ 
 &  & 0.352 & -0.283 & 1.242 & 2.780 & -0.269 & 2.962 &  & 0.352 & -0.283 & 1.242 & 2.780 & -0.269 & 2.962 &  \\ 
 & 0.25 & 0.135 & -0.030 & 0.454 & 0.802 & -0.475 & 0.656 & 25.627 & 0.135 & -0.030 & 0.454 & 0.802 & -0.475 & 0.656 & 25.627 \\ 
 &  & 0.391 & -0.090 & 0.631 & 2.895 & -0.598 & 3.399 &  & 0.391 & -0.090 & 0.631 & 2.895 & -0.598 & 3.399 &  \\
 
 \cmidrule(r){2-9} \cmidrule(r){10-16}
  
\multirow{12}{*}{\shortstack[c]{60 years \\ (267 assets)}}  & 0.01 & 0.107 & 0.161 & 0.345 & -0.211 & 0.244 & 0.852 & 29.418 & 0.107 & 0.161 & 0.345 & -0.211 & 0.244 & 0.852 & 29.418 \\ 
 &  & 0.368 & 1.350 & 2.368 & -1.760 & 0.912 & 4.611 &  & 0.368 & 1.350 & 2.368 & -1.760 & 0.912 & 4.611 &  \\ 
 & 0.05 & 0.135 & 0.093 & 1.171 & 0.041 & 0.612 & 0.671 & 28.588 & 0.135 & 0.093 & 1.171 & 0.041 & 0.612 & 0.671 & 28.588 \\ 
 &  & 0.486 & 0.308 & 3.716 & 0.167 & 1.493 & 2.934 &  & 0.486 & 0.308 & 3.716 & 0.167 & 1.493 & 2.934 &  \\ 
 & 0.1 & 0.192 & -0.205 & 0.839 & 0.516 & 0.345 & 0.509 & 28.620 & 0.192 & -0.205 & 0.839 & 0.516 & 0.345 & 0.509 & 28.620 \\ 
 &  & 0.711 & -0.637 & 2.096 & 1.836 & 0.674 & 2.183 &  & 0.711 & -0.637 & 2.096 & 1.836 & 0.674 & 2.183 &  \\ 
 & 0.15 & 0.240 & 0.252 & 0.657 & 0.257 & 0.362 & 0.638 & 28.943 & 0.240 & 0.252 & 0.657 & 0.257 & 0.362 & 0.638 & 28.943 \\ 
 &  & 0.867 & 0.888 & 1.313 & 1.101 & 0.656 & 3.119 &  & 0.867 & 0.888 & 1.313 & 1.101 & 0.656 & 3.119 &  \\ 
 & 0.2 & 0.296 & -0.236 & 0.536 & 0.721 & -0.344 & 0.588 & 28.443 & 0.296 & -0.236 & 0.536 & 0.721 & -0.344 & 0.588 & 28.443 \\ 
 &  & 1.055 & -0.827 & 1.053 & 3.062 & -0.589 & 3.021 &  & 1.055 & -0.827 & 1.053 & 3.062 & -0.589 & 3.021 &  \\ 
 & 0.25 & 0.313 & -0.060 & 0.404 & 0.598 & -0.605 & 0.686 & 29.091 & 0.313 & -0.060 & 0.404 & 0.598 & -0.605 & 0.686 & 29.091 \\ 
 &  & 1.105 & -0.233 & 0.733 & 2.763 & -1.035 & 3.705 &  & 1.105 & -0.233 & 0.733 & 2.763 & -1.035 & 3.705 &  \\
 
 \cmidrule(r){2-9} \cmidrule(r){10-16}
 
  \multirow{12}{*}{\shortstack[c]{50 years \\ (528 assets)}}  & 0.01 & 0.110 & 0.003 & 0.367 & -0.086 & 0.418 & 0.873 & 29.683 & 0.110 & 0.003 & 0.367 & -0.086 & 0.418 & 0.873 & 29.683 \\ 
 &  & 0.471 & 0.038 & 2.938 & -0.951 & 2.285 & 4.883 &  & 0.471 & 0.038 & 2.938 & -0.951 & 2.285 & 4.883 &  \\ 
 & 0.05 & 0.142 & 0.005 & 1.167 & -0.045 & 0.753 & 0.790 & 29.039 & 0.142 & 0.005 & 1.167 & -0.045 & 0.753 & 0.790 & 29.039 \\ 
 &  & 0.610 & 0.018 & 4.423 & -0.206 & 2.372 & 3.612 &  & 0.610 & 0.018 & 4.423 & -0.206 & 2.372 & 3.612 &  \\ 
 & 0.1 & 0.275 & -0.120 & 0.628 & 0.276 & 0.164 & 0.695 & 29.396 & 0.275 & -0.120 & 0.628 & 0.276 & 0.164 & 0.695 & 29.396 \\ 
 &  & 1.192 & -0.385 & 1.978 & 1.168 & 0.397 & 3.189 &  & 1.192 & -0.385 & 1.978 & 1.168 & 0.397 & 3.189 &  \\ 
 & 0.15 & 0.284 & 0.446 & 0.642 & -0.134 & 0.034 & 0.831 & 29.516 & 0.284 & 0.446 & 0.642 & -0.134 & 0.034 & 0.831 & 29.516 \\ 
 &  & 1.228 & 1.631 & 1.568 & -0.643 & 0.076 & 4.176 &  & 1.228 & 1.631 & 1.568 & -0.643 & 0.076 & 4.176 &  \\ 
 & 0.2 & 0.304 & -0.002 & 0.546 & 0.296 & -0.332 & 0.760 & 29.626 & 0.304 & -0.002 & 0.546 & 0.296 & -0.332 & 0.760 & 29.626 \\ 
 &  & 1.305 & -0.007 & 1.206 & 1.381 & -0.695 & 3.923 &  & 1.305 & -0.007 & 1.206 & 1.381 & -0.695 & 3.923 &  \\ 
 & 0.25 & 0.299 & 0.033 & 0.781 & 0.255 & -0.133 & 0.789 & 29.797 & 0.299 & 0.033 & 0.781 & 0.255 & -0.133 & 0.789 & 29.797 \\ 
 &  & 1.271 & 0.147 & 1.549 & 1.356 & -0.276 & 4.293 &  & 1.271 & 0.147 & 1.549 & 1.356 & -0.276 & 4.293 &  \\
   
   \bottomrule
\end{tabular}
	\label{tab:coef_hr_dr}
\end{sidewaystable}

Finally, Table \ref{tab:coef_hr_ff3} reports regressions including additional betas from the three-factor model of \cite{fama1993common}.\footnote{We have to include only 2 additional betas as the market beta is already included in our full model.} This model is not explicitly related to the asymmetric features of market or volatility risk, but as we show in the Section \ref{sec:qs_risk}, these factors may be just capturing market risk in different horizons in specific parts of the joint distribution of market and asset returns, so we should check whether they are not superior in describing these kind of risks. As in the case of other horse race regressions, the additional risk factors do not drive out the QS measures, which repeat the same pattern as in the cases without the additional variables.

\begin{table}[ht!]
\centering
\scriptsize
\caption{\textit{Estimated Coefficients of the TR, EVR and Full Models controlled for \cite{fama1973risk} factors.} Displayed are prices of risk of full models also including either HML and SMB betas of \cite{fama1993common}. We use CRSP database between July 1926 and December 2015. Models are estimated for various values of thresholds given by $\tau$. We employ 3 samples with varying number of minimum years. Long horizon is given by frequencies corresponding to 3-year cycle and longer. Below the coefficients, we include Fama-MacBeth $t$-statistics.}
\begin{tabular}{cccccccccc}
\toprule
 & $\tau$ & $\lambda^{SMB}$ & $\lambda^{HML}$ & $\lambda^{\text{TR}}_{\text{\text{long}}}$ & $\lambda^{\text{TR}}_{\text{short}}$ & $\lambda^{\text{EV}}_{\text{long}}$ & $\lambda^{\text{EV}}_{\text{short}}$ & $\lambda^{\text{CAPM}}$ & RMSPE \\ 
\cmidrule(r){2-10} 
 
 \multirow{12}{*}{\shortstack[c]{70 years \\ (142 assets)}}  & 0.01 & 0.035 & -0.050 & 0.133 & 0.636 & -0.217 & -0.174 & 0.838 & 26.281 \\ 
 &  & 0.266 & -0.280 & 0.960 & 3.276 & -1.558 & -0.569 & 4.206 &  \\ 
 & 0.05 & -0.073 & -0.181 & 0.401 & 1.108 & -0.238 & 0.093 & 0.893 & 26.349 \\ 
 &  & -0.550 & -1.034 & 1.238 & 3.072 & -0.989 & 0.186 & 4.007 &  \\ 
 & 0.1 & -0.009 & -0.200 & 0.187 & 0.888 & 0.276 & 0.076 & 0.674 & 26.119 \\ 
 &  & -0.066 & -1.188 & 0.562 & 1.777 & 1.101 & 0.112 & 3.093 &  \\ 
 & 0.15 & 0.001 & -0.165 & 0.376 & 0.612 & 0.231 & 0.289 & 0.702 & 25.969 \\ 
 &  & 0.010 & -0.949 & 1.232 & 1.068 & 1.044 & 0.416 & 3.504 &  \\ 
 & 0.2 & 0.090 & -0.150 & 0.016 & 0.535 & 0.712 & -0.259 & 0.609 & 25.052 \\ 
 &  & 0.685 & -0.869 & 0.050 & 0.827 & 2.965 & -0.341 & 3.107 &  \\ 
 & 0.25 & 0.067 & -0.150 & 0.118 & 0.171 & 0.753 & -0.731 & 0.691 & 25.349 \\ 
 &  & 0.517 & -0.873 & 0.359 & 0.251 & 2.759 & -0.962 & 3.604 &  \\ 
   
   \cmidrule(r){2-10} 
  
\multirow{12}{*}{\shortstack[c]{60 years \\ (267 assets)}}  & 0.01 & -0.149 & 0.040 & 0.198 & 0.389 & -0.233 & 0.281 & 0.851 & 29.218 \\ 
  & & -1.184 & 0.252 & 1.657 & 2.622 & -1.979 & 1.053 & 4.481 &  \\ 
 & 0.05 & -0.175 & -0.016 & 0.291 & 1.216 & -0.204 & 0.552 & 0.827 & 28.377 \\ 
 &  & -1.434 & -0.101 & 1.035 & 3.859 & -1.018 & 1.410 & 3.858 &  \\ 
 & 0.1 & -0.121 & -0.051 & 0.063 & 0.996 & 0.255 & 0.341 & 0.650 & 28.578 \\ 
 &  & -0.978 & -0.333 & 0.216 & 2.589 & 1.152 & 0.676 & 3.045 &  \\ 
 & 0.15 & -0.143 & -0.052 & 0.414 & 0.837 & 0.025 & 0.442 & 0.751 & 28.833 \\ 
 &  & -1.182 & -0.340 & 1.561 & 1.831 & 0.137 & 0.821 & 3.926 &  \\ 
 & 0.2 & -0.060 & -0.054 & -0.151 & 0.709 & 0.640 & -0.261 & 0.618 & 28.464 \\ 
 &  & -0.492 & -0.359 & -0.559 & 1.550 & 3.120 & -0.448 & 3.258 &  \\ 
 & 0.25 & -0.102 & -0.059 & 0.014 & 0.623 & 0.505 & -0.534 & 0.726 & 29.051 \\ 
 &  & -0.850 & -0.388 & 0.055 & 1.243 & 2.506 & -0.940 & 4.000 &  \\ 

\cmidrule(r){2-10} 

  \multirow{12}{*}{\shortstack[c]{50 years \\ (528 assets)}}  & 0.01 & -0.087 & 0.006 & -0.005 & 0.457 & -0.081 & 0.493 & 0.874 & 29.354 \\ 
  & & -0.761 & 0.041 & -0.063 & 3.577 & -0.921 & 2.582 & 4.805 &  \\ 
 & 0.05 & -0.088 & -0.051 & 0.135 & 1.172 & -0.213 & 0.708 & 0.905 & 28.909 \\ 
 &  & -0.806 & -0.361 & 0.604 & 4.502 & -1.349 & 2.251 & 4.504 &  \\ 
 & 0.1 & -0.044 & -0.096 & 0.022 & 0.791 & 0.137 & 0.120 & 0.774 & 29.182 \\ 
 &  & -0.394 & -0.697 & 0.092 & 2.507 & 0.781 & 0.289 & 3.917 &  \\ 
 & 0.15 & -0.095 & -0.126 & 0.612 & 0.730 & -0.347 & 0.153 & 0.950 & 29.086 \\ 
 &  & -0.854 & -0.919 & 2.645 & 1.926 & -2.262 & 0.344 & 5.144 &  \\ 
 & 0.2 & -0.042 & -0.084 & 0.091 & 0.682 & 0.201 & -0.306 & 0.804 & 29.402 \\ 
 &  & -0.376 & -0.613 & 0.422 & 1.710 & 1.202 & -0.652 & 4.429 &  \\ 
 & 0.25 & -0.055 & -0.090 & 0.101 & 0.914 & 0.163 & -0.166 & 0.835 & 29.519 \\ 
 &  & -0.499 & -0.662 & 0.476 & 2.004 & 1.033 & -0.354 & 4.793 &  \\
 
   \bottomrule
\end{tabular}
\label{tab:coef_hr_ff3}
\end{table}

\section{Different Definition of Long horizon - 1.5 years}
\label{app:diff_long}

\begin{sidewaystable}[ph!]
\scriptsize
\centering
\caption{\textit{Estimated Coefficients of the TR, EVR and Full Models.} Prices of risk estimated on monthly stock data from CRSP database sampled between July 1926 and December 2015. Models are estimated for various values of thresholds given by $\tau$. We employ three samples with varying number of minimum years. Long horizon is given by frequencies corresponding to 1.5-year cycle and longer. Below the coefficients, we include Fama-MacBeth $t$-statistics.}
\begin{tabular}{cccccccccccccccc}
 \toprule
	 	& & \multicolumn{4}{c}{Tail market risk} & \multicolumn{4}{c}{Extreme volatility risk} & \multicolumn{6}{c}{Full model}\\
		\cmidrule(r){2-6} \cmidrule(r){7-10} \cmidrule(r){11-16}
  & $\tau$ & $\lambda_{\text{long}}^{\text{TR}}$ & $\lambda_{\text{short}}^{\text{TR}}$ & $\lambda^{\text{CAPM}}$ & RMSPE & $\lambda_{\text{long}}^{\text{EV}}$ & $\lambda_{\text{short}}^{\text{EV}}$ & $\lambda^{\text{CAPM}}$ & RMSPE & $\lambda^{\text{TR}}_{\text{\text{long}}}$ & $\lambda^{\text{TR}}_{\text{short}}$ & $\lambda^{\text{EV}}_{\text{long}}$ & $\lambda^{\text{EV}}_{\text{short}}$ & $\lambda^{\text{CAPM}}$ & RMSPE \\ 
  \cmidrule(r){2-6} \cmidrule(r){7-10} \cmidrule(r){11-16}
\multirow{12}{*}{\shortstack[c]{70 years \\ (142 assets)}} & 0.01 & -0.045 & 0.644 & 0.751 & 26.672 & -0.084 & -0.258 & 0.946 & 28.576 & 0.135 & 0.610 & -0.213 & -0.150 & 0.832 & 26.436 \\ 
  & & -0.476 & 3.303 & 3.910 &  & -0.883 & -0.853 & 4.885 &  & 0.896 & 3.085 & -1.366 & -0.495 & 4.151 &  \\ 
  & 0.05 & 0.149 & 1.272 & 0.716 & 26.801 & 0.242 & 0.378 & 0.689 & 28.212 & 0.160 & 1.258 & -0.024 & 0.297 & 0.738 & 26.755 \\ 
  & & 0.699 & 3.485 & 3.861 &  & 1.413 & 0.774 & 3.047 &  & 0.431 & 3.519 & -0.081 & 0.602 & 3.046 &  \\ 
  & 0.1 & 0.432 & 1.153 & 0.735 & 27.109 & 0.519 & 0.441 & 0.549 & 27.226 & -0.005 & 1.004 & 0.442 & 0.287 & 0.553 & 26.567 \\ 
  & & 1.673 & 2.265 & 4.119 &  & 2.588 & 0.649 & 2.461 &  & -0.011 & 1.964 & 1.357 & 0.425 & 2.269 &  \\ 
 & 0.15 & 0.611 & 0.842 & 0.730 & 26.651 & 0.557 & 0.522 & 0.592 & 27.025 & 0.312 & 0.832 & 0.344 & 0.501 & 0.615 & 26.280 \\ 
 & & 2.493 & 1.453 & 4.050 &  & 2.836 & 0.756 & 2.817 &  & 0.898 & 1.436 & 1.210 & 0.719 & 2.816 &  \\ 
 & 0.2 & 0.732 & 0.224 & 0.780 & 26.944 & 0.750 & -0.313 & 0.590 & 25.765 & 0.050 & 0.753 & 0.709 & -0.139 & 0.580 & 25.417 \\ 
  & & 2.726 & 0.326 & 4.432 &  & 3.604 & -0.411 & 2.950 &  & 0.137 & 1.127 & 2.463 & -0.183 & 2.784 &  \\ 
 & 0.25 & 0.786 & -0.165 & 0.808 & 27.241 & 0.869 & -0.563 & 0.631 & 25.703 & 0.045 & 0.418 & 0.844 & -0.420 & 0.628 & 25.608 \\ 
  & & 2.898 & -0.230 & 4.631 &  & 3.695 & -0.727 & 3.238 &  & 0.130 & 0.589 & 2.751 & -0.544 & 3.165 &  \\ 
\cmidrule(r){2-6} \cmidrule(r){7-10} \cmidrule(r){11-16}

\multirow{12}{*}{\shortstack[c]{60 years \\ (267 assets)}} & 0.01 & -0.036 & 0.430 & 0.759 & 29.723 & -0.083 & 0.263 & 0.934 & 30.514 & 0.168 & 0.380 & -0.227 & 0.250 & 0.865 & 29.496 \\ 
  & & -0.491 & 2.685 & 4.130 &  & -1.105 & 1.002 & 5.093 &  & 1.342 & 2.388 & -1.784 & 0.948 & 4.630 &  \\ 
 & 0.05 & 0.225 & 1.187 & 0.661 & 28.681 & 0.256 & 0.611 & 0.655 & 29.996 & 0.175 & 1.241 & 0.016 & 0.646 & 0.675 & 28.617 \\ 
  & & 1.211 & 3.659 & 3.574 &  & 1.493 & 1.545 & 2.878 &  & 0.562 & 3.921 & 0.060 & 1.626 & 2.819 &  \\ 
 & 0.1 & 0.368 & 0.958 & 0.715 & 29.239 & 0.550 & 0.371 & 0.501 & 29.193 & -0.109 & 0.906 & 0.538 & 0.402 & 0.490 & 28.678 \\ 
  & & 1.549 & 2.407 & 3.999 &  & 2.516 & 0.753 & 2.200 &  & -0.318 & 2.265 & 1.730 & 0.815 & 2.015 &  \\ 
 & 0.15 & 0.559 & 0.729 & 0.706 & 29.244 & 0.488 & 0.495 & 0.593 & 29.614 & 0.358 & 0.770 & 0.250 & 0.398 & 0.622 & 29.089 \\ 
  & & 2.326 & 1.482 & 3.940 &  & 2.349 & 0.923 & 2.805 &  & 1.191 & 1.573 & 0.968 & 0.739 & 2.918 &  \\ 
 & 0.2 & 0.538 & 0.236 & 0.762 & 29.736 & 0.677 & -0.251 & 0.581 & 28.726 & -0.166 & 0.657 & 0.757 & -0.296 & 0.555 & 28.607 \\ 
  & & 2.139 & 0.450 & 4.343 &  & 3.146 & -0.440 & 2.913 &  & -0.560 & 1.309 & 2.870 & -0.517 & 2.739 &  \\ 
 & 0.25 & 0.523 & 0.208 & 0.782 & 30.035 & 0.669 & -0.526 & 0.645 & 29.286 & -0.041 & 0.539 & 0.680 & -0.506 & 0.641 & 29.244 \\ 
  & & 2.114 & 0.375 & 4.491 &  & 3.065 & -0.895 & 3.382 &  & -0.154 & 0.996 & 2.755 & -0.870 & 3.356 &  \\   
\cmidrule(r){2-6} \cmidrule(r){7-10} \cmidrule(r){11-16}
   
\multirow{12}{*}{\shortstack[c]{50 years \\ (528 assets)}} & 0.01 & -0.087 & 0.436 & 0.825 & 29.726 & -0.088 & 0.463 & 0.969 & 30.286 & -0.007 & 0.407 & -0.073 & 0.422 & 0.871 & 29.689 \\ 
  & & -1.535 & 2.981 & 4.634 &  & -1.507 & 2.500 & 5.396 &  & -0.082 & 2.829 & -0.798 & 2.305 & 4.787 &  \\ 
 & 0.05 & 0.001 & 1.162 & 0.762 & 29.233 & 0.071 & 0.636 & 0.807 & 30.075 & 0.060 & 1.242 & -0.055 & 0.765 & 0.790 & 29.054 \\ 
  & & 0.004 & 3.980 & 4.238 &  & 0.542 & 1.991 & 3.795 &  & 0.209 & 4.422 & -0.236 & 2.446 & 3.474 &  \\ 
 & 0.1 & 0.195 & 0.783 & 0.784 & 29.756 & 0.316 & 0.069 & 0.672 & 29.669 & -0.060 & 0.770 & 0.302 & 0.155 & 0.662 & 29.400 \\ 
  & & 0.972 & 2.401 & 4.418 &  & 1.836 & 0.168 & 3.152 &  & -0.187 & 2.416 & 1.165 & 0.381 & 2.932 &  \\ 
 & 0.15 & 0.404 & 0.811 & 0.765 & 29.583 & 0.168 & -0.003 & 0.776 & 30.089 & 0.532 & 0.782 & -0.150 & 0.084 & 0.810 & 29.552 \\ 
  & & 2.010 & 1.968 & 4.288 &  & 0.991 & -0.006 & 3.817 &  & 1.840 & 1.923 & -0.637 & 0.187 & 3.893 &  \\ 
 & 0.2 & 0.324 & 0.631 & 0.808 & 29.957 & 0.377 & -0.316 & 0.725 & 29.730 & 0.038 & 0.692 & 0.332 & -0.342 & 0.727 & 29.650 \\ 
  & & 1.599 & 1.363 & 4.594 &  & 2.045 & -0.687 & 3.694 &  & 0.144 & 1.535 & 1.359 & -0.731 & 3.611 &  \\ 
 & 0.25 & 0.303 & 0.775 & 0.819 & 30.024 & 0.373 & -0.116 & 0.752 & 29.926 & 0.046 & 0.912 & 0.335 & -0.136 & 0.750 & 29.817 \\ 
  & & 1.509 & 1.546 & 4.690 &  & 2.019 & -0.247 & 3.980 &  & 0.199 & 1.833 & 1.536 & -0.287 & 3.958 &  \\
   \bottomrule
\end{tabular}
\label{tab:coef_3factor_alt}
\end{sidewaystable}



\newpage
\clearpage
\section{Specification of the Competing Models}
\label{app:comp_models}

In this section, we briefly describe the specification of the models we use in the Appendix \ref{sec:robust}. We denote market excess return as $r_m$ and its mean and variance as $\mu_m$ and $\sigma^2_m$, respectively. Excess return of an asset is denoted as $r_i$ with mean $\mu_i$ and variance $\sigma^2_i$.

We present how we estimate betas in the first-stage regression. The second-stage regression is the same for all the models and is performed via OLS by regressing the average asset returns on their betas. This then leads to the estimated values of RMSPE.

\subsection{Downside Risk Models}
We follow two specifications of the downside risk models. First, we use specification of \cite{ang2006downside} and estimate their relative downside risk betas as
\begin{align}
\beta^{DR1}_i \equiv \beta^-_{i, \mu_m} - \beta_i = \frac{\mathbb{C}ov(r_i, r_m | r_m < \mu_m)}{\mathbb{V}ar(r_m | r_m < \mu_m)} - \frac{\mathbb{C}ov(r_i, r_m)}{\mathbb{V}ar(r_m)} .
\end{align}
 Downside risk beta specification of \cite{lettau2014conditional} is then obtained as
\begin{align}
\beta^{DR2}_i \equiv \beta^-_{i, \delta} - \beta_i = \frac{\mathbb{C}ov(r_i, r_m | r_m < \delta)}{\mathbb{V}ar(r_m | r_m < \delta)} - \frac{\mathbb{C}ov(r_i, r_m)}{\mathbb{V}ar(r_m)}
\end{align}
where we define the threshold value as $\delta \equiv \mu_m - \sigma_m$.

\subsection{Generalized Disappointment Aversion Models}
We employ specification of Generalized Disappointment Aversion (GDA) models of \cite{farago2017downside} and estimate two main versions of their cross-sectional models. Their models are based on disappointment events $\mathcal{D}_t$.	
	\subsubsection{GDA3}
	First model is their three-factor model, which does not contain volatility-related factors.  The betas posses the following form
	\begin{align}
	\beta_{i,m} &\equiv \frac{\mathbb{C}ov(r_i, r_m)}{\mathbb{V}ar(r_m)} \\
	\beta_{i, \mathcal{D}} &\equiv \frac{\mathbb{C}ov(r_i, I(\mathcal{D}))} {\mathbb{V}ar(I(\mathcal{D}))} \\
	\beta_{i, m\mathcal{D}} &\equiv \frac{\mathbb{C}ov(r_i, r_m I(\mathcal{D}))}{\mathbb{V}ar(r_m I(\mathcal{D}))}
	\end{align}
	where we follow the specification and set $\mathcal{D}_t = \{ r_{m,t} < b \}$ where $b = -0.03$ and $I$ is an indicator function.
	
	\subsubsection{GDA5}
	Five-factor specification of the GDA model contains, in addition to the betas from the three-factor model, the following betas
	\begin{align}
	\beta_{i,X} &\equiv \frac{\mathbb{C}ov(r_i, \Delta \sigma^2_{m})}{\mathbb{V}ar(\Delta \sigma^2_{m})} \\
	\beta_{i, X\mathcal{D}} &\equiv \frac{\mathbb{C}ov(r_i, \Delta \sigma^2_{m} I(\mathcal{D}))}{\mathbb{V}ar(\Delta \sigma^2_{m} I(\mathcal{D}))}
	\end{align}
	where the disappointment events are given by $\mathcal{D}_t = \Big\{ r_{m,t} -  a \frac{\sigma_m}{\sigma_X } \Delta \sigma^2_{m,t}  < b \Big\}$ where $\Delta \sigma^2_{m,t}$ are increments of market volatility, $\sigma^2_X = \mathbb{V}ar (\Delta \sigma^2_{m})$,  $a=0.5$ and $b=-0.03$.

\subsection{Coskewness and Cokurtosis}
Following work of \cite{10.2307/2326275, harvey2000conditional, https://doi.org/10.1111/1540-6261.00425, ang2006downside}, we estimate the coskewness and cokurtosis as
\begin{align}
CSK_i &\equiv \frac{\mathbb{E}[(r_i - \mu_i) (r_m - \mu_m)^2]} {\sqrt{\mathbb{E}[(r_i - \mu_i)^2]} \mathbb{E}[(r_m - \mu_m)^2]}, \\
CKT_i &\equiv \frac{\mathbb{E}[(r_i - \mu_i) (r_m - \mu_m)^3]} {\sqrt{\mathbb{E}[(r_i - \mu_i)^2]} \mathbb{E}[(r_m - \mu_m)^{3/2}]} \, .
\end{align}

\subsection{Fama-French Three-Factor Model}
Betas of the three-factor model of \cite{fama1993common} are estimated via time-series regression of excess asset return on three factors: SMB (obtained by sorting stocks based on their size),  HML (obtained by sorting stocks based on their book-to-market vale) and MKT (market factor)
\begin{align}
r_{i,t} = \alpha_i + \beta_i^{SMB} SMB_t + \beta_i^{HML} HML_t + \beta_i^{MKT} MKT_t + e_{i,t} .
\end{align}
Factor data were obtained from \href{http://mba.tuck.dartmouth.edu/pages/faculty/ken.french/data_library.html}{Kenneth French's online data library}.


\newpage
\clearpage
\section{Detailed Description of the Portfolio Results}
\label{app:portfolios_detailed}

\subsection{Fama-French Portfolios}

In this section, we employ two sets of Fama-French portfolios. First set contains two samples: 25 portfolios double-sorted on size and value and 30 industry portfolios. These two datasets were chosen because they posses the longest history available across all the Fama-French portfolios. Their time span ranges between July 1926 and April 2020. Second set contains three datasets of portfolios sorted on the following characteristics: operating profit, investment and book-to-market. Portfolios sorted on operating profit and investment posses significantly shorter history of observations between July 1963 and March 2020.

Regarding the first dataset, the results are summarized in Table \ref{tab:coef_ff}. In the case of portfolios double sorted on size and value, the short component of QS and short component of EVR risks are priced. Regarding the industry sorted portfolios, only the short term TR is consistently priced across the model specifications. For those investors who fear the high volatility states, these results suggest that the more appropriate strategy involves investing based on the industries rather than size and value, as you do not have to pay a premium for portfolios that posses low EVR betas - portfolios whose extreme negative returns are less probable to co-occur with extreme positive increments of market volatility.

\begin{sidewaystable}[ph!]
\scriptsize
\centering
\caption{\textit{Fama-French Long History Portfolios.} Prices of risk estimated on monthly return data of 30 industry portfolios and portfolios double sorted on size and book-to-market. Sample period covers time interval between July 1926 and April 2020. Long horizon is given by frequencies corresponding to 3-year cycle and longer. Below the coefficients, we include Fama-MacBeth $t$-statistics.}
\begin{tabular}{cccccccccccccccc}
 \toprule
	 	& & \multicolumn{4}{c}{Tail market risk} & \multicolumn{4}{c}{Extreme volatility risk} & \multicolumn{6}{c}{Full model}\\
		\cmidrule(r){2-6} \cmidrule(r){7-10} \cmidrule(r){11-16}
  & $\tau$ & $\lambda_{\text{long}}^{\text{TR}}$ & $\lambda_{\text{short}}^{\text{TR}}$ & $\lambda^{\text{CAPM}}$ & RMSPE & $\lambda_{\text{long}}^{\text{EV}}$ & $\lambda_{\text{short}}^{\text{EV}}$ & $\lambda^{\text{CAPM}}$ & RMSPE & $\lambda^{\text{TR}}_{\text{\text{long}}}$ & $\lambda^{\text{TR}}_{\text{short}}$ & $\lambda^{\text{EV}}_{\text{long}}$ & $\lambda^{\text{EV}}_{\text{short}}$ & $\lambda^{\text{CAPM}}$ & RMSPE \\ 
\cmidrule(r){2-6} \cmidrule(r){7-10} \cmidrule(r){11-16}
\multirow{12}{*}{\shortstack[c]{25 portfolios \\ sorted on size and value}} & 0.01 & 0.060 & 0.502 & 0.645 & 2.557 & 0.047 & 0.425 & 0.707 & 2.602 & -0.002 & 0.483 & 0.076 & 0.320 & 0.615 & 2.545 \\ 
   &  & 0.461 & 1.249 & 3.618 &  & 0.395 & 0.950 & 3.251 &  & -0.010 & 1.438 & 0.354 & 1.393 & 2.722 &  \\ 
   & 0.05 & -0.486 & 3.170 & 0.704 & 2.258 & 0.231 & -2.263 & 0.586 & 2.520 & -0.569 & 3.041 & -0.093 & -0.848 & 0.831 & 2.230 \\ 
   &  & -1.257 & 6.276 & 3.990 &  & 0.685 & -3.315 & 1.460 &  & -1.875 & 5.590 & -0.291 & -1.088 & 2.491 &  \\ 
   & 0.1 & 0.597 & 0.270 & 0.607 & 2.535 & -0.252 & 5.020 & 0.843 & 2.496 & 0.505 & 0.445 & 0.016 & 3.783 & 0.523 & 2.422 \\ 
   &  & 1.259 & 0.243 & 3.536 &  & -0.511 & 3.067 & 1.771 &  & 1.152 & 0.415 & 0.043 & 3.358 & 1.496 &  \\ 
   & 0.15 & -0.954 & 3.584 & 0.826 & 2.473 & -0.892 & 6.249 & 1.142 & 2.011 & -1.015 & 2.112 & -0.946 & 6.342 & 1.318 & 1.866 \\ 
   &  & -2.112 & 4.345 & 4.607 &  & -2.326 & 5.690 & 3.224 &  & -2.366 & 2.549 & -2.636 & 6.463 & 4.177 &  \\ 
   & 0.2 & -0.349 & 1.421 & 0.739 & 2.615 & -0.606 & 9.894 & 0.858 & 2.001 & -0.850 & 2.598 & -0.606 & 10.489 & 0.860 & 1.922 \\ 
   &  & -0.827 & 1.054 & 4.140 &  & -1.592 & 5.712 & 2.824 &  & -1.920 & 1.958 & -1.486 & 6.393 & 2.935 &  \\ 
   & 0.25 & -0.271 & 1.630 & 0.717 & 2.594 & -0.216 & 5.347 & 0.770 & 2.390 & -0.932 & 1.242 & 0.065 & 6.623 & 0.674 & 2.319 \\ 
   &  & -0.800 & 1.143 & 4.137 &  & -0.626 & 2.398 & 2.839 &  & -1.952 & 1.106 & 0.172 & 4.443 & 2.573 &  \\
\cmidrule(r){2-6} \cmidrule(r){7-10} \cmidrule(r){11-16}

\multirow{12}{*}{\shortstack[c]{30 industry \\ portfolios}} & 0.01 & -0.132 & 0.874 & 0.712 & 1.411 & -0.208 & -0.740 & 1.007 & 1.897 & 0.020 & 0.823 & -0.155 & -0.402 & 0.781 & 1.373 \\ 
   &  & -1.077 & 3.823 & 3.231 &  & -1.611 & -1.909 & 4.160 &  & 0.086 & 3.356 & -0.597 & -1.011 & 2.833 &  \\ 
   & 0.05 & 0.036 & 0.903 & 0.675 & 1.898 & 0.445 & -2.141 & 0.443 & 1.878 & -0.162 & 0.791 & 0.390 & -1.728 & 0.447 & 1.792 \\ 
   &  & 0.101 & 1.975 & 3.274 &  & 1.357 & -2.558 & 1.296 &  & -0.326 & 1.878 & 0.690 & -2.027 & 1.022 &  \\ 
   & 0.1 & 0.208 & 1.044 & 0.651 & 1.848 & 0.722 & -2.173 & 0.261 & 1.720 & -0.504 & 0.545 & 0.875 & -1.972 & 0.209 & 1.683 \\ 
   &  & 0.592 & 1.680 & 3.286 &  & 3.348 & -1.969 & 0.986 &  & -0.838 & 0.809 & 1.825 & -1.751 & 0.604 &  \\ 
   & 0.15 & 0.561 & 1.135 & 0.607 & 1.704 & 0.930 & -2.461 & 0.246 & 1.612 & 0.147 & 0.774 & 0.647 & -2.147 & 0.358 & 1.564 \\ 
   &  & 1.843 & 1.594 & 3.044 &  & 3.644 & -2.162 & 0.929 &  & 0.278 & 1.122 & 1.303 & -1.947 & 1.080 &  \\ 
   & 0.2 & 0.602 & 0.442 & 0.665 & 1.917 & 0.840 & -1.770 & 0.384 & 1.784 & 0.177 & 0.431 & 0.674 & -1.970 & 0.430 & 1.767 \\ 
   &  & 1.714 & 0.568 & 3.388 &  & 2.398 & -1.535 & 1.437 &  & 0.414 & 0.555 & 1.434 & -1.688 & 1.468 &  \\ 
   & 0.25 & 0.823 & 0.926 & 0.648 & 1.734 & 0.652 & -2.961 & 0.550 & 1.806 & 0.879 & 0.883 & 0.046 & -2.418 & 0.673 & 1.611 \\ 
   &  & 2.093 & 1.221 & 3.239 &  & 1.993 & -2.469 & 2.191 &  & 2.152 & 1.157 & 0.123 & -1.984 & 2.628 &  \\   
%
   \bottomrule
\end{tabular}
\label{tab:coef_ff}
\end{sidewaystable}

The second set of portfolios include equities sorted on operating profit, investments and book-to-market. The results are given in the Table \ref{tab:coef_ff_other}. Generally, short TR is priced across these portfolios with the expected sign. On the other hand, using the portfolios sorted on investment, there is a strong negative relation between long TR and asset returns, which may seem unintuitive. Regarding the EVR, its short term part is priced across investment portfolios and book-to-market portfolios.

\begin{sidewaystable}[ph!]
\scriptsize
\centering
\caption{\textit{Fama-French Portfolios.} Prices of risk estimated on monthly return data of portfolios sorted on operating profit, investment and book-to-market. Sample period covers time interval between July 1963 (July 1926 for book-to-market portfolios) and March 2020. Long horizon is given by frequencies corresponding to 3-year cycle and longer. Below the coefficients, we include Fama-MacBeth $t$-statistics.}
\begin{tabular}{cccccccccccccccc}
 \toprule
	 	& & \multicolumn{4}{c}{Tail market risk} & \multicolumn{4}{c}{Extreme volatility risk} & \multicolumn{6}{c}{Full model}\\
		\cmidrule(r){2-6} \cmidrule(r){7-10} \cmidrule(r){11-16}
  & $\tau$ & $\lambda_{\text{long}}^{\text{TR}}$ & $\lambda_{\text{short}}^{\text{TR}}$ & $\lambda^{\text{CAPM}}$ & RMSPE & $\lambda_{\text{long}}^{\text{EV}}$ & $\lambda_{\text{short}}^{\text{EV}}$ & $\lambda^{\text{CAPM}}$ & RMSPE & $\lambda^{\text{TR}}_{\text{\text{long}}}$ & $\lambda^{\text{TR}}_{\text{short}}$ & $\lambda^{\text{EV}}_{\text{long}}$ & $\lambda^{\text{EV}}_{\text{short}}$ & $\lambda^{\text{CAPM}}$ & RMSPE \\ 
\cmidrule(r){2-6} \cmidrule(r){7-10} \cmidrule(r){11-16}
\multirow{12}{*}{\shortstack[c]{Operating profit}} & 0.01 & 0.266 & 0.844 & 0.337 & 0.417 & 0.353 & -0.329 & 0.467 & 0.663 & 0.097 & 1.012 & 0.185 & -0.388 & 0.244 & 0.265 \\ 
   &  & 1.217 & 1.929 & 1.462 &  & 1.603 & -1.014 & 1.668 &  & 0.333 & 2.738 & 0.466 & -1.264 & 0.700 &  \\ 
   & 0.05 & -0.301 & 1.308 & 0.669 & 0.920 & 0.684 & -2.969 & 0.422 & 0.723 & -0.355 & 0.306 & 0.782 & -1.790 & 0.317 & 0.625 \\ 
   &  & -0.692 & 2.328 & 3.255 &  & 2.673 & -2.070 & 1.617 &  & -0.728 & 0.618 & 2.164 & -1.951 & 0.987 &  \\ 
   & 0.1 & 0.119 & -1.008 & 0.747 & 0.976 & 0.844 & -1.697 & 0.142 & 0.701 & 0.018 & 0.394 & 0.986 & -2.376 & 0.037 & 0.688 \\ 
   &  & 0.315 & -1.461 & 3.608 &  & 1.493 & -0.783 & 0.318 &  & 0.037 & 0.945 & 1.939 & -1.705 & 0.093 &  \\ 
   & 0.15 & 0.262 & -0.933 & 0.709 & 1.041 & 0.803 & -1.173 & 0.263 & 0.986 & -0.807 & -0.974 & 1.804 & -2.295 & -0.104 & 0.820 \\ 
   &  & 0.696 & -1.041 & 3.335 &  & 2.318 & -1.426 & 1.008 &  & -0.751 & -1.458 & 2.108 & -2.814 & -0.255 &  \\ 
   & 0.2 & -1.191 & 0.226 & 0.884 & 0.886 & 1.284 & -4.962 & 0.147 & 0.849 & -1.646 & 0.742 & 1.662 & -3.192 & 0.121 & 0.579 \\ 
   &  & -1.379 & 0.340 & 4.024 &  & 3.769 & -2.240 & 0.577 &  & -2.360 & 1.415 & 4.102 & -1.850 & 0.471 &  \\ 
   & 0.25 & -0.947 & 1.579 & 0.725 & 0.914 & 0.195 & -2.627 & 0.599 & 0.589 & -0.332 & -0.049 & 0.356 & -2.294 & 0.541 & 0.575 \\ 
   &  & -0.913 & 1.638 & 3.511 &  & 0.511 & -1.525 & 2.166 &  & -0.433 & -0.075 & 1.418 & -2.138 & 2.185 &  \\
\cmidrule(r){2-6} \cmidrule(r){7-10} \cmidrule(r){11-16}

\multirow{12}{*}{\shortstack[c]{Investment}} & 0.01 & 0.234 & 2.347 & -0.145 & 1.219 & 1.093 & -0.047 & -0.114 & 3.332 & 0.768 & 2.230 & -0.774 & 0.729 & 0.196 & 0.933 \\ 
   &  & 2.228 & 7.966 & -0.626 &  & 5.265 & -0.251 & -0.405 &  & 1.009 & 3.649 & -1.126 & 3.077 & 0.374 &  \\ 
   & 0.05 & -1.855 & 5.496 & 0.668 & 1.333 & -1.674 & -1.749 & 1.995 & 2.535 & -2.088 & 5.045 & 0.035 & 1.831 & 0.593 & 1.279 \\ 
   &  & -4.548 & 5.866 & 3.130 &  & -9.143 & -1.320 & 7.719 &  & -4.508 & 8.124 & 0.151 & 2.755 & 2.487 &  \\ 
   & 0.1 & -4.729 & 4.393 & 1.274 & 2.007 & 0.950 & 7.075 & -0.330 & 2.960 & -5.437 & 3.225 & -0.552 & 0.004 & 1.902 & 1.984 \\ 
   &  & -5.902 & 3.724 & 5.048 &  & 1.964 & 9.647 & -0.699 &  & -7.711 & 4.848 & -1.215 & 0.007 & 3.911 &  \\ 
   & 0.15 & -2.556 & 7.013 & 0.811 & 2.068 & -2.463 & 16.677 & 1.543 & 2.440 & -5.158 & 8.173 & 4.022 & 14.513 & -1.794 & 1.366 \\ 
   &  & -6.042 & 7.252 & 3.908 &  & -8.222 & 7.779 & 5.729 &  & -5.133 & 9.396 & 4.581 & 7.545 & -3.767 &  \\ 
   & 0.2 & -4.208 & 14.158 & 0.559 & 1.746 & -3.982 & 10.646 & 2.676 & 2.933 & -1.998 & 13.989 & -2.161 & 10.860 & 1.174 & 1.301 \\ 
   &  & -8.644 & 8.388 & 2.439 &  & -7.310 & 6.644 & 6.950 &  & -1.360 & 6.574 & -1.248 & 2.824 & 1.387 &  \\ 
   & 0.25 & -5.272 & 2.109 & 1.236 & 1.770 & 0.216 & 7.497 & 0.527 & 3.296 & -4.753 & 3.109 & -0.416 & 2.632 & 1.341 & 1.681 \\ 
   &  & -9.910 & 1.765 & 4.718 &  & 0.428 & 4.862 & 1.257 &  & -9.098 & 3.275 & -0.863 & 1.710 & 3.095 &  \\   
\cmidrule(r){2-6} \cmidrule(r){7-10} \cmidrule(r){11-16}
   
\multirow{12}{*}{\shortstack[c]{Book-to-market}}  & 0.01 & 0.783 & 0.378 & 0.249 & 1.537 & 1.045 & 0.830 & -0.323 & 2.178 & 2.016 & -1.162 & -1.923 & -1.355 & 1.793 & 1.141 \\ 
   &  & 4.365 & 1.142 & 1.220 &  & 4.243 & 2.022 & -1.130 &  & 3.866 & -2.861 & -3.316 & -3.587 & 3.971 &  \\ 
   & 0.05 & 1.909 & 2.898 & 0.061 & 1.386 & -0.682 & -6.138 & 1.661 & 2.020 & 2.821 & 2.012 & -0.957 & 2.715 & 0.669 & 1.148 \\ 
   &  & 3.552 & 2.593 & 0.274 &  & -1.912 & -4.116 & 3.755 &  & 3.917 & 2.384 & -2.543 & 2.847 & 1.886 &  \\ 
   & 0.1 & 2.997 & -1.702 & 0.276 & 1.449 & -0.672 & 5.498 & 1.240 & 2.791 & 3.098 & -3.078 & -1.303 & -3.900 & 1.558 & 1.159 \\ 
   &  & 4.274 & -2.004 & 1.368 &  & -1.457 & 2.332 & 2.571 &  & 4.323 & -3.434 & -2.526 & -2.110 & 3.180 &  \\ 
   & 0.15 & 2.979 & 1.100 & 0.147 & 1.504 & -1.844 & 0.164 & 2.095 & 2.378 & 5.562 & -6.472 & -0.957 & 8.492 & 0.252 & 1.063 \\ 
   &  & 4.180 & 0.971 & 0.693 &  & -3.473 & 0.147 & 4.427 &  & 4.697 & -4.100 & -2.493 & 3.918 & 0.772 &  \\ 
   & 0.2 & 1.940 & 5.670 & 0.272 & 2.043 & -1.110 & 8.519 & 1.166 & 2.151 & 1.055 & 4.724 & -0.066 & 4.433 & 0.335 & 1.999 \\ 
   &  & 3.102 & 4.096 & 1.333 &  & -2.942 & 3.797 & 4.017 &  & 1.799 & 3.161 & -0.133 & 2.907 & 0.908 &  \\ 
   & 0.25 & -0.046 & 8.320 & 0.442 & 2.212 & -0.956 & 9.493 & 1.193 & 2.160 & -1.799 & 6.455 & -0.477 & 8.514 & 0.893 & 2.006 \\ 
   &  & -0.072 & 3.362 & 2.283 &  & -2.841 & 3.578 & 4.198 &  & -2.185 & 3.050 & -1.778 & 3.276 & 3.536 &  \\ 
   \bottomrule
\end{tabular}
\label{tab:coef_ff_other}
\end{sidewaystable}

\subsection{Other Portfolios}

In this section, we provide analysis of QS risk performed on other widely used datasets. The estimated models are reported in Table \ref{tab:coef_various}. First, we focus on portfolios employed in \cite{lettau2014conditional}. This dataset contains portfolios formed across multiple asset classes. First, the dataset contains 6 currency portfolios sorted on interest rate differential (we exclude high inflation currencies similar to the approach of \cite{lettau2014conditional}). Second, we have 5 commodity futures portfolios sorted on basis. Third, we include returns on 5 corporate bond portfolios sorted on credit spread. And fourth, we have equity portfolios sorted on various characteristics (6 double sorted on size and value, 5 on CAPM beta, 5 on industry, 6 double sorted on momentum and size).\footnote{We do not include option portfolios because they have short history starting in 1986, which is not suitable for our analysis.} Here, we present results for the aggregated dataset. This dataset was introduced to show the usefulness of downside risk beta for pricing. From the results we can conclude that the short component of TR for most $\tau$ threshold values is priced using the aggregated dataset. Its long component is significant for some medium values of $\tau$. Regarding the EVR, its short term component for lower values of $\tau$ is priced as well.

Second, we look at the equity portfolios sorted on cash flow duration proposed in \cite{WEBER2018486}. The results can be found in the second section of Table \ref{tab:coef_various}. Similarly as in the previous case, short term part of TR is priced across these portfolios. On the other hand, its long term part is negatively priced across these assets, which may be counterintuitive. The EVR is priced using its both components.

Finally, we use returns on factors constructed from various asset classes from \cite{ilmanen2021factor}. This dataset was chosen because of its long history and because it spans many asset classes including U.S. and international equities, fixed income assets, currencies and commodities using value, momentum, carry, defensive and multi-style type of investment strategy. We report the results in the third panel of Table \ref{tab:coef_various}. We can see that using the TR model, the long term TR is priced, and both parts of EVR are priced. But if we look at the results of the Full model, only the EVR coefficients remain consistently significant.

\begin{sidewaystable}[ph!]
\scriptsize
\centering
\caption{\textit{Various Portfolios.} Prices of risk estimated on monthly data of various datasets. Models are estimated for various values of thresholds given by $\tau$. Long horizon is given by frequencies corresponding to 3-year cycle and longer. Below the coefficients, we include Fama-MacBeth $t$-statistics.}
\begin{tabular}{cccccccccccccccc}
 \toprule
	 	& & \multicolumn{4}{c}{Tail market risk} & \multicolumn{4}{c}{Extreme volatility risk} & \multicolumn{6}{c}{Full model}\\
		\cmidrule(r){2-6} \cmidrule(r){7-10} \cmidrule(r){11-16}
  & $\tau$ & $\lambda_{\text{long}}^{\text{TR}}$ & $\lambda_{\text{short}}^{\text{TR}}$ & $\lambda^{\text{CAPM}}$ & RMSPE & $\lambda_{\text{long}}^{\text{EV}}$ & $\lambda_{\text{short}}^{\text{EV}}$ & $\lambda^{\text{CAPM}}$ & RMSPE & $\lambda^{\text{TR}}_{\text{\text{long}}}$ & $\lambda^{\text{TR}}_{\text{short}}$ & $\lambda^{\text{EV}}_{\text{long}}$ & $\lambda^{\text{EV}}_{\text{short}}$ & $\lambda^{\text{CAPM}}$ & RMSPE \\ 
\cmidrule(r){2-6} \cmidrule(r){7-10} \cmidrule(r){11-16}
\multirow{12}{*}{\shortstack[c]{\cite{lettau2014conditional}}} & 0.01 & 0.744 & 0.126 & 0.455 & 8.739 & 0.562 & 0.115 & 0.112 & 9.116 & 0.953 & -0.042 & -0.135 & -0.420 & 0.600 & 8.702 \\ 
   &  & 3.703 & 0.225 & 1.792 &  & 3.080 & 0.225 & 0.399 &  & 3.238 & -0.074 & -0.484 & -0.855 & 1.894 &  \\ 
   & 0.05 & 0.428 & 4.432 & 0.590 & 8.724 & 0.738 & 1.779 & 0.264 & 9.634 & 0.059 & 4.368 & 0.268 & 0.378 & 0.501 & 8.708 \\ 
   &  & 0.956 & 3.801 & 2.294 &  & 1.838 & 1.575 & 0.968 &  & 0.126 & 3.565 & 0.515 & 0.333 & 1.603 &  \\ 
   & 0.1 & 0.784 & 3.342 & 0.602 & 9.988 & 1.029 & 0.109 & 0.248 & 9.935 & -0.847 & 4.122 & 1.468 & -1.817 & 0.202 & 9.693 \\ 
   &  & 1.344 & 2.138 & 2.359 &  & 1.945 & 0.071 & 0.858 &  & -1.824 & 2.579 & 2.593 & -1.154 & 0.668 &  \\ 
   & 0.15 & 0.627 & 7.934 & 0.578 & 9.408 & 1.213 & 1.165 & 0.293 & 10.037 & -0.117 & 8.497 & 0.906 & -1.273 & 0.392 & 9.367 \\ 
   &  & 1.243 & 3.951 & 2.241 &  & 1.783 & 0.657 & 0.999 &  & -0.257 & 3.551 & 1.116 & -0.622 & 1.263 &  \\ 
   & 0.2 & 0.990 & 11.846 & 0.487 & 8.925 & 0.519 & 2.222 & 0.481 & 10.623 & 2.392 & 10.492 & -2.392 & 2.494 & 0.954 & 8.525 \\ 
   &  & 1.871 & 4.994 & 1.939 &  & 0.746 & 1.004 & 1.639 &  & 2.514 & 5.140 & -2.560 & 1.176 & 2.799 &  \\ 
   & 0.25 & 1.187 & 10.080 & 0.459 & 9.502 & 0.700 & 1.154 & 0.455 & 10.705 & 1.842 & 9.434 & -1.674 & 3.161 & 0.767 & 9.295 \\ 
   &  & 2.208 & 3.700 & 1.809 &  & 0.956 & 0.503 & 1.541 &  & 2.538 & 3.594 & -1.944 & 1.544 & 2.353 &  \\
\cmidrule(r){2-6} \cmidrule(r){7-10} \cmidrule(r){11-16}

\multirow{12}{*}{\shortstack[c]{\cite{WEBER2018486}}} & 0.01 & 0.611 & 1.702 & 0.008 & 1.428 & 0.826 & 8.684 & -0.796 & 1.460 & 0.852 & 1.246 & -0.795 & 6.825 & -0.123 & 0.695 \\ 
   &  & 2.809 & 4.083 & 0.033 &  & 3.741 & 3.537 & -2.297 &  & 1.779 & 2.466 & -1.568 & 2.545 & -0.281 &  \\ 
   & 0.05 & -2.640 & 4.875 & 1.027 & 2.864 & 1.953 & -3.555 & -0.249 & 4.277 & -2.562 & 5.166 & 2.319 & -3.806 & -0.211 & 1.318 \\ 
   &  & -4.897 & 7.174 & 4.416 &  & 4.587 & -4.026 & -0.696 &  & -4.832 & 7.206 & 5.026 & -4.633 & -0.641 &  \\ 
   & 0.1 & -4.297 & 3.906 & 1.361 & 3.989 & 0.827 & 9.226 & -0.426 & 3.270 & 0.316 & 3.719 & 2.061 & 8.453 & -1.611 & 3.190 \\ 
   &  & -5.230 & 3.027 & 5.394 &  & 1.262 & 6.569 & -0.669 &  & 0.704 & 3.318 & 3.698 & 6.560 & -2.939 &  \\ 
   & 0.15 & 0.257 & 8.307 & 0.155 & 3.620 & -0.829 & 10.041 & 0.840 & 3.099 & -0.568 & 5.073 & 0.858 & 11.976 & -0.447 & 2.906 \\ 
   &  & 0.783 & 7.019 & 0.696 &  & -2.711 & 5.768 & 2.299 &  & -1.005 & 5.208 & 1.369 & 4.526 & -0.885 &  \\ 
   & 0.2 & -2.453 & 9.974 & 0.580 & 3.804 & -0.838 & 6.443 & 1.052 & 4.413 & -2.801 & 15.421 & 3.088 & 13.817 & -1.799 & 2.779 \\ 
   &  & -4.107 & 6.689 & 2.658 &  & -1.926 & 6.203 & 2.645 &  & -4.146 & 6.730 & 4.330 & 7.357 & -3.280 &  \\ 
   & 0.25 & -3.435 & 8.407 & 0.658 & 3.870 & 1.904 & 2.376 & -0.463 & 4.693 & -5.417 & 8.747 & 4.081 & 8.589 & -1.818 & 3.097 \\ 
   &  & -5.066 & 6.948 & 2.989 &  & 5.773 & 2.107 & -1.514 &  & -6.571 & 7.086 & 6.850 & 6.427 & -4.362 &  \\    
\cmidrule(r){2-6} \cmidrule(r){7-10} \cmidrule(r){11-16}
   
\multirow{12}{*}{\shortstack[c]{\cite{ilmanen2021factor}}} & 0.01 & 2.401 & 1.279 & -0.528 & 20.445 & 2.279 & 3.100 & -0.440 & 19.912 & 1.087 & 1.338 & 1.086 & 2.842 & -0.424 & 19.407 \\ 
   &  & 11.322 & 0.988 & -1.714 &  & 11.091 & 2.741 & -1.484 &  & 1.076 & 0.991 & 1.094 & 2.455 & -1.420 &  \\ 
   & 0.05 & 6.482 & -0.905 & 0.058 & 19.846 & 4.840 & 5.149 & -0.779 & 20.155 & 4.160 & -1.209 & 1.709 & 3.082 & -0.283 & 19.650 \\ 
   &  & 12.922 & -0.367 & 0.196 &  & 10.085 & 2.232 & -2.446 &  & 2.146 & -0.477 & 0.952 & 1.342 & -0.541 &  \\ 
   & 0.1 & 7.660 & 1.560 & 0.377 & 24.118 & 5.830 & 6.750 & -0.729 & 22.877 & 4.715 & 0.885 & 1.916 & 6.173 & -0.038 & 23.208 \\ 
   &  & 11.398 & 0.480 & 1.487 &  & 10.703 & 1.834 & -2.309 &  & 2.781 & 0.266 & 1.304 & 1.657 & -0.102 &  \\ 
   & 0.15 & 8.689 & -0.554 & 0.438 & 24.783 & 6.698 & 12.464 & -0.763 & 23.232 & 3.026 & -0.202 & 4.250 & 10.625 & -0.335 & 23.133 \\ 
   &  & 11.734 & -0.167 & 1.803 &  & 9.759 & 2.889 & -2.407 &  & 2.416 & -0.059 & 3.628 & 2.430 & -1.045 &  \\ 
   & 0.2 & 9.876 & -2.295 & 0.368 & 27.105 & 10.214 & 1.875 & -0.756 & 22.503 & -0.278 & -4.394 & 10.922 & 1.825 & -1.075 & 22.170 \\ 
   &  & 11.540 & -0.632 & 1.487 &  & 10.411 & 0.374 & -2.359 &  & -0.259 & -1.188 & 9.470 & 0.346 & -3.447 &  \\ 
   & 0.25 & 8.151 & 3.455 & 0.557 & 30.032 & 10.378 & 20.757 & -0.502 & 28.317 & 1.925 & 0.215 & 8.146 & 19.229 & -0.268 & 27.793 \\ 
   &  & 10.336 & 0.925 & 2.305 &  & 9.569 & 4.311 & -1.594 &  & 1.781 & 0.057 & 6.879 & 3.642 & -0.969 &  \\
   \bottomrule
\end{tabular}
\label{tab:coef_various}
\end{sidewaystable}

%

\end{document}